\documentclass{article}
\usepackage{filecontents}
\usepackage{geometry}
\usepackage{xr-hyper}
\usepackage{hyperref}
\usepackage[utf8]{inputenc}
\usepackage{amsmath}
\usepackage{amssymb}
\usepackage{mathtools}

\usepackage[abs]{overpic}
\usepackage{subfig}

\makeatletter
\newcommand*{\addFileDependency}[1]{
  \typeout{(#1)}
  \@addtofilelist{#1}
  \IfFileExists{#1}{}{\typeout{No file #1.}}
}
\makeatother

\newcommand*{\myexternaldocument}[1]{%
    \externaldocument{#1}%
    \addFileDependency{#1.tex}%
    \addFileDependency{#1.aux}%
}

\myexternaldocument{Supplementary}

\def \ve{\varepsilon}

\begin{document}

\title{Discrete-to-continuum models of pre-stressed cytoskeletal filament networks}

\author{J.~Köry$^{1,*}$, N.~A.~Hill$^{1}$, X.~Y.~Luo$^{1}$ and P.~S.~Stewart$^{1}$}


\date{%
    $^1$School of Mathematics and Statistics, University of Glasgow, University Place, Glasgow G12 8QQ, UK\\
    $^*$Corresponding author: jakub.koery@glasgow.ac.uk\\[2ex]
    \today
}

\maketitle

\providecommand{\keywords}[1]{\textbf{\textit{Keywords: }} #1}

\begin{abstract}
We introduce a mathematical model for the mechanical behaviour of the eukaryotic cell cytoskeleton. This discrete model involves a regular array of pre-stressed protein filaments that exhibit resistance to enthalpic stretching, joined at crosslinks to form a network. Assuming that the inter-crosslink distance is much shorter than the lengthscale of the cell, we upscale the discrete force balance to form a continuum system of governing equations and deduce the corresponding macroscopic stress tensor. We use these discrete and continuum models to analyse the imposed displacement of a bead placed in the domain, characterising the cell rheology through the force-displacement curve. We further derive an analytical approximation to the stress and strain fields in the limit of small bead radius, predicting the net force required to generate a given deformation and elucidating the dependency on the microscale properties of the filaments. We apply these models to networks of the intermediate filament vimentin and demonstrate good agreement between predictions of the discrete, continuum and analytical approaches. In particular, our model predicts that the network stiffness increases sublinearly with the filament pre-stress and scales logarithmically with the bead size.
\end{abstract}




\section{Introduction}

Eukaryotic cells exhibit a complicated rheology in response to mechanical stimuli, arising primarily through deformation of their cytoskeleton, a complex network of crosslinked filamentous proteins including actin filaments, microtubules and intermediate filaments (e.g. vimentin). In addition to stretching of the filaments themselves, the system also dissipates energy both through transport of viscous fluid through this network and through transient crosslink (CL) dynamics \cite{Mofrad2006,Alberts2017,Vaziri2008}. Depending on the rate at which the deformation is applied, cells have been shown to behave as visco-elastic, soft-glassy, or poro-elastic materials \cite{Sato1996,Fabry2001,Moeendarbary2013, hu2017}. This complex rheology underpins a wide variety of cellular behaviour including migration and growth. In particular, epithelial cells can undergo an epithelial-mesenchymal transition, where these cells disassemble their cytoskeleton to become migratory \cite{Alberts2017}. Such transitions underpin healthy growth and development during embryogenesis and tissue repair \cite{Ahmed2015,Serrano2021}, but also accompany progression of tumour cells towards more aggressive (i.e. invasive) phenotypes. Hence, a thorough knowledge of cell rheology (and in particular its mechanical properties) is a likely pre-requisite for successful anti-cancer treatments \cite{Alberts2017,Thiery2009}. 




Tensegrity models of the cell cytoskeleton postulate that certain elements are pre-stretched which must be balanced by other elements under compression \cite{Ingber2003}. It is now well established that both actin and vimentin filaments \emph{in vivo} are pre-stretched (i.e. under tension) \cite{Mofrad2006,Charras2002}.
On the other hand, microtubules have been shown to bear significant compressive loads \cite{Wang2001,brangwynne2006microtubules} due to their large bending stiffness. Although actin and microtubules have generally attracted more attention in the literature, the intermediate filament vimentin also greatly impacts cell mechanics due to its capacity to withstand very large strains (especially in comparison with actin and microtubules) \cite{Hu2019,Patteson2020}. 

Most models describing the mechanical behaviour of individual cytoskeletal filaments have been derived using the theory of semi-flexible polymer chains \cite{meng2017}, incorporating not only their elastic stretching and bending, but also uncoiling of their undulations under an applied stress \cite{pritchard2014}. 
As result, the distance between two ends of the filament differs from its stress-free contour length, so models relate the axial force applied to one end of the filament to the end-to-end distance normalized with respect to the contour length \cite{meng2017}. Similar relationships have also been derived based on the theory of Cosserat rods \cite{holzapfel2011,holzapfel2013}. 

Due to the complexity of cell cytoplasm \emph{in vivo}, \emph{in silico} approaches are useful to elucidate the mechanisms underlying the mechanical behaviour on the network scale. Existing mathematical models typically fall into two categories. Discrete models of cell mechanics (including molecular dynamics simulations) enable the inclusion of detailed biophysics on the microscale derived from first principles, but also contain large number of discrete elements and their interactions which makes them computationally expensive \cite{Kim2009a, Lee2009,Muller2015,Muller2016,Kim2018}. On the other hand, continuum models of cell mechanics are typically much less computationally demanding, allowing fast parameter sweeps, but, because they are proposed to match macroscopic (i.e. cell-scale) phenomena, the manner in which microscale (molecular-scale) parameters and processes influence the macroscale response is often unclear \cite{Vaziri2008}.

The mechanical response of crosslinked networks of semiflexible filaments (e.g. actin or collagen) subject to various loading configurations has been studied using discrete network models elucidating key length and energy scales \cite{Head2003,Head2005,Han2018,Berthier2022}. 
Under bulk deformations (uniaxial or shear strain), the dominant modes of deformation -- material stretching, entropic stretching and bending -- have been linked to the regions of affine and non-affine deformations in the parameter space consisting of the filament length and the crosslink density \cite{Head2003}. A similar approach has been subsequently used to mimic localized perturbations in cytoskeletal networks via point forces applied at a single crosslink \cite{Head2005}. Local deformations were further explored in recent years, modelling the stress stiffening of extracellular matrices induced by contractile cells pulling on the adjacent fibers \cite{Han2018,Berthier2022}. However, at high filament densities encountered \emph{in vivo}, the discrete simulations become computationally expensive \cite{Head2003} and as the networks are typically highly disordered, there is no simple and reliable way to derive the corresponding continuum  (computationally faster) model. Furthermore, scaling arguments do not account for the \emph{in vivo} network pre-stress discussed above which makes direct utilization of the deduced power laws impossible.

The vast majority of macroscale continuum models are inferred by ensemble averaging based on polymer physics \cite{unterberger2014advances,meng2017}. The models stemming from rubber elasticity form the oldest and largest group, including chain, full-network and microsphere models \cite{Flory1943,Wang1952,Arruda1993,Treloar1979,Miehe2004}. The latter have been applied to actin networks resulting in hyper-elastic and visco-elastic constitutive models \cite{Unterberger2013Hyperelastic,Unterberger2013viscoelasticity,Holzapfel2014}. Other approaches utilized Doi--Edwards theory \cite{Storm2005} or the effective medium approach \cite{Broedersz2008}. Discrete lattice models have also been employed, but to the best of our knowledge, rigorous upscaling techniques have not been used to derive a macroscale model. It is also worth noting that these discrete lattices often have unrealistic topologies - models using triangular lattices with coordination number $6$ are not representative of crosslinked cytoskeletal networks and further care is needed to achieve a biologically realistic node connectivity \cite{Broedersz2011a,Broedersz2012}. Efforts involving more rational and rigorous mathematical methods (such as discrete-to-continuum upscaling or homogenization) to systematically bridge between these two approaches are still largely missing. This problem also pertains to collagen networks where predictions of discrete and continuum models often disagree \cite{chandran2006affine,stracuzzi2022risky}. 

Rational mathematical modelling has been successfully applied to study dynamic aspects of cytoskeletal reorganization during cell motility, including the dynamics of actin, myosin and other crosslinking proteins at the leading edge. This approach leads to mathematical formulations that are often amenable to analytical study and can provide explicit solutions, e.g. predicting the dependency of cell velocity on properties of the substrate \cite{Mogilner2002,Othmer2019}. However, such rational techniques have seldom been applied to study mechanics of crosslinked networks.

Recent research has focused on the effective transport properties of cytoplasm as a porous medium \cite{Haspinger2021,Oelz2021}; as a result, the forces generated within the cytoskeleton as it is deformed by the transported object remain incompletely understood. The force required to move a spherical object (bead) inside a living cell was recently measured using the optical tweezers, elucidating dependence on key parameters such as bead size and pulling velocity \cite{hu2017,Hu2019}. The primary goal of current study is to formalize these dependencies using a theoretical model built from first principles. To this end, we develop a multiscale framework for mechanical response during prescribed motion of an internal organelle or bead which rationally encodes a state-of-the-art microscale constitutive law for the axial stretching of individual semi-flexible filaments.

The paper is organized as follows. First, in Section \ref{sec:SecDiscreteModel} we introduce a discrete model of the cell cytoskeleton consisting of a two-dimensional crosslinked network with prescribed displacement of a set of CLs. In Section \ref{sec:Upscaling} we upscale this discrete force balance using discrete-to-continuum asymptotics, arrive at a macroscale continuum model equipped with appropriate boundary conditions and infer the corresponding stress tensor and strain-energy density. In Section \ref{sec:ResultsDiscrAndCont} we compare simulations of the discrete and continuum models and numerically explore how net force exerted on the transported bead depends on key model parameters. In Section \ref{sec:SmallDefsSmallBead} we consider the limit of small deformations in the continuum problem and compute an asymptotic approximation to the net force as a function of bead displacement, valid whenever the bead size is much smaller than the macroscopic length scale. 



\section{Discrete model and nondimensionalization}
\label{sec:SecDiscreteModel}

\subsection{Initial network}
\label{sec:SubsecReferenceNetwork}

\subsubsection{Geometry}
\begin{figure}[ht!]
\centering
\includegraphics[width=1.0\textwidth]{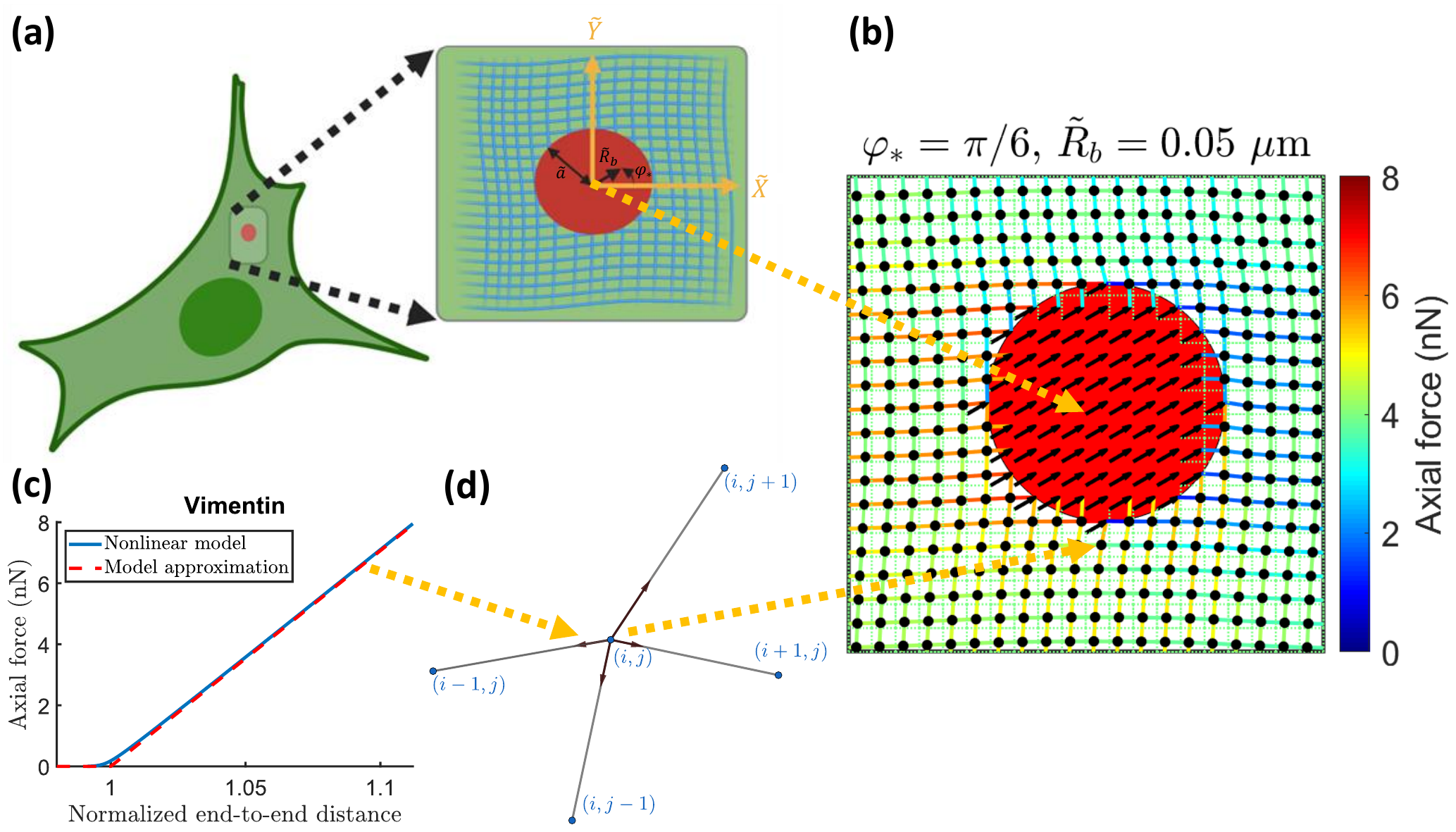}
\caption{Panel (a) shows a cell schematic with a small inserted bead (red). Zooming onto the bead, we idealize the undeformed cytoskeleton as a regular grid of curved filaments. Displacing the bead by a distance $\tilde{R}_{b}$ at an angle $\varphi_*$, we compute the locations of all crosslinks (black dots) in the perturbed network, as shown in panel (b). The calculation is based on a realistic microscale constitutive law for axial response of individual FSs (panel c) and assumes local force balance at CL $(i,j)$ (panel d) with contributing forces drawn as black arrows. Panel (c) also documents that the equation \eqref{eqn:LinearSpringsDimlessSmallTau1NoCompression} provides an excellent approximation to model \eqref{eqn:Blundell} for forces below tensile strength  using default parameters for vimentin as estimated in Supplementary Section \ref{sec:SuppDefaultDimensionalParametersActinVimentin}.}
\label{fig:DiscreteModelSetUp}
\end{figure}

We consider a planar square region within a eukaryotic cell of fixed side length $\tilde{D}$, well away from the nucleus and the cell membrane (Figure \ref{fig:DiscreteModelSetUp}a). This region is parameterized by coordinates $\tilde{X}$ and $\tilde{Y}$, along the two edges of the square with origin at the centre. Focusing on mesh-forming crosslinking proteins (e.g. filamin) we propose a simple model assuming that the cytoskeleton can be modelled as a square grid of semi-flexible filaments. Although this arrangement is highly idealized, it facilitates a formal upscaling. Initially the filaments are assumed to be equally spaced and are oriented (after averaging out microscale fluctuations) parallel to either the $\tilde{X}$ or $\tilde{Y}$ axes (blue lines in Figure \ref{fig:DiscreteModelSetUp}a), with crosslinks (CLs) at their intersections, forming a regular two-dimensional grid. These CLs divide each filament into $N$ filament segments (FSs). Initially these crosslinks are a distance ${\tilde R} = \tilde{D}/N$ apart, so crosslink $(i,j)$ is located at
\begin{equation}\label{eqn:referenceDimensional}
\tilde{\boldsymbol{X}}_{i,j} = \left(\tilde{X}_{i}, \tilde{Y}_{j} \right)  = \left(i,j \right) \tilde{R}, \quad \text{ where } i,j = - \tfrac{1}{2}N, -\tfrac{1}{2}N+1, ..., \tfrac{1}{2}N-1, \tfrac{1}{2}N;
\end{equation}
we assume that $N$ is even for simplicity.
Note that throughout this work, tildes denote dimensional variables and parameters.

At subcellular scales, thermal effects play an important role causing undulations in cytoskeletal filaments even in the absence of external force. As result, a FS connecting arbitrary two neighbouring CLs need not be straight and its end-to-end distance need not be equal to its contour length (or arclength). For simplicity, we assume that all filaments are of the same stress-free contour length $\tilde{L}$, with the stress-free contour length of FSs being $\tilde{\Lambda} = \tilde{L}/N$, noting that these two quantities are typically distinct from the domain size $\tilde{D}$ and inter-CL distance $\tilde{R}$. Our model contains a relatively large number of parameters; for convenience, our notation is summarized in Section \ref{sec:SuppNotation} of Supplementary Material.

\subsubsection{Pre-stretch}

In later sections, we specialize our modelling framework to actin and vimentin networks; tensegrity models of the cytoskeleton postulate that these elements are typically pre-stretched \cite{Ingber2003}. For a fixed $\tilde{R}$, the filament pre-stretch is controlled by the normalized end-to-end distance
\begin{equation}\label{eqn:DefOfXi}
\xi= \frac{\tilde{R}}{\tilde{\Lambda}} = \frac{\tilde{D}}{\tilde{L}},
\end{equation}
which generates an axial force due to pre-stress denoted $\tilde{f}_p$. 
Although the macroscale pre-stress has been measured experimentally \cite{Wang2001,Wang2002,Stamenovic2004}, the complexity of cytoskeleton \emph{in vivo} (the number of different filaments and crosslinks and their interactions) makes it difficult to estimate $\tilde{f}_p$ and so this will be considered a free parameter (similar to previous studies, e.g. \cite{Coughlin2003}). The corresponding values of $\tilde{\Lambda}$ and $\xi$ then follow from the microscale constitutive law for the axial force discussed in the next section.

In experiments, the macroscale pre-stress is usually estimated by measuring the total force exerted on a particular surface within the cell, and then normalizing by the cross-sectional area of that surface \cite{Wang2001}. Applying an analogous method to the boundary of our square domain, we estimate the macroscale pre-stress of our filament networks by summing the force exerted by each of the adjoining filaments on that boundary and dividing by the boundary length. In this way, we estimate the total macroscale pre-stress as
$$ \tilde{\sigma}_p = \frac{(N-1) \tilde{f}_p}{\tilde{D}}. $$

\subsection{Deformed network}

\subsubsection{Applied deformation}

As a model for optical tweezers experiments \cite{hu2017, Hu2019}, we consider the motion of circular bead of radius $\tilde{a}$ initially placed at the origin of the domain (Figure \ref{fig:DiscreteModelSetUp}b). In this paper we restrict attention to quasi-static deformations, neglecting inertia and assuming zero net force on every CL for all time. In this simple model, we assume that CLs are free to rotate with no unfolding, unbinding, breakage or slippage. Thus, the energy supplied by the prescribed motion of the bead is stored as elastic energy in the filament network. The deformed coordinates of CL $(i,j)$ are denoted as $ \tilde{\boldsymbol{x}}_{i,j}= (\tilde{x}_{i,j}, \tilde{y}_{i,j})$. 

\subsubsection{Implicit microscale constitutive law for axial force in a filament segment}
\label{sec:SubSecTensileForces}

We denote $\tilde{r}$ as the distance between CLs after deformation. In this study we follow models for semi-flexible filaments, and let the axial force ${\tilde f}$ in each FS be a function of the end-to-end (straight-line) distance between its two end points normalized by its stress-free contour length $r = {\tilde r}/{\tilde \Lambda}$ (Figure \ref{fig:DiscreteModelSetUp}c) \cite{holzapfel2013,meng2017}. Thus, tortuosity of individual FSs is accounted for implicitly. We use a well-established constitutive law for a single semi-flexible filament under tension, which includes the interplay between thermal undulations, bending stiffness and material extensibility \cite{Blundell2009, meng2017}, in the form
\begin{equation}\label{eqn:Blundell}
\frac{\tilde{r}}{\tilde{\Lambda}} = r(\tilde{f};\tilde{\Lambda}) = \left( 1 + \frac{\tilde{f}}{\pi \tilde{Y} \tilde{b}^2} \right) \left( 1 - \sqrt{\frac{\tilde{k}_B \tilde{T}}{\pi  \tilde{\Lambda}_p \left(\tilde{f} + \left(\pi^2 \tilde{k}_B \tilde{T} \tilde{\Lambda}_p/\tilde{\Lambda}^2 \right) \right)}} \right) 
\end{equation}
is the Euler buckling threshold force, $\tilde{k}_B \approx 1.38 \times 10^{-23} \text{m}^2 \text{kg s}^{-2} \text{K}^{-1}$ is the Boltzmann constant, $\tilde{T} = 300$ K is the absolute temperature, $\tilde{Y}$ is the Young's modulus, $\tilde{\Lambda}_p$ is the persistence length and $\tilde{b}$ is the radius of the filament under consideration. The constitutive law \eqref{eqn:Blundell} for an individual filament assumes that the stress-free contour length $\tilde{\Lambda}$ and the end-to-end distance $\tilde{r}$ are comparable (i.e. the normalized end-to-end distance $r$ is close to $1$, \cite{pritchard2014}). The first factor in \eqref{eqn:Blundell} accounts for extensibility of the material while the second factor constitutes a model for an inextensible filament balancing thermal effects with its bending stiffness. Fixing all material parameters and substituting the initial values for $\tilde{r} = \tilde{R}$ and $\tilde{f} = \tilde{f}_p$ provides an implicit relationship between $\tilde{f}_p$ and $\tilde{\Lambda}$. Note that for extensible filaments, direct inversion to obtain $\tilde{f}$ as a function of $r$ is cumbersome \cite{holzapfel2013,meng2017}. For a detailed description of the energy stored in individual FSs, see Supplementary Section \ref{sec:SuppAxialForcesDetails}.


\subsection{Force balance at a crosslink}

The local force balance at each CL requires that the net force (Figure \ref{fig:DiscreteModelSetUp}d) must be zero \cite{Heussinger2007,Blundell2011}. 
As the forces equilibrate at every CL, it follows that the total moment of forces about any CL is also zero. Note that apart from the axial forces, one would typically also need to introduce restoring forces due to the resistance of filaments to bending \cite{Blundell2011}. However due to the combination of high filament density and the imposed pre-stretch of actin and vimentin in our model, the response will be dominated by the elastic stretching and the bending can be neglected \cite{Head2003,pritchard2014,Broedersz2011a,Broedersz2011b}. 



\subsection{Boundary conditions}

All CLs on the outer boundary of the domain are assumed to be pinned, mimicking attachment to the membrane, nucleus or some other organelle. The bead is assumed to be at least as large as the mesh size (typically much larger, in line with the optical tweezers experiments \cite{hu2017}) and therefore a hole of appropriate shape and size must be extracted from the discrete network. To mimick a rigid body translation, we model the bead motion via an imposed displacement of all CLs 
within the initial outline of the bead by a distance $\tilde{R}_{b}$ at a pulling angle $\varphi_*$ measured anti-clockwise from the $\tilde{X}$ axis.

\subsection{Baseline parameter values}
\label{sec:BaselineParValues}

We identify baseline parameter values representative of the cytoskeleton and denote these with the subscript $c$. For instance, we choose a baseline filament spacing as $\tilde{R}_{c} = 0.05$ $\mu$m which, fixing the domain size as $\tilde{D} = 5$ $\mu$m, means that every filament is divided into $N_{c}=100$ FSs \cite{hu2017}. All other model parameters are listed and the corresponding values representative of the cytoskeleton are estimated in Supplementary Material (Section \ref{sec:SuppDefaultDimensionalParametersActinVimentin}). 

To ensure consistency as we vary the number of filaments, in simulations we hold the domain size and the total volume of filaments fixed to the baseline values by adjusting the mesh spacing and the filament radius according to
$$\tilde{R} = \frac{N_{c}}{N}  \tilde{R}_{c}, \qquad \tilde{b} = \sqrt{\frac{N_{c}}{N}} \tilde{b}_{c}.$$
Similarly, we hold the macroscale pre-stress fixed by adjusting the filament pre-stress and analogously rescale the axial force at arbitrary $r$ according to
$$\tilde{f}_p = \frac{N_c}{N} \tilde{\mathcal{F}}_p, \qquad \tilde{f}(r) = \frac{N_{c}}{N} \tilde{\mathcal{F}}(r),$$
where $\tilde{\mathcal{F}}(\xi) = \tilde{\mathcal{F}}_p$.

\subsection{Nondimensionalization}
\label{sec:DiscrToCont}

We nondimensionalize all lengths based on the domain side length $\tilde{D}$, and forces (including $\tilde{\mathcal{F}}_p$) with respect to the enthalpic (elastic) force $\pi \tilde{Y} \tilde{b}_{c}^2$. We denote as $l_{i \pm 1/2, j}$ and $l_{i,j \pm 1/2}$ the deformed lengths of FSs connecting CL $(i,j)$ to CLs $(i \pm 1, j)$ and $(i, j \pm 1)$, respectively. At CL $(i,j)$, we define unit vectors pointing in the directions of the four adjacent FSs as
$$ \hat{\boldsymbol{r}}_{i \pm \frac12, j} =  \frac{ \left( x_{i \pm 1,j} - x_{i,j}, y_{i \pm 1,j} - y_{i,j} \right)}{l_{i \pm \frac12,j}}, \qquad \hat{\boldsymbol{r}}_{i, j \pm \frac12} = \frac{ \left( x_{i,j \pm 1} - x_{i,j}, y_{i,j \pm 1} - y_{i,j} \right) }{l_{i,j \pm \frac12}}, $$
and upon multiplying by $\varepsilon_{c} N$, where $\varepsilon_{c} = N_{c}^{-1}$, the dimensionless force balance takes the form
\begin{equation}\label{eqn:DimlessForceBalance}
\begin{aligned}
\boldsymbol{0} = & 
\mathcal{F} \left( \xi N  l_{i-\frac12,j} \right) \hat{\boldsymbol{r}}_{i - \frac12, j} + \mathcal{F} \left( \xi N l_{i+\frac12,j}\right) \hat{\boldsymbol{r}}_{i + \frac12, j} + 
\mathcal{F} \left( \xi N l_{i,j-\frac12} \right) \hat{\boldsymbol{r}}_{i, j - \frac12} + \mathcal{F} \left( \xi N l_{i,j+\frac12}\right) \hat{\boldsymbol{r}}_{i, j + \frac12}.
\end{aligned}
\end{equation}
The dimensionless 
magnitude of the bead displacement is denoted as $R_{b}:= \tilde{R}_{b}/\tilde{D}$.

\subsection{Analysis of dimensionless microscale constitutive law}

The dimensionless constitutive law for an individual filament \eqref{eqn:Blundell} becomes
\begin{equation}\label{eqn:dimlessBlundellBest}
r(\mathcal{F}; \mathcal{T}_1, \mathcal{T}_2, \xi, \varepsilon_{c}, N) = \left(1+ \mathcal{F} \right) \left( 1 - \sqrt{\frac{\mathcal{T}_1}{\mathcal{F}/(\varepsilon_{c} N) + 4 \pi^3 \left( \varepsilon_{c} \xi N  \mathcal{T}_2 \right)^2 \mathcal{T}_1}} \right),
\end{equation}
where
\begin{equation}\label{eqn:dimlessBlundellBestParametersMacroscale}
\mathcal{T}_1 = \frac{\tilde{\mathcal{F}}_{\mathrm{entropic}}}{\tilde{\mathcal{F}}_{\mathrm{enthalpic}}}  = \frac{\tilde{k}_B \tilde{T}}{\pi^2 \tilde{Y} \tilde{b}_{c}^2 \tilde{\Lambda}_p} \qquad \text{and} \qquad \mathcal{T}_2 = \frac{\tilde{\Lambda}_p}{2 \tilde{R}_{c}}
\end{equation}
are the dimensionless ratios of the entropic force ($\tilde{\mathcal{F}}_{\mathrm{entropic}} = \tilde{k}_B \tilde{T}/(\pi \tilde{\Lambda}_p)$) to the enthalpic force ($\tilde{\mathcal{F}}_{\mathrm{enthalpic}} = \pi \tilde{Y} \tilde{b}_{c}^2$) and one half of the ratio of the persistence length to the end-to-end distance, respectively\footnote{The factor of $1/2$ was chosen in line with previous studies so that our $\mathcal{T}_2$ is a direct analogue of the so-called normalized filament stiffness \cite{meng2017}. Note further that the $r$ introduced in \eqref{eqn:Blundell} should be regarded as a distance normalized with respect to the stress-free contour length and even though without units, this quantity is distinct from the nondimensionalized end-to-end distance (with respect to the macroscale).}. 
Note that all dimensionless parameters featuring in \eqref{eqn:dimlessBlundellBest} are independent of the force due to pre-stress $\mathcal{F}_p$ and $N$, with the exception of $\xi$. Given that $\mathcal{F}(\xi)=\mathcal{F}_p$, we obtain
\begin{equation}\label{eqn:dimlessBlundellBestReference}
\xi = \left\{1+\mathcal{F}_p\right\} \left\{ 1 - \left( \ve_{c} N \right)^{-1} \left( \mathcal{F}_p  \left( \varepsilon_{c} N \right)^{-3} \mathcal{T}_1^{-1} + 4 \pi^3  \mathcal{T}_2^2  \xi^2 \right)^{-1/2} \right\},
\end{equation}
which provides a quartic polynomial for pre-stretch $\xi$ as function of $\mathcal{F}_p$, which cannot easily be inverted analytically. However, for vimentin filaments we compute $\mathcal{T}_1 \approx 1.9 \times 10^{-8}$ and $\mathcal{T}_2 \approx 10$ (based on parameters listed in Table \ref{table_parameters} in Supplementary Material) and so provided $\ve_{c} N =O(1)$ and $\mathcal{F}_p \gg \mathcal{T}_1$ (i.e. the force due to pre-stress is much greater than the entropic force; for $\ve_{c} N \gg 1$ we do not need any additional conditions) we approximate
\begin{equation}\label{eqn:TwoTermApproximationPreStretchPrestress}
\xi = 1 + \mathcal{F}_p.
\end{equation}
The approximation \eqref{eqn:TwoTermApproximationPreStretchPrestress} is not sufficiently accurate for actin, since $\mathcal{T}_1 \approx 10^{-9}$ but $\mathcal{T}_2 \approx 170$ (i.e. the persistence length of actin is much larger than the representative cytoskeletal mesh size), and an expansion in powers of $\mathcal{T}_2^{-1}$ is required (see Supplementary Material, Section \ref{sec:SubSecExplicitPreStretchPreStress}).

In the main text we focus attention on networks of vimentin filaments. We further consider networks of actin filaments in Supplementary Material (Section \ref{sec:SubsecBrittlenessActin}), although here the critical stretch for filament breakage is typically very low and so the networks quickly disassemble.

In summary, the dimensionless problem is governed by eight dimensionless parameters ($\mathcal{T}_1$, $\mathcal{T}_2$, $\mathcal{F}_p$, $\varepsilon_{c}$, $N$, $R_{b}$, $\varphi_*$, $a$) and the microscale constitutive law \eqref{eqn:dimlessBlundellBest}, where $\xi$ is given by \eqref{eqn:TwoTermApproximationPreStretchPrestress}. Model parameters and their default values are listed in Supplementary Material Section \ref{sec:SuppDefaultDimensionalParametersActinVimentin}.

\subsection{Negligible response to compression and simplified microscale constitutive law}
\label{sec:SubsecLinearizedMicroscaleBehaviour}

Neither actin nor vimentin filaments can sustain large compressive stresses due to their low bending stiffness \cite{meng2017}. We therefore assume that the response to compression is negligible, 
similar to previous studies for actin networks \cite{Unterberger2013Hyperelastic,Holzapfel2014} (vimentin filaments have even lower bending stiffness). Given the weak response of filaments to compression and also the smallness of $\mathcal{T}_1$ discussed in the previous section, we can neglect the square root term in \eqref{eqn:dimlessBlundellBest} and using \eqref{eqn:TwoTermApproximationPreStretchPrestress} derive a simplified microscale constitutive law \eqref{eqn:dimlessBlundellBest} in the form

\begin{equation}\label{eqn:LinearSpringsDimlessSmallTau1NoCompression}
\mathcal{F} = 
\Big\{
    \begin{array}{lr}
        \mathcal{F}_p + \left(r- \xi \right)= r-1, & \text{if } r > 1\\ [5pt]
        0, & \text{if } r<1,
    \end{array}
\end{equation}
which is continuous at $r=1$, i.e. when the filament is straightened out to its full contour length ($\tilde{r}=\tilde{\Lambda}$), as \eqref{eqn:TwoTermApproximationPreStretchPrestress} holds. Note that in the case of vimentin, this linearized expression was not obtained via Taylor expansion of the full model \eqref{eqn:dimlessBlundellBest} about $r=\xi$, but was instead derived rationally based on the smallness of $\mathcal{T}_1$; it is analogous to previous models studying mechanics of pre-stressed filament networks \cite{Coughlin2003}.
Equation \eqref{eqn:LinearSpringsDimlessSmallTau1NoCompression} provides a very good approximation to \eqref{eqn:Blundell} using parameters pertinent to the intermediate filament vimentin (Figure \ref{fig:DiscreteModelSetUp}c) across all values of $r$. The model \eqref{eqn:LinearSpringsDimlessSmallTau1NoCompression} will be used in the sections of this paper where we present discrete and continuum simulations for vimentin. Note that it is possible, in principle, to simulate networks where filaments are modelled using \eqref{eqn:Blundell} in its full form, but numerical simulations take significantly longer due to its implicit form. The multiscale continuum framework developed in Section \ref{sec:Upscaling} below can account for arbitrary microscale constitutive law relating the axial force to the end-to-end distance. In the following section we will assume that axial stretching on the microscale is governed by a general constitutive law $\mathcal{F}(r)$.


\section{Upscaling and continuum model}
\label{sec:Upscaling}

\subsection{Upscaling the force balance}

We now define a small parameter $\ve \equiv N^{-1} \ll 1$, the (dimensionless) undeformed CL-to-CL distance. We upscale the discrete model \eqref{eqn:DimlessForceBalance} in the limit $\ve \to 0$ to form a continuum model. We assume that there exist smooth functions $x(X,Y)$ and $y(X,Y)$ defined on the square domain $- \tfrac{1}{2}< X,Y < \tfrac{1}{2}$ such that for all $i,j$ we have $g(X_i, Y_j) = g_{i,j}$ where $g$ is either $x$ or $y$. Assuming $x$, $y$ and $\mathcal{F}$ are sufficiently smooth, we Taylor expand the discrete equations \eqref{eqn:DimlessForceBalance} (centering about $(X_i,Y_j)$) and rationally derive a continuum model \cite{Barry2022}. Further details of the derivation can be found in Supplementary Material (Section \ref{sec:appDiscreteToCont}). The first non-trivial balance in the momentum equations gives
\begin{equation}\label{eqn:EffectiveGeneralNonlinearFilamentNoBending}
\left(\mathcal{F} \left( \xi \sqrt{x_X^2 +y_X^2} \right)  \frac{(x_X,y_X)}{\sqrt{x_X^2 + y_X^2}} \right)_X + \left( \mathcal{F} \left( \xi \sqrt{x_Y^2 +y_Y^2} \right)  \frac{(x_Y,y_Y)}{\sqrt{x_Y^2 + y_Y^2}}\right)_Y = \boldsymbol{0} ,
\end{equation}
where subscripts denote partial derivatives. 
This system of two coupled nonlinear equations in the divergence form constitutes the upscaled problem. Notice that in the (continuum) $N \to \infty$ limit, the constitutive law \eqref{eqn:dimlessBlundellBest} converges to $\mathcal{F} = r - 1$ which is identical to \eqref{eqn:LinearSpringsDimlessSmallTau1NoCompression}; the resulting equations under the linearized microscale constitutive law are deduced in Supplementary Section \ref{sec:SuppDeduceContinuumLinearSprings}.

The momentum balance equations \eqref{eqn:EffectiveGeneralNonlinearFilamentNoBending} are consistent with other classical results in continuum mechanics (see Supplementary Material, Section \ref{sec:StrainEnergyFiberReinforced}, for details). Since these equations are expressed in divergence form, we can define $F_{kl} = \partial x_k/\partial X_l$ to be the components of the corresponding deformation gradient tensor and immediately deduce the nominal stress tensor in the form
\begin{gather}\label{eqn:StressTensor1}
\tilde{\boldsymbol{S}} = \frac{1}{\tilde{R}_{c}}
\begin{pmatrix}
  \displaystyle \tilde{\mathcal{F}} \left( \xi \sqrt{F_{11}^2 + F_{21}^2} \right) \frac{F_{11}}{\sqrt{F_{11}^2 + F_{21}^2}}  &
  \displaystyle  \tilde{\mathcal{F}} \left( \xi \sqrt{F_{11}^2 + F_{21}^2} \right) \frac{F_{21}}{\sqrt{F_{11}^2 + F_{21}^2}}  \\
  \displaystyle \tilde{\mathcal{F}} \left( \xi \sqrt{F_{12}^2 + F_{22}^2} \right) \frac{F_{12}}{\sqrt{F_{12}^2 + F_{22}^2}}  &
  \displaystyle \tilde{\mathcal{F}} \left( \xi \sqrt{F_{12}^2 + F_{22}^2} \right) \frac{F_{22}}{\sqrt{F_{12}^2 + F_{22}^2}}
\end{pmatrix}.
\end{gather}
This formulation is a special case (reflecting the particular geometry of the undeformed configuration) of the stress tensor derived for an arbitrary distribution of filament directions using the Doi--Edwards construction \cite{Storm2005}. In the initial configuration, $\boldsymbol{F}= \boldsymbol{I}$ and therefore $\tilde{\boldsymbol{S}} = \tilde{\mathcal{F}}_p/\tilde{R}_c \boldsymbol{I}$, consistent with our prediction of macroscale pre-stress in Section \ref{sec:SubsecReferenceNetwork}. Similarly, we conclude that the dimensional strain energy density in the deformed configuration is
\begin{equation}\label{eqn:StrainEnergyAsFunctionOfCInvariants}
\tilde{W}(\boldsymbol{C})= \frac{ \tilde{\mathcal{E}} \left( \xi \sqrt{I_4(\boldsymbol{C})} \right) + \tilde{\mathcal{E}} \left( \xi \sqrt{I_6(\boldsymbol{C})} \right) }{\tilde{R}_{c}^2},
\end{equation}
where $\tilde{\mathcal{E}}$ denotes the energy stored in elastic stretching of the filaments (see Supplementary Material, Section \ref{sec:SuppAxialForcesDetails}), $\boldsymbol{C}$ is the right Cauchy--Green deformation tensor and $\sqrt{I_4(\boldsymbol{C})}$ and $\sqrt{I_6(\boldsymbol{C})}$ represent local stretches in $X$ and $Y$ directions, reflecting the underlying square-grid geometry of the cytoskeleton with two preferred filament directions. Such anisotropic contributions to the strain energy are often proposed in phenomenological models for fiber-reinforced materials (e.g. \cite{Spencer2014}).

\subsection{Boundary conditions}

The pinning of the outer layer of CLs in the discrete model gives in the continuum limit
\begin{equation}\label{eqn:ContinuumBCOld}
x(X,Y) = X, \quad y(X,Y) = Y
\end{equation}
along all boundaries characterized by $X = \pm \tfrac{1}{2}$ or $Y = \pm \tfrac{1}{2}$. In the continuum model, the bead is represented by a disc of radius $\tilde{a}$ cut out from the domain, initially centred at $(\tilde{X},\tilde{Y}) = (0,0)$ and displaced by $\tilde{R}_{b}$ at a pulling angle $\varphi_*$. Note that in the dimensionless setting, we must have $a = \tilde{a}/\tilde{D} = O(1)$. The bead boundary condition is written for $-\pi < \varphi \leq \pi$ as
\begin{equation}\label{eqn:ContinuumCentral}
x(a \cos{(\varphi)},a \sin{(\varphi)}) = a \cos{(\varphi)} + R_{b} \cos{(\varphi_*)}, \qquad y(a \cos{(\varphi)},a \sin{(\varphi)}) = a \sin{(\varphi)} + R_{b} \sin{(\varphi_*)}.
\end{equation}

\section{Discrete and continuum simulations}
\label{sec:ResultsDiscrAndCont}

To facilitate direct comparison between the discrete and continuum predictions, we return to the dimensional variables and introduce the continuum displacement fields
\begin{equation}\label{eqn:displacementsContinuum}
\tilde{u}(\tilde{X},\tilde{Y}) = \tilde{x}(\tilde{X},\tilde{Y}) - \tilde{X} \qquad \tilde{v}(\tilde{X},\tilde{Y}) = \tilde{y}(\tilde{X},\tilde{Y}) - \tilde{Y},
\end{equation}
as well as their discrete counterparts
\begin{equation}\label{eqn:displacements}
\tilde{u}_{i,j} = \tilde{x}_{i,j} - \tilde{X}_{i} \qquad \tilde{v}_{i,j} = \tilde{y}_{i,j} - \tilde{Y}_{j}
\end{equation}
for all $i$ and $j$. Unless stated otherwise, all lengths (including those indicated in colorbars) are given in microns and all forces in nanonewtons.

In quasi-static simulations of the discrete model \eqref{eqn:DimlessForceBalance}, we use numerical continuation from the initial configuration to find steady-state solutions for a variety of bead displacements. In order to avoid pulling along the initial direction of one of the filaments or exactly along the diagonal, we choose a default pulling angle as $\varphi_* = \pi/6$. To avoid FSs crossing each other, we only displace the bead up to a maximal distance equal to the undeformed mesh size, i.e. $0 \leq \tilde{R}_{b} \leq \tilde{R}$. For every $\tilde{R}_{b}$, we solve for the locations of CLs outside the bead using \texttt{fsolve} toolbox in \textsc{MATLAB} (based on Newton's method) and then calculate the resultant force acting on the bead by summing up tensile forces from all attached FSs.

The continuum problem \eqref{eqn:EffectiveGeneralNonlinearFilamentNoBending} is solved in FEniCS using Newton solver and we employed Lagrange finite elements of degree $1$ \cite{logg2012}. As the difference in the predicted force on the bead using default model parameters, maximum displacement and domain resolutions (the minimum number of elements across the square in both $\tilde{X}$ and $\tilde{Y}$ directions) equal to $200$ and $400$ was less than $0.3 \%$ of the value at $200$, we conclude that the resolution $400$ provides us with a sufficiently fine mesh giving trustworthy force estimates. We use this value as default from now onwards.
In the continuum model, the net force acting on the bead is then found by numerically integrating the traction ($\tilde{\boldsymbol{S}}^T \boldsymbol{N}$ where $\boldsymbol{N}$ is the unit normal to the bead) over the bead boundary.

\subsection{Simulation with default parameters for vimentin}

\begin{figure}[ht!]
    \centering
    \includegraphics[width=1.0\textwidth]{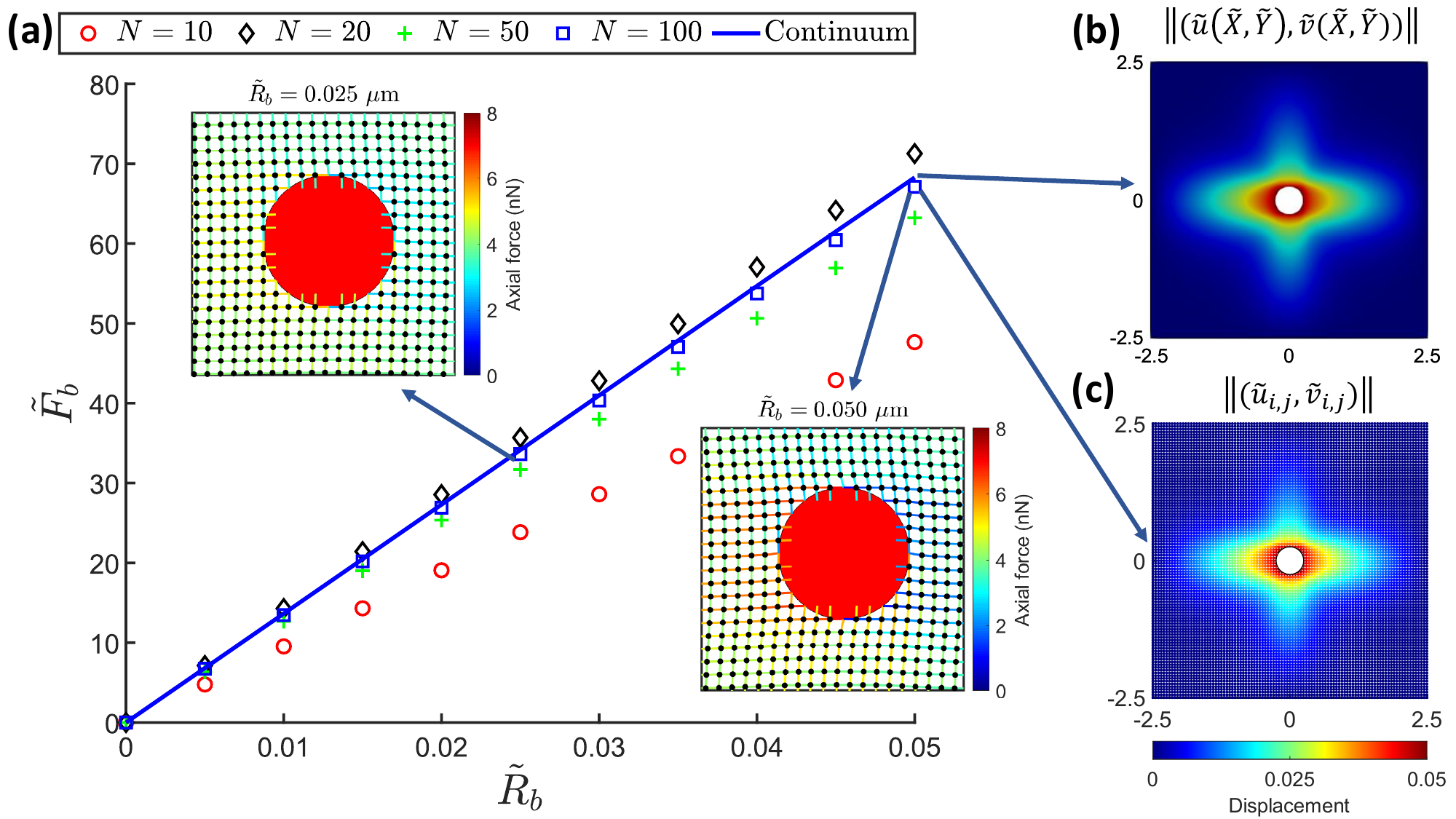}
    \caption{(a) The force-displacement graphs for increasing $N$ in the discrete model (symbols) with default parameters converge to that for the continuum limit (solid blue). The insets depict solution profiles for $\tilde{R}_{b} = 0.025 \mu$m and $0.05 \mu$m. Panels on the right show the magnitude of the displacement in the undeformed configuration in the continuum (b) and discrete (c) model (the latter visualized as a scatter plot).}
\label{fig:IncreaseNDefaultParsWithDisplacement}
\end{figure}

In order to assess the convergence of discrete simulations to the continuum as $N \to \infty$, in Figure \ref{fig:IncreaseNDefaultParsWithDisplacement} we plot the force-displacement graphs for various $N$, fixing all other parameters at their default value (including $\varepsilon_{c}=1/100$). In each case the graph of the magnitude of the force as a function of bead displacement (termed the force-displacement curve) is almost perfectly linear because we restrict attention to (small) deformations up to a single mesh size. For every given displacement the discrete and continuum predictions of the force approach one another as $N$ becomes large (Figure \ref{fig:IncreaseNDefaultParsWithDisplacement}a) and the results are almost indistinguishable for $N=1/R_{c}=100$. Note that the convergence is not monotonic for small $N$, but this is an artifact caused by the relatively small number of FSs attached to the bead in such cases. In order to elucidate how the steady state force distribution changes with increasing bead displacement, in the insets of Figure \ref{fig:IncreaseNDefaultParsWithDisplacement}(a) we show the accumulation of tension in the wake of the moving bead. The solution profiles for such dense network ($N=100$) are not easy to visualize and throughout this work we will therefore zoom onto a small region in the vicinity of the bead where the perturbation is localized. The magnitude of the continuum displacement field $\left|\left|\left(\tilde{u}, \tilde{v}\right) \right| \right|$ (Figure \ref{fig:IncreaseNDefaultParsWithDisplacement}b) shows good agreement with its discrete counter-part (Figure \ref{fig:IncreaseNDefaultParsWithDisplacement}c). Note that the near-perfect symmetries of these fields with respect to the $\tilde{X}$ and $\tilde{Y}$ axes can be explained by the smallness of the deformation: while the nonlinear system \eqref{eqn:EffectiveContinuumDivergenceDimensional_Displacements} does not suggest any symmetry, the structure of the small-deformations limit \eqref{eqn:EffectiveGeneralNonlinearFilamentNoBendingXODeltaK=0}-\eqref{eqn:SmallDefCentralDisplHole} derived in Section \ref{sec:SmallDefsSmallBead} (together with the symmetries of the domain under consideration) indicate that both components of the displacement field must be even functions of $\tilde{X}$ ($\tilde{Y}$) for a fixed $\tilde{Y}$ ($\tilde{X}$). In summary, this figure shows that the discrete and continuum predictions are in excellent agreement with one another as the mesh spacing reduces.

\subsection{Effect of model parameters}
\label{sec:Subsec_EffectOfPars}
In this section we explore dependency on model parameters, namely the pulling angle $\varphi_*$ (Section \ref{eqn:SubSubDependencyPullingAngle}) and the force due to pre-stress $\tilde{\mathcal{F}}_p$ (Section \ref{eqn:SubSubDependencyPreStress}).

\subsubsection{Pulling angle $\varphi_*$}
\label{eqn:SubSubDependencyPullingAngle}

\begin{figure}[ht!]
\centering
\includegraphics[width=1.0\textwidth]{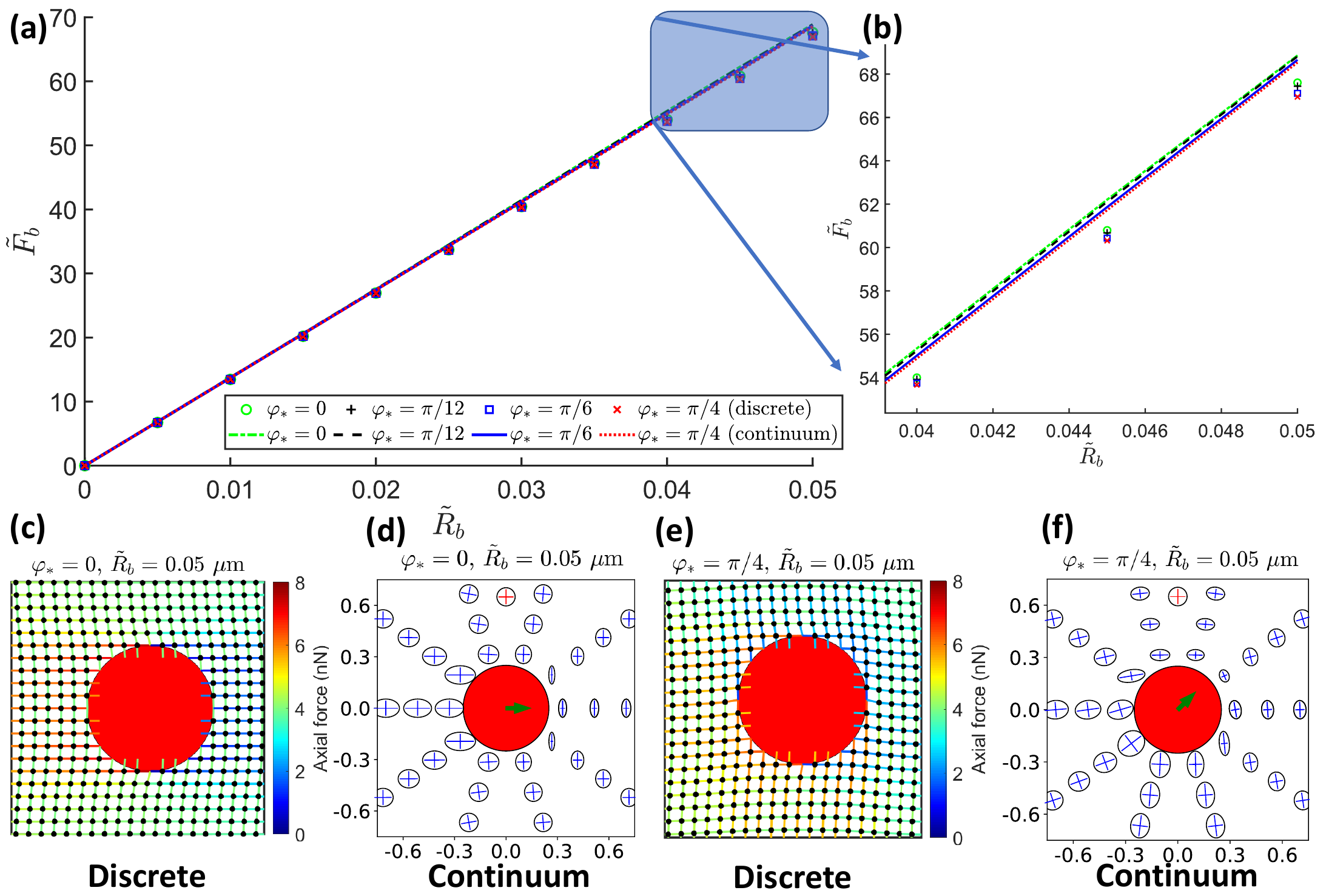}
\caption{In panel (a), discrete (symbols) and continuum (lines) force-displacement graphs are plotted for default model parameters and for pulling angles $0$ (green), $\pi/12$ (black), $\pi/6$ (blue) and $\pi/4$ (red) radians. As the resulting curves lie very close to one another for both models (and any fixed $\varphi_*$), to make the differences between various pulling angles visible, we zoom onto the maximum bead displacement in panel (b). Note that the force-displacement graphs for $\varphi_{**} \in (\pi/4, \pi/2)$ will mirror those for $\varphi_* = \pi/2 - \varphi_{**}$ due to the square shape of the macroscopic domain; in other words, due to the symmetry upon swapping $X$ and $Y$. Panels (c) and (e) show the discrete solution profiles (zoomed-in onto the bead) at the largest displacement $\tilde{R}_{b} = \tilde{R} =$ $0.05 \mu$m for two extreme values of the pulling angle $\varphi_* = 0$ (c) and $\pi/4$ (e) radians - the response is stiffest when one pulls in the direction of one of the two filament families and softest when pulling along the diagonal. The corresponding principal stresses and directions of the continuum stress tensor ($\boldsymbol{S}^{T}$) are plotted using ellipses at selected points near the bead in panels (d) and (f). Note that the continuum results are plotted using the undeformed variables with the corresponding pre-stress shown via red crossheads inside circles located at the top, that the green arrows indicate the direction of bead's motion and that the principal stresses were all normalized with respect to the same value chosen so that the ellipses do not overlap yet are large enough to be clearly seen.}
\label{fig:VaryPullingAngleVimentinDefaultPreStress}
\end{figure}

In order to assess the anisotropy of the force-displacement curves induced by our assumption of a regular array of filaments, in Figure \ref{fig:VaryPullingAngleVimentinDefaultPreStress} we examine the dependency on the pulling angle across its entire range. Amongst both the discrete and continuum simulations, the force-displacement curves remain within 1\% of one another for the full range of pulling angles (Figure \ref{fig:VaryPullingAngleVimentinDefaultPreStress}a,b). Furthermore, this difference remains small across the entire range of bead sizes considered (data not shown), consistent with the predictions of the continuum model in the limit of small deformations (see Section \ref{sec:SmallDefsSmallBead} below). As before, we observe good agreement between the continuum and discrete model predictions. However, despite the force exerted on the bead being almost independent of the pulling angle, we note that the overall stress profile is qualitatively different for different pulling angles (Figure \ref{fig:VaryPullingAngleVimentinDefaultPreStress}c-f): the more aligned the direction of movement is with the initial direction of the filament, 
the greater the increase (decrease) in tension in the wake (at the front) of the moving bead.
In summary, this figure shows that while the force-displacement curve is approximately independent of the direction of bead movement, the stress profile within the material is sensitive to the direction of pulling.

\subsubsection{Force due to pre-stress $\tilde{\mathcal{F}}_p$}
\label{eqn:SubSubDependencyPreStress}

\begin{figure}[ht!]
\centering
\begin{overpic}[width=1.0\textwidth,tics=10]{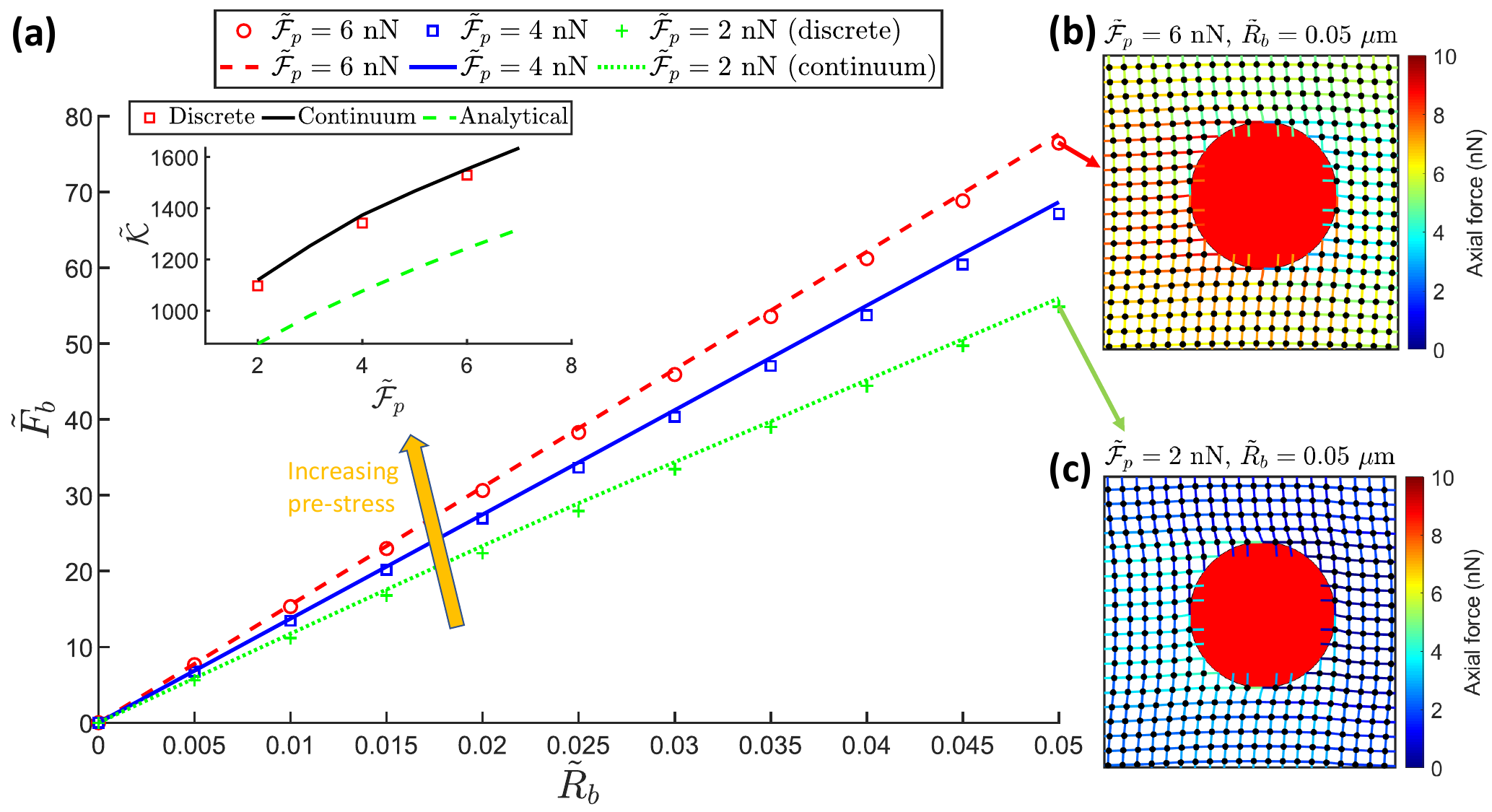}
\end{overpic}
\caption{Force-displacement graphs for discrete (symbols) and continuum (lines) models for default model parameters and $\tilde{\mathcal{F}}_p = 2$ (green), $4$ (blue) and $6$ (red) nN (a). The inset shows the dependence of the effective stiffness $\tilde{\mathcal{K}} = \tilde{F}_b^{max}/\tilde{R}_b^{max}$ (evaluated at the maximum displacement) as function of $\tilde{\mathcal{F}}_p$. Panels on the right show solution profiles at $\tilde{R}_{b} = 0.05 \mu$m for $\tilde{\mathcal{F}}_p = 6$ (b) and $2$ nN (c).}
\label{fig:ForceDisplacementVaryPrestressVimentin}
\end{figure}

In order to assess the importance of the filament pre-stress (since this is not known experimentally), in Figure \ref{fig:ForceDisplacementVaryPrestressVimentin}(a) we study force response for increasing $\tilde{\mathcal{F}}_p$ and default parameters otherwise. As might be expected, with increasing (tensile) pre-stress in the filaments, the response gets stiffer, i.e. the gradient of the force displacement curve increases. Due to the smallness of the deformations, the deviations from linear behaviour of the force-displacement curves are negligible in all studied cases which allows us to introduce a scalar measure of the network stiffness $\tilde{\mathcal{K}} = \mathrm{d} \tilde{F}_b/ \mathrm{d} \tilde{R}_b$ which we approximate as $\text{max}(\tilde{F}_b)/\text{max}(\tilde{R}_b)$ evaluated at the largest bead displacement. The network stiffness increases with the pre-stress in a slightly sublinear manner (see the inset in Figure \ref{fig:ForceDisplacementVaryPrestressVimentin}a). 
As expected, the overall force distribution within the network scales with the amount of pre-stress (Figure \ref{fig:ForceDisplacementVaryPrestressVimentin}b,c).

\section{Small-deformation and small-bead analysis}
\label{sec:SmallDefsSmallBead}
To provide further insight into the force-displacement relationship, and in particular the dependency on the model parameters, we investigate the limit $R_{b} \ll 1$, i.e. the limit of small macroscale deformations. Assuming that the bead displacement is small, it is natural to assume that all components of the deformation gradient tensor are small everywhere in the macroscopic domain. Note that the small deformations assumption is consistent with our restriction to bead displacements up to one inter-CL distance in the discrete model. We analyze small deformations by substituting
\begin{equation}\label{eqn:SmallDeform}
x(X,Y) = X + R_{b} \hat{x}(X,Y) + O(R_{b}^2), \qquad y(X,Y) = Y + R_{b} \hat{y}(X,Y) + O(R_{b}^2),
\end{equation}
with $R_{b} \ll 1$ into the continuum problem. 
Following Section \ref{sec:appDeriveSmallDeform} of the Supplementary Material, we arrive at the macroscale equations at $O(R_{b})$
\begin{equation}\label{eqn:EffectiveGeneralNonlinearFilamentNoBendingXODelta}
\left(\xi \mathcal{F}'(\xi) \hat{x}_{X}\right)_{X} + \left(\mathcal{F}(\xi) \hat{x}_{Y}\right)_{Y} = 0,
\end{equation}
\begin{equation}\label{eqn:EffectiveGeneralNonlinearFilamentNoBendingYODelta}
\left(\mathcal{F}(\xi) \hat{y}_{X}\right)_{X} + \left(\xi \mathcal{F}'(\xi) \hat{y}_{Y}\right)_{Y} = 0.
\end{equation}
Note that equations \eqref{eqn:EffectiveGeneralNonlinearFilamentNoBendingXODelta} and \eqref{eqn:EffectiveGeneralNonlinearFilamentNoBendingYODelta} are decoupled. Since the constitutive law for the force in the FS is always monotonically increasing as a function of end-to-end distance (i.e. $\mathcal{F}'(\xi)>0)$, we can divide both equations by $ \xi \mathcal{F}'(\xi)$ to obtain
\begin{equation}\label{eqn:EffectiveGeneralNonlinearFilamentNoBendingXODeltaK=0}
\hat{x}_{XX} + \omega \hat{x}_{YY} = 0,
\end{equation}
\begin{equation}\label{eqn:EffectiveGeneralNonlinearFilamentNoBendingYODeltaK=0}
\omega \hat{y}_{XX} + \hat{y}_{YY} = 0,
\end{equation}
where $\omega := \mathcal{F}(\xi)/(\xi \mathcal{F}'(\xi))>0$. For our particular FS constitutive law, $\mathcal{F} = r-1$. These equations are subject to boundary conditions
\begin{equation}\label{eqn:SmallDefBCsK=0}
\hat{x} = \hat{y} = 0,
\end{equation}
evaluated on the outer boundary of the domain. Similarly, on the boundary of the bead (circle of radius $a$) we impose for any $-\pi < \varphi  \leq \pi$ that
\begin{equation}\label{eqn:SmallDefCentralDisplHole}
\hat{x}(a \cos{(\varphi)},a \sin{(\varphi)}) = \cos{(\varphi_*)}, \qquad \hat{y}(a \cos{(\varphi)},a \sin{(\varphi)}) = \sin{(\varphi_*)}.
\end{equation}
For our choice of FS constitutive law we deduce $\omega = 1-1/\xi < 1$ which will be used in the elliptical transformation below (Figure \ref{fig:ellipticalGeo}). To the best of our knowledge it is not possible to solve \eqref{eqn:EffectiveGeneralNonlinearFilamentNoBendingXODeltaK=0}-\eqref{eqn:SmallDefCentralDisplHole} exactly. However, under the assumption $a \ll 1$, it is possible to find an asymptotic approximation valid in the inner region (i.e. close to $X^2 + Y^2 = a^2$). This assumption can easily be justified, as the beads used in optical tweezers experiments are typically small compared to the cell size \cite{hu2017}.


\subsection{Solution in the limit $a \ll 1$}

\begin{figure}[ht!]
\centering
\begin{overpic}[width=1.0\textwidth,tics=10]{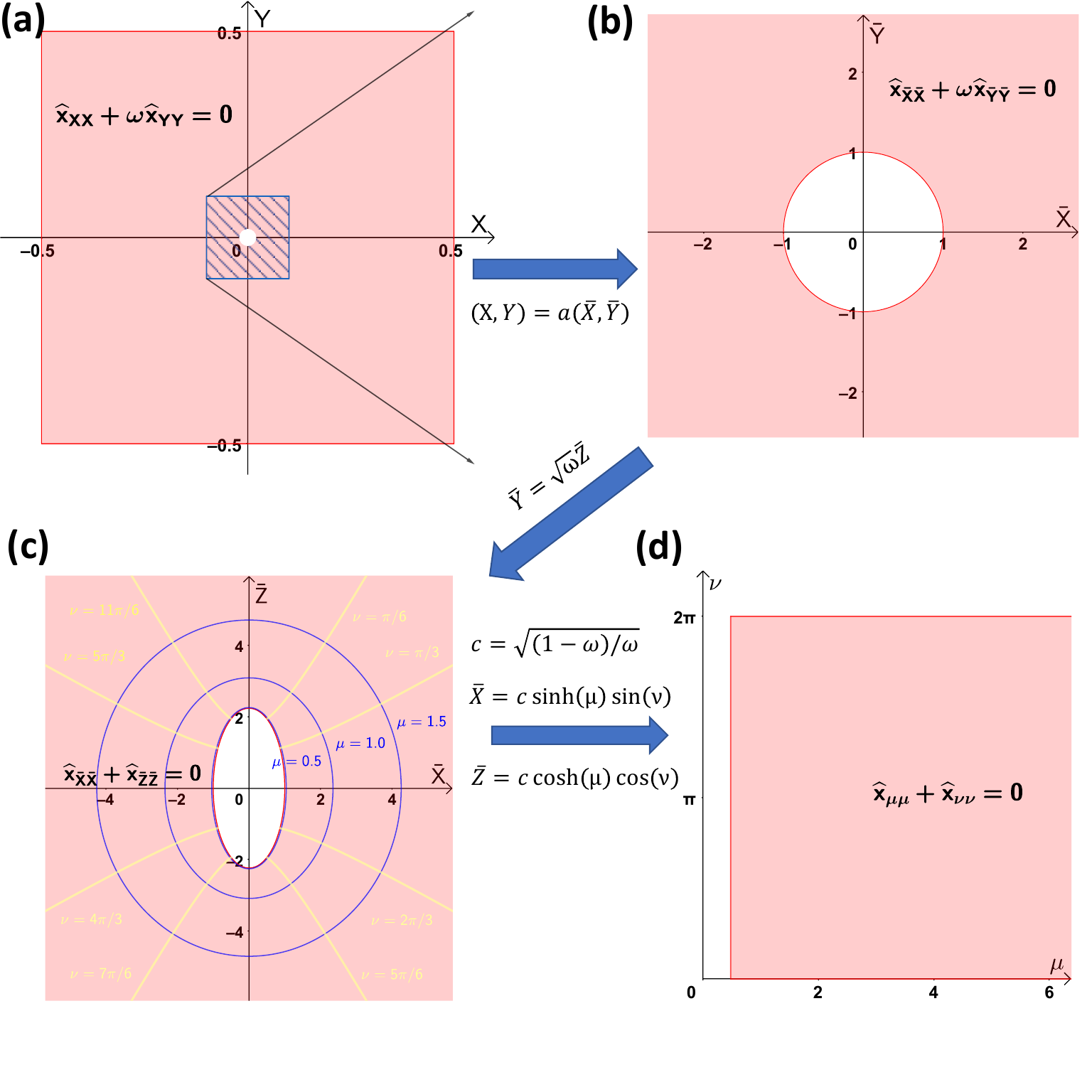}
\end{overpic}
\caption{Demonstration of key steps in the solution process using $a=0.02$ and $\omega=0.2$. Starting from the macroscale variables $(X,Y)$ (a) and assuming $a \ll 1$, we rescale to the inner layer (b). Then, we stretch the $\bar{Y}$ coordinate by the means of which we transform the governing equation into Laplace's equation which is to be solved subject to Dirichlet boundary conditions at an elliptical inner boundary in $(\bar{X},\bar{Z})$ (c). Using elliptical coordinates - note in panel (c) that the blue curves correspond to $\mu= \text{constant}$ while yellow to $\nu=\text{constant}$, the $\mu = \cosh^{-1}{\left( ( 1- \omega)^{-1/2} \right)} \approx 0.5$ representing the inner boundary -  we can finally transform this non-trivial geometry unto a rectangular one in $(\mu,\nu)$ while keeping the governing equation same (d).}
\label{fig:ellipticalGeo}
\end{figure}

As the two equations are decoupled, we solve them separately. The technical details are presented in Supplementary Material (Section \ref{sec:App_SmallBeadAsymptotics}). The solution strategy for $\hat{x}$ ($\hat{y}$ problem is dealt with analogously) is summarized in Figure \ref{fig:ellipticalGeo}: we study the outer problem \eqref{eqn:EffectiveGeneralNonlinearFilamentNoBendingXODeltaK=0} subject to the outer boundary conditions together with the inner problem obtained by rescaling $(\bar{X},\bar{Y}) = (X,Y)/a$, which localizes the problem to the neighbourhood of the bead (Figure \ref{fig:ellipticalGeo}a,b). In the inner region, we then need to transform the $\bar{Y}$ coordinate to $\bar{Z} = \bar{Y}/\sqrt{\omega}$ which transforms the governing equation into Laplace's equation on a (stretched) domain with an elliptical (inner) boundary (Figure \ref{fig:ellipticalGeo}c). Elliptical coordinates \eqref{eqn:ElliptCoordinates} then allow us to transform this problem onto a semi-infinite strip while keeping the same governing equation so that an analytical solution can be found easily (Figure \ref{fig:ellipticalGeo}d).
Undetermined constants in the inner solution are obtained by transforming back to Cartesian coordinates, writing in outer variables and matching with the outer $\hat{x}$. Eventually, we conclude the inner approximation (denoted with superscript $I$)
\begin{equation}\label{eqn:Smalla_InnerApproximationx}
\hat{x}^I = \cos(\varphi_*) \left\{1 + \frac{ 2 \cosh^{-1}((1-\omega)^{-\frac{1}{2}})  - \ln{(1 - 2 q + 2 \sqrt{q^2 - q})}}{2 \ln{(1/a)} + \ln{(4 \omega/(1-\omega))} - 2\cosh^{-1}((1-\omega)^{-\frac{1}{2}}) } \right\} + O\left(a^2\right),
\end{equation}
where
\begin{equation}\label{eqn:qAsFtionOfXY}
q(\bar{X},\bar{Y}) = \frac{-\omega \bar{X}^2 - \bar{Y}^2 + (1- \omega) - \sqrt{(\omega \bar{X}^2 + \bar{Y}^2 - (1-\omega))^2 + 4(1-\omega) \omega \bar{X}^2}}{2(1-\omega)}.
\end{equation}
Similarly, to find an inner approximation for $\hat{y}$, we first transform $\bar{X}$ to $ \bar{W} = \bar{X}/\sqrt{\omega}$ and then use elliptical coordinates \eqref{eqn:ElliptCoordinates_yProblem}, where we derive
\begin{equation}\label{eqn:Smalla_InnerApproximationy}
\hat{y}^I = \sin(\varphi_*) \left\{1 + \frac{ 2 \cosh^{-1}((1-\omega)^{-\frac{1}{2}})  - \ln{(1 - 2 q_2 + 2 \sqrt{q_2^2 - q_2})}}{2 \ln{(1/a)} + \ln{(4 \omega/(1-\omega))} - 2\cosh^{-1}((1-\omega)^{-\frac{1}{2}}) }  \right\} + O\left(a^2 \right),
\end{equation}
where
\begin{equation}\label{eqn:q2AsFtionOfXY}
q_2(\bar{X},\bar{Y}) = \frac{- \bar{X}^2 - \omega \bar{Y}^2 + (1- \omega) - \sqrt{( \bar{X}^2 + \omega \bar{Y}^2 - (1-\omega))^2 + 4(1-\omega) \omega \bar{Y}^2}}{2(1-\omega)}.
\end{equation}
Differentiating \eqref{eqn:Smalla_InnerApproximationx} and \eqref{eqn:Smalla_InnerApproximationy} with respect to $X$ and $Y$ we deduce leading-order approximations for the strain fields away from the bead (see Supplementary Material, Section \ref{sec:SuppLeadingOrderStrainFields}).

\subsection{Stress field and net force exerted on the bead}

Substituting the small-deformations ansatz \eqref{eqn:SmallDeform} into the stress tensor \eqref{eqn:StressTensor1} we further expand using $R_{b} \ll 1$ to obtain in the inner layer
\begin{gather}\label{eqn:NominalStressSmallDef}
\tilde{\boldsymbol{S}} = \frac{\pi \tilde{Y} \tilde{b}_{c}^2}{\varepsilon_{c} \tilde{D}} \left\{
\begin{pmatrix}
  \displaystyle \mathcal{F}(\xi) &
  \displaystyle 0 \\
  \displaystyle 0  &
  \displaystyle \mathcal{F}(\xi)
\end{pmatrix} + R_b 
\begin{pmatrix}
  \displaystyle \xi \mathcal{F}'(\xi)  \hat{x}_X^I &
  \displaystyle \mathcal{F}(\xi)  \hat{y}_X^I \\
  \displaystyle \mathcal{F}(\xi)  \hat{x}_Y^I  &
  \displaystyle \xi \mathcal{F}'(\xi)  \hat{y}_Y^I
\end{pmatrix} + O(R_{b}^2) \right\} .
\end{gather}
As in the full continuum problem, the net force exerted on the bead is calculated by integrating $\tilde{\boldsymbol{S}}^T \boldsymbol{N}$ over the boundary of the bead $\bar{X}^2+ \bar{Y}^2 = 1$ using the displacement profiles \eqref{eqn:Smalla_InnerApproximationx} and \eqref{eqn:Smalla_InnerApproximationy} in the inner layer. Performing these calculations and using the constitutive law \eqref{eqn:LinearSpringsDimlessSmallTau1NoCompression} (details are included in Supplementary Material, Section \ref{sec:SuppEvaluateNetForce}), we derive an analytical expression for the force response of the material to the bead being pulled through it valid asymptotically (accurate up to $O(a)$ error) of the form
\begin{equation}\label{eqn:SmallBeadSmallDefAsymptoticsForceVector}
\tilde{\boldsymbol{F}}_b \approx - (\cos{(\varphi_*)}, \sin{(\varphi_*)}) \tilde{F}_b^0,
\end{equation}
where
\begin{equation}\label{eqn:magnitudeOfNetForceLinearSprings}
\begin{aligned}
\tilde{F}_b^0 = \frac{ 2 \pi  R_{b}/\varepsilon_{c} \sqrt{\mathcal{F}_p \left( 1 + \mathcal{F}_p \right)}}{\displaystyle \ln{\left(2 \left( \sqrt{\mathcal{F}_p(1+\mathcal{F}_p)}-\mathcal{F}_p\right)/a\right)}} \pi \tilde{Y} \tilde{b}_c^2.
\end{aligned}
\end{equation}
Note that this force on the bead is in the direction opposite to that of the pulling, as expected. Equation \eqref{eqn:magnitudeOfNetForceLinearSprings} elucidates how the force-displacement curve depends on key model parameters, namely filament's pre-stress $\mathcal{F}_p$, Young's modulus $\tilde{Y}$ and radius $\tilde{b}_c$, mesh spacing $\varepsilon_{c}$ and bead radius $a$. Finally, we deduce an analytical formula for the (dimensional) effective network stiffness
\begin{equation}\label{eqn:EffectiveNetworkStiffnessFormula}
\tilde{\mathcal{K}} = \frac{\mathrm{d} \tilde{F}_b}{\mathrm{d} \tilde{R}_b} \approx \frac{\tilde{F}_b^0}{\tilde{R}_b} = \frac{\pi \tilde{Y} \tilde{b}_{c}^2}{\tilde{R}_{c}} \frac{2 \pi \sqrt{\mathcal{F}_p \left( 1 + \mathcal{F}_p \right)}}{\displaystyle \ln{\left(2 \left( \sqrt{\mathcal{F}_p(1+\mathcal{F}_p)}-\mathcal{F}_p\right)/a\right)}}.
\end{equation}
Note from the inset in panel (a) of Figure \ref{fig:ForceDisplacementVaryPrestressVimentin} that for the default bead size, the analytical result is already in good qualitative agreement with the discrete and continuum models.

\subsection{Dependence of force response on the bead size}

\begin{figure}[ht!]
\centering
\begin{overpic}[width=1.0\textwidth,tics=10]{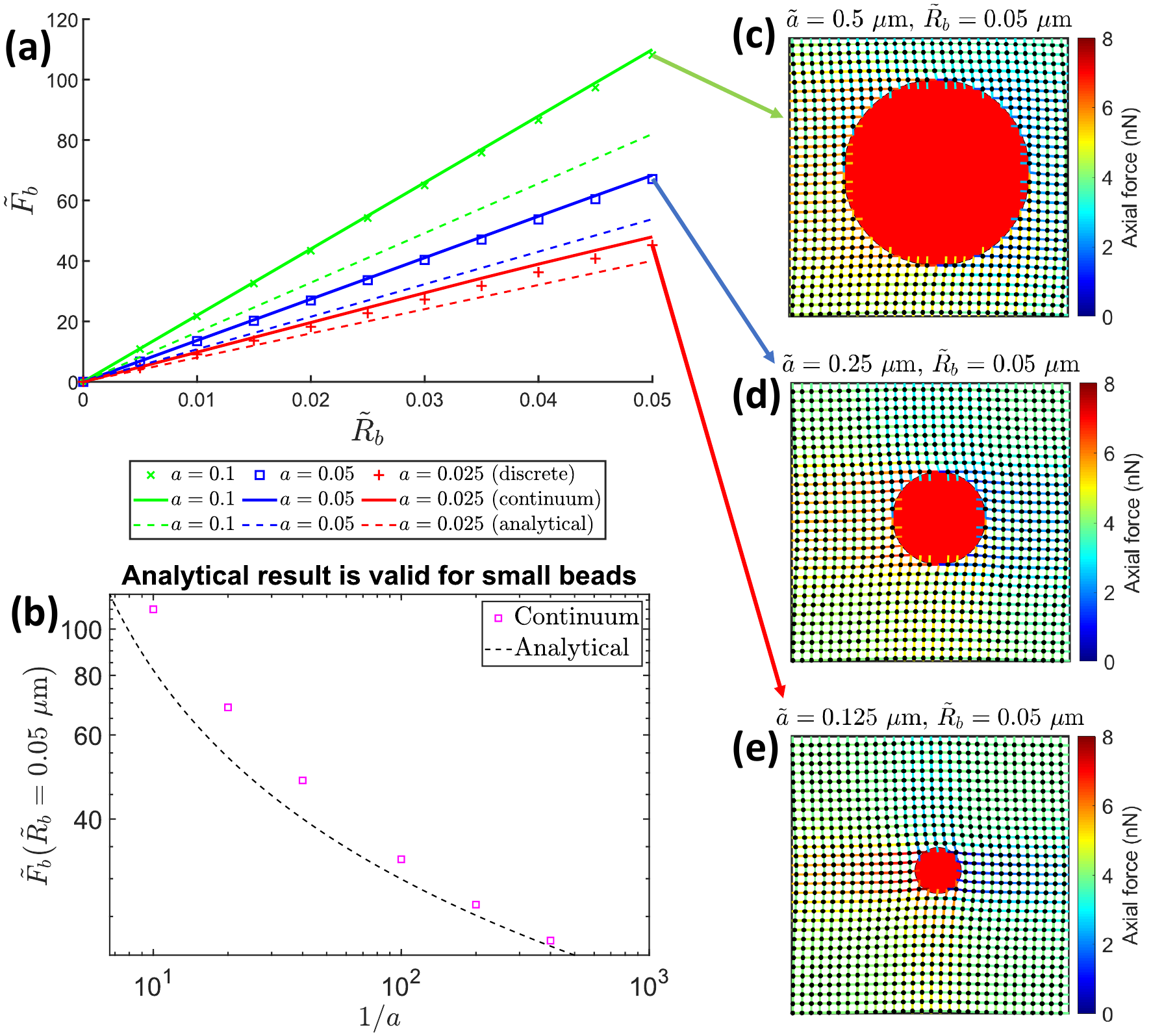}
\end{overpic}
\caption{(a) Force-displacement curves for default model parameters and the bead radius equal to twice ($0.5 \mu$m; green) and half ($0.125 \mu$m; red) the default value ($0.25 \mu$m; blue), with panels (c), (e) and (d) showing the corresponding solution profiles at the maximum displacement, $\tilde{R}_{b} = 0.05 \mu$m. Panel (b) demonstrates the convergence of the continuum simulations onto the prediction of the $a \ll 1$ asymptotics.}
\label{fig:Rheology}
\end{figure}

In order to assess the comparison between discrete, continuum and analytical approaches, in Figure \ref{fig:Rheology}(a) we plot force-displacement curves for a number of values of $a$. The discrete and continuum model results agree well for all considered values and, as expected, the larger the bead is the greater the force required for its transport. Moreover, as $a$ decreases, our asymptotic result \eqref{eqn:magnitudeOfNetForceLinearSprings} approaches simulation results of the continuum model. More specifically, as $a$ is reduced from $0.1$ to $0.025$, the absolute (relative) approximation error at the maximum displacement ($\tilde{R}_b = 0.05 \mu$m) decreases from roughly $28$ nN to $8$ nN. Figure \ref{fig:Rheology}(b) confirms the increasing agreement between the direct numerical simulations of the continuum model and our analytical approximation as $a$ is further reduced. When plotted using logarithmic scales on both axes, the continuum model predictions do not collapse onto a straight line indicating that the net force does not scale with $a$ according to a power law but behaves in the logarithmic manner instead (c.f. Equation \eqref{eqn:magnitudeOfNetForceLinearSprings}). Note that the continuum and analytical results almost overlap for $a=1/400$. The discrete solution profiles at the maximum displacement value $\tilde{R}_b = 0.05 \mu$m for varying bead radius $a$ are presented in Figures \ref{fig:Rheology}(c,d,e). With decreasing $a$, the number of FSs exerting force on the bead decreases linearly, but their individual stretches (and hence forces) are larger. In summary, this figure confirms that discrete and continuum predictions converge to the analytical formula \eqref{eqn:magnitudeOfNetForceLinearSprings} thus establishing it as a useful predictor of the net force exerted on a small bead.


\section{Discussion}

In this paper we have developed a multiscale framework for modelling the mechanical response of the eukaryotic cell cytoskeleton to internal motion of a small internal bead or organelle, mimicking recent rheological tests using optical tweezers \cite{hu2017}. In particular, we have developed a discrete model of the cell cytoskeleton by assuming a planar regular square grid of cytoskeletal filaments, using a microscale constitutive law for the mechanical response of each filament segment (Figure \ref{fig:DiscreteModelSetUp}) \cite{meng2017} . This model is highly idealized, ignoring the complex irregular geometry of the cytoskeletal network, including three-dimensional effects, and representing the structure by just one type of filament (in this case the intermediate filament vimentin). However, the simplicity of our framework allows a rational upscaling of the discrete model, from which we can construct a macroscale continuum model which encodes the microscale properties of the individual filaments. This continuum model provides an excellent match to discrete simulations across the parameter space at a fraction of the computational cost. Furthermore, in the limit of small bead displacements the continuum model can be solved asymptotically for small bead size by stretching the geometry of the (bead) boundary region, transforming to elliptical coordinates and matching with the outer region (Figure  \ref{fig:ellipticalGeo}), from which it is possible to construct a closed form expression \eqref{eqn:magnitudeOfNetForceLinearSprings} for the net force acting on the bead as a function of its size, the Young's modulus and radius of the filaments, the angle at which the bead is pulled through the network relative to the filaments, the network pre-stress and its spacing. In future, expression \eqref{eqn:magnitudeOfNetForceLinearSprings} could in principle be used to infer an estimate of a microscale filament pre-stress from the macroscale force-displacement data. The option of having both discrete and continuum formulation allows us to consider a variety of sizes of transported objects. For example, cell organelles are often much larger than the mesh size so that the continuum description for cytoskeleton
is justified and computationally inexpensive (and one can make use of our analytical result \eqref{eqn:magnitudeOfNetForceLinearSprings}). Conversely, the discrete simulations without a hole would form an appropriate model for transport of small cytoplasmic molecules which are usually smaller than the mesh spacing \cite{hu2017}.

A unified picture emerges from solving these discrete, continuum and analytical models: the system predicts an approximately linear relationship between the force on the bead and its displacement, and the gradient of this curve provides an estimate of the network stiffness (Figure \ref{fig:IncreaseNDefaultParsWithDisplacement}). In particular, we show that this network stiffness is approximately independent of the angle at which the bead is pulled through the structure (Figure \ref{fig:VaryPullingAngleVimentinDefaultPreStress}), consistent with the optical tweezers experiments \cite{hu2017}. The net force $F_b$ increases sublinearly with the filament pre-stress across the studied range although the deviation from linear behaviour is small (Figure \ref{fig:ForceDisplacementVaryPrestressVimentin}) and decreases in a logarithmic manner ($F_b  \propto \left(\ln{(1/a)} + \text{const} \right)^{-1}$) as the radius of the bead ($a$) reduces (Figure \ref{fig:Rheology}). We note that a linear increase in network stiffness with increasing pre-stress is found in tensegrity studies of cell mechanics, even though such linearity is typically established under bulk (shearing) deformations as opposed to local perturbations studied here \cite{Mofrad2006}. 


Numerical simulations in the absence of pre-stress take significantly longer than their pre-stretched counterparts. Initially stress-free networks thus appear to be the borderline case beyond which (pre-compressed filaments) neither discrete (MATLAB) solver nor continuum (FEniCS) solver converge. By analogy with the literature on central force networks we therefore expect that the problem with initially stress-free network suffers from ill-posedness issues associated with the so-called stiffness percolation (positive elastic modulus at zero strain) \cite{pritchard2014}. The present work has also restricted attention to quasi-static deformations. However eukaryotic cells are known to exhibit a complicated rheology involving additional dissipative factors such as transient crosslink binding/unbinding, sliding and unfolding, giving rise to viscoelastic behaviour at the macroscale \cite{Lieleg2009BiophysJ,Lieleg2010SoftMatter,Oosterwyck2013,Lee2009}. Furthermore, in its current form the model neglects the mechanical role of the cytosol fluid in which the filament network is immersed \cite{Moeendarbary2013}. Future work will extend this formulation to include these additional features, modelling the cell as a poro-visco-elastic continuum, allowing exploration of how these different mechanical responses manifest in different cell types.

This discrete-to-continuum modelling approach is not restricted to cytoskeletal networks and could similarly be applied to other crosslinked networks of semi-flexible filaments such as collagen \cite{stracuzzi2022risky}. Furthermore, our approach could be modified to model cells migrating through (and interacting with) extra-cellular matrix \cite{Preziosi2016,Kim2018,Tsingos2023}.

\vskip6pt

\enlargethispage{20pt}


\providecommand{\dataaccess}[1]{\textbf{\textit{Data accessibility: }} #1}

\dataaccess{This article has no experimental data. Numerical scripts for the discrete model were written in Matlab version R2021a, those solving the continuum model were written in python using FEniCS version 2019.2.0.dev0--, and can be accessed at \url{http://dx.doi.org/10.5525/gla.researchdata.1443}}

\providecommand{\funding}[1]{\textbf{\textit{Funding: }} #1}

\funding{J.K., N.A.H., X.Y.L. and P.S.S. acknowledge funding from EPSRC grant no. EP/S030875/1.}

\providecommand{\ack}[1]{\textbf{\textit{Acknowledgements: }} #1}

\ack{We thank Mr. Gordon McNicol, Drs. Namshad Thekkethil and Yangkun Du (University of Glasgow) and Profs. Ming Guo and Roger Kamm (MIT) for valuable discussions.}

\bibliography{references} 
\bibliographystyle{plain} 

\makeatletter\@input{xxSupp.tex}\makeatother

\end{document}


\title{Discrete-to-continuum models of pre-stressed cytoskeletal filament networks \\ \vspace{0.5cm}
\Large Supplementary Material}

\author{
J. Köry$^{1}$, N. A. Hill$^{1}$, X. Luo$^{1}$ and P. S. Stewart$^{1}$}

\affil{$^{1}$School of Mathematics and Statistics, University of Glasgow, Mathematics and Statistics Building, University Place, Glasgow G12 8QQ, UK}

\maketitle



This Supplementary Material is organized as follows. Section \ref{sec:SuppNotation} summarizes notation adopted throughout this article. Model parameters together with their default values are listed in Section \ref{sec:SuppDefaultDimensionalParametersActinVimentin}. Section \ref{sec:SuppDetailsOfModelDerivation} contains detailed derivations pertaining to the discrete model and Section \ref{sec:UpscalingAndContinuum} to the upscaling and the resulting continuum model. Finally, Section \ref{sec:SmallDefAndSmallBeadDetails} presents calculations relating to the small-deformations and small-bead limits, including that of the net force acting on the bead.

\section{Summary of notation}
\label{sec:SuppNotation}

Below we list notation adopted in this work, stating the symbols and their definitions. We note that this table is not exhaustive but with its help one can easily deduce all notation adopted in this work. For example, the mesh spacing representative of the cytoskeleton $\varepsilon_{c}$ is obtained by adding subscript $c$ to the mesh spacing $\varepsilon$.

\nomenclature[A, 01]{\(FS\)}{Abbreviation for filament segment}
\nomenclature[A, 02]{\(CL\)}{Abbreviation for crosslink}
\nomenclature[A, 03]{\(\sim\)}{Dimensional variable (above the symbol)}
\nomenclature[A, 04]{\(i,j\)}{Indices of the discrete network (as subscript)}
\nomenclature[A, 05]{\(kl\)}{Indices of tensors attaining value 1 or 2 for the two spatial dimensions (as subscript)}
\nomenclature[A, 06]{\(c\)}{Value representative of the cytoskeleton (as subscript)}
\nomenclature[A, 07]{\(I/O\)}{Pertaining to inner/outer region (as superscript)}

\nomenclature[B ,01]{\(\boldsymbol{X} = \left( X,Y \right) \)}{Initial configuration variables}
\nomenclature[B, 02]{\(\boldsymbol{x} = \left( x,y \right) \)}{Deformed configuration variables}
\nomenclature[B, 03]{\(\left( u,v \right) \)}{Components of the displacement field}
\nomenclature[B, 04]{\(\left( \hat{x},\hat{y} \right) \)}{Small-deformations variables}
\nomenclature[B, 05]{\(\left( \bar{X},\bar{Y} \right) \)}{Initial configuration variables rescaled to the bead boundary region}
\nomenclature[B, 06]{\( \bar{Z} \)}{Stretched $\bar{Y}$ coordinate}
\nomenclature[B, 07]{\(\left( \mu, \nu \right) \)}{Elliptical coordinates}
\nomenclature[B, 08]{\(\tilde{r}\)}{End-to-end distance (straight-line distance between two ends of a filament segment)}
\nomenclature[B, 09]{\(r\)}{End-to-end distance normalized with respect to the stress-free contour length}
\nomenclature[B, 10]{\(f\)}{Axial force in a filament segment (scales with $N$)}
\nomenclature[B, 11]{\(\mathcal{F}\)}{Axial force in a filament segment ($N=N_c$)}
\nomenclature[B, 12]{\(e\)}{Energy stored in a filament segment (scales with $N$)}
\nomenclature[B, 13]{\(\mathcal{E}\)}{Energy stored in a filament segment ($N=N_c$)}
\nomenclature[B, 14]{\( \varphi \)}{Polar angle}

\nomenclature[C, 01]{\(\tilde{D}\)}{Domain length}
\nomenclature[C, 02]{\(N\)}{Number of filament segments belonging to one filament}
\nomenclature[C, 03]{\(\tilde{R}\)}{Initial mesh spacing}
\nomenclature[C, 04]{\(\varepsilon\)}{Dimensionless initial mesh spacing}
\nomenclature[C, 05]{\(\tilde{L}\)}{Stress-free contour length of a filament}
\nomenclature[C, 06]{\(\tilde{\Lambda}\)}{Stress-free contour length of a filament segment}
\nomenclature[C, 07]{\(\xi\)}{Initial mesh spacing normalized with respect to the stress-free contour length}
\nomenclature[C, 08]{\(f_p\)}{Force in a filament segment due to pre-stress (scales with $N$)}
\nomenclature[C, 09]{\(\mathcal{F}_p\)}{Force in a filament segment due to pre-stress ($N=N_c$)}
\nomenclature[C, 10]{\( \omega \)}{Dimensionless parameter of the small-deformations problem}
\nomenclature[C, 11]{\(\tilde{\sigma}_p\)}{Macroscale pre-stress}
\nomenclature[C, 12]{\(a\)}{Bead radius}

\nomenclature[D, 01]{\(\tilde{Y}\)}{Young's modulus of a filament}
\nomenclature[D, 02]{\(\tilde{b}\)}{Radius of a filament}
\nomenclature[D, 03]{\(\tilde{k}_{B}\)}{Boltzmann constant}
\nomenclature[D, 04]{\(\tilde{T}\)}{Absolute temperature}
\nomenclature[D, 05]{\(\tilde{\Lambda}_p\)}{Persistence length of a filament}
\nomenclature[D, 06]{\(\tilde{\mathcal{F}}_{\text{entropic}}\)}{Entropic force}
\nomenclature[D, 07]{\(\tilde{\mathcal{F}}_{\text{enthalpic}}\)}{Enthalpic force}
\nomenclature[D, 08]{\(\mathcal{T}_1\)}{Ratio of the entropic force to the enthalpic force}
\nomenclature[D, 09]{\(\mathcal{T}_2\)}{One half of the ratio of the persistence length to the initial end-to-end distance}

\nomenclature[E, 01]{\(\varphi_*\)}{{Pulling angle}}
\nomenclature[E, 02]{\(R_b\)}{Magnitude of the bead displacement}
\nomenclature[E, 03]{\(F_b\)}{Magnitude of the net force acting on the bead}
\nomenclature[E, 04]{\(\mathcal{K}\)}{Scalar measure of network stiffness}
\nomenclature[E, 05]{\(\hat{\boldsymbol{r}}_{i \pm \tfrac{1}{2},j} / \hat{\boldsymbol{r}}_{i,j \pm \tfrac{1}{2}}  \)}{Unit vectors pointing in the directions of filament segments adjacent to node (i,j)}
\nomenclature[E, 06]{\(l_{i \pm \tfrac{1}{2},j} / l_{i,j \pm \tfrac{1}{2}}  \)}{Deformed lengths of the filament segments}
\nomenclature[E, 07]{\(\boldsymbol{I}\)}{Identity tensor}
\nomenclature[E, 08]{\(\boldsymbol{F}\)}{Deformation gradient tensor}
\nomenclature[E, 09]{\(\boldsymbol{C}\)}{Right Cauchy-Green deformation tensor}
\nomenclature[E, 10]{\(I_{4/6}\left(\boldsymbol{C}\right) \)}{Invariants of the right Cauchy-Green deformation tensor}
\nomenclature[E, 11]{\(\boldsymbol{S}\)}{Nominal stress tensor}
\nomenclature[E, 12]{\(W\)}{Strain energy density}

\vspace{-1cm}
\printnomenclature

\section{Model parameters}
\label{sec:SuppDefaultDimensionalParametersActinVimentin}

\subsection{Summary of the dimensional and dimensionless models}

The discrete model is only representative of a cytoskeletal mesh with spacing $\tilde{R}_{c}$ provided one takes $N = N_{c} = 1/\varepsilon_{c}= \tilde{D}/\tilde{R}_{c}$ - increasing $N$ only facilitates convergence to the continuum model. Subject to the microscale constitutive law \eqref{eqn:Blundell}, the dimensional model is then governed by ten parameters - three parameters describing the initial geometry with the bead ($\tilde{D}$, $\tilde{R}_{c}$ and $\tilde{a}$), four parameters describing mechanical properties of the filaments under consideration ($\tilde{Y}$, $\tilde{b}_{c}$, $\tilde{\Lambda}_p$ and $\tilde{\mathcal{F}}_p$), two parameters governing the bead displacement ($\tilde{R}_{b}$ and $\varphi_*$) and temperature $\tilde{T}$. Remaining parameters can be deduced from these. Note in particular that the stress-free contour length $\tilde{\Lambda}_{c}(\tilde{\mathcal{F}}_p)$ can be found numerically upon substituting $\tilde{b}=\tilde{b}_{c}$, $\tilde{r} = \tilde{R}_{c}$ and $\tilde{f} = \tilde{\mathcal{F}}_p$ (provided $N=N_{c}$) into \eqref{eqn:Blundell}; alternatively one can use an explicit approximation \eqref{eqn:TwoTermApproximationPreStretchPrestressDimensional} derived in Section \ref{sec:SubSecExplicitPreStretchPreStress}.

Subject to the microscale constitutive law \eqref{eqn:dimlessBlundellBest}, the dimensionless model (for $N=1/\varepsilon_{c}$) is governed by $7$ parameters - $\varepsilon_{c}$, $a$, $\mathcal{T}_{1}$, $\mathcal{T}_2$, $\mathcal{F}_p$, $R_{b}$ and $\varphi_*$ - and $\xi$ must be found by numerically solving \eqref{eqn:dimlessBlundellBestReference}. Alternatively, one can impose the explicit approximations \eqref{eqn:LinearSpringsDimlessSmallTau1NoCompression} (vimentin) or \eqref{eqn:ActinLargeCalT2ApproximationForceDistance} (actin; $\xi$ is the given by \eqref{eqn:ActinLargeCalT2ApproximationPrestressPrestretch}) for the microscale constitutive law, in which case the dimensionless model ($N=1/\varepsilon_{c}$) needs $5$ parameters for both actin and vimentin - $\varepsilon_{c}$, $a$, $\mathcal{F}_p$, $R_{b}$ and $\varphi_*$ - and actin requires one more parameter ($\mathcal{T}_2$) for full specification. Recall that in each of the above cases, it is assumed that filaments cannot withstand any compressive loads.

\subsection{Default dimensional parameters}

Recall the default values $\tilde{D} = 5 \mu$m, $\tilde{R}_{c} = 0.05 \mu$m, $N=(N_{c}=)100$ (estimated in Section \ref{sec:SubsecReferenceNetwork}), $\varphi_* = \pi/6$ and $0 \leq \tilde{R}_{b} \leq \tilde{R}_{c}$ (Section \ref{sec:ResultsDiscrAndCont}). We further use as default values $\tilde{T} = 300$ K for the absolute temperature and $\tilde{a} = 0.25$ $\mu$m for the bead radius \cite{hu2017Supp}. The standard value of the Boltzmann constant is $\tilde{k}_B \approx 1.38 \times 10^{-23} \text{m}^2 \text{kg s}^{-2} \text{K}^{-1}$.

Mechanical behaviour of actin filaments subject to tension has been widely studied experimentally. The microscale constitutive law \eqref{eqn:Blundell} has been shown to be equivalent to another constitutive model which in turn reproduced the experimental data well \cite{Blundell2009Supp,Holzapfel2013}. The following estimates for material parameters from \eqref{eqn:Blundell} pertain to actin filaments \emph{in vivo}: $\tilde{Y} = 2$ GPa, $\tilde{\Lambda}_p = 17 \mu$m and $\tilde{b} = 3.5$ nm \cite{meng2017Supp}.

Even though vimentin has only gained significant attention from the scientific community relatively recently, much is already known about its tensile behaviour. Using atomistic simulations, three distinct regimes in force extension diagram of single vimentin dimer under tension were uncovered and the underlying changes in its molecular structure identified \cite{Qin2009}. Tensile behaviour of single vimentin filaments was measured \cite{Block2017Supp}, confirming three distinct regimes reported previously \cite{Qin2009}, and showing good agreement between optical trap and atomic force microscopy experiments. Unfortunately, it is very unclear how these results could be translated to vimentin FSs of various contour lengths. For simplicity, we thus assume that a tensile response of a single vimentin FS can be modelled using Equation \eqref{eqn:Blundell} with an appropriate choice of model parameters. Vimentin filaments are about $10$ nm in diameter ($\tilde{b}_{c} = 5$ nm) and their Young's modulus was measured to be about $\tilde{Y}=0.9$ GPa \cite{Guzman2006Supp}. The persistence length of vimentin is approximately $\tilde{\Lambda}_p = 1 \mu$m \cite{meng2017Supp}.

To allow as large deformations as possible without breaking the actin filaments, we take the tensile strength $\tilde{\mathcal{F}}_{\text{max}}$ of actin to be the upper bound of the values found in the literature, i.e. $600$ pN \cite{tsuda1996}, and the value $8$ nN provides us with a lower bound for the tensile strength of vimentin \cite{Block2017Supp}. The only unknown parameter is then the value of the force due to pre-stress $\tilde{\mathcal{F}}_p$. As discussed in the main body of this paper, we have not managed to find any estimate for microscale force due to pre-stress (and, equivalently, pre-stretch $\xi$ or stress-free contour length $\tilde{\Lambda}$ values) representative of cells \emph{in vivo}. Therefore, $\tilde{\mathcal{F}}_p$ will here be considered a free parameter, with the default value equal to one half of the tensile strength estimates from above.

For completeness we also list the default values for the mechanical parameters of the dimensionless model to be
$$ \mathcal{T}_1^{\text{actin}} \approx 1 \times 10^{-9} \qquad \mathcal{T}_2^{\text{actin}} \approx 170 \qquad \mathcal{T}_1^{\text{vimentin}} \approx 1.9 \times 10^{-8} \qquad \mathcal{T}_2^{\text{vimentin}} \approx 10.$$
Table \ref{table_parameters} lists key model parameters together with their default values.

\begin{table}
\caption{Default values of discrete model parameters representative of cytoskeleton ($N=1/\varepsilon_{c}=100$). Note that whenever applicable, the default value is stated for vimentin with the value for actin in parantheses.}
\vspace{0.3cm}
\label{table_parameters}
\centering
\begin{tabular}{|c|c|c|c|c|}\hline
Parameter & Symbol & Default value & Units & Source \\ \hline 
 \hline
Cell region size & $\tilde{D}$ & $5$ & $\mu$m & biologically plausible  \\ \hline 
Inter-crosslink distance & $\tilde{R}_{c}$ & $0.05$ & $\mu$m & \cite{hu2017Supp} \\ \hline
\# of FSs per filament & $N_{c}$ & $100$ & - & $\tilde{D}/\tilde{R}_{c}$ \\ \hline 
Absolute temperature & $\tilde{T}$ & $300$ & K & \emph{in vivo} (rounded) \\ \hline
Bead radius & $\tilde{a}$ & $0.25$ & $\mu$m & \cite{hu2017Supp} \\ \hline
Bead displacement - maximum magnitude & $\tilde{R}_{b}$ & $0.05$ & $\mu$m & $\tilde{R}_{b} = \tilde{R}_{c}$  \\ \hline
Bead displacement - angle & $\varphi_*$ & $\pi/6$ & - & generic case \\ \hline
Young's modulus & $\tilde{Y}$ & $0.9$ ($2.0$) & GPa & \cite{Guzman2006Supp,meng2017Supp} \\ \hline
Filament radius & $\tilde{b}_{c}$ & $0.005$ ($0.0035$) & $\mu$m & \cite{Guzman2006Supp,meng2017Supp}   \\ \hline
Persistence length & $\tilde{\Lambda}_p$ & $1.0$ ($17.0$) & $\mu$m &  \cite{Guzman2006Supp,meng2017Supp}  \\ \hline
Tensile strength & $\tilde{\mathcal{F}}_{\text{max}}$ & $8.0$ ($0.6$) & nN & Supplementary Section \ref{sec:SuppDefaultDimensionalParametersActinVimentin} \\ \hline
Force due to pre-stress & $\tilde{\mathcal{F}}_p$ & $4.0$ ($0.3$) & nN & Supplementary Section \ref{sec:SuppDefaultDimensionalParametersActinVimentin}  \\ \hline 
\end{tabular}
\end{table}

\section{Discrete model}
\label{sec:SuppDetailsOfModelDerivation}

\subsection{Stored elastic energy in undeformed and deformed configurations}
\label{sec:SuppAxialForcesDetails}
The total (elastic) energy of the discrete network, denoted $\tilde{e}_T$ (introducing subscript $T$ for total), is the sum of contributions due to axial filament stretching. First let us note that the microscale constitutive law \eqref{eqn:Blundell} is parameterized by the stress-free contour length $\tilde{\Lambda}$ which in turn scales as $O(1/N)$ in the $N \to \infty$ limit. Therefore, we simply write $\tilde{f} = \tilde{f}(\tilde{r}/\tilde{\Lambda};N)$. We then express the elastic energy stored in an individual FS for general $N$ as
\begin{equation}\label{eqn:ExprEnergyFS}
\tilde{e} (r;N) = \int\limits_{\tilde{R}_{sf}}^{r \tilde{\Lambda}} \tilde{f} \left(\frac{\tilde{s}}{\tilde{\Lambda}};N \right) d \tilde{s},
\end{equation}
where $\tilde{R}_{sf}$ denotes the stress-free end-to-end distance. Defining the deformed lengths of the FSs between neighbouring CLs as
\begin{equation}\label{eqn:DeformedDimensionalLengthsDef}
\tilde{l}_{i \pm \frac{1}{2},j} = \sqrt{ \left( \tilde{x}_{i \pm 1,j} - \tilde{x}_{i,j}  \right)^2 + \left( \tilde{y}_{i \pm 1,j} - \tilde{y}_{i,j} \right)^2} \qquad \tilde{l}_{i ,j \pm \frac{1}{2}} = \sqrt{ \left( \tilde{x}_{i,j \pm 1} - \tilde{x}_{i,j}  \right)^2 + \left( \tilde{y}_{i,j \pm 1} - \tilde{y}_{i,j} \right)^2},
\end{equation}
where the usage of the index $\pm \frac{1}{2}$ arises naturally between any two neighbouring CLs, the total energy $\tilde{e}_{T}$ can be obtained by summing up these energies for all FSs, i.e.
\begin{equation}\label{eqn:FSEnergyDefined}
\begin{aligned}
\tilde{e}_T = &  \displaystyle \sum\limits_{i=-N/2+1}^{N/2-1} \sum\limits_{j=-N/2+1}^{N/2-1} \left[ \tilde{e} \left( \frac{\tilde{l}_{i-\frac{1}{2},j}}{\tilde{\Lambda}}; N \right) + \tilde{e} \left( \frac{\tilde{l}_{i,j-\frac{1}{2}}}{\tilde{\Lambda}}; N \right) \right] + \\ 
& \sum\limits_{j=-N/2+1}^{N/2-1} \tilde{e} \left( \frac{\tilde{l}_{\frac{N}{2}-\frac{1}{2},j}}{\tilde{\Lambda}} ; N \right)  +  \sum\limits_{i=-N/2+1}^{N/2-1} \tilde{e} \left( \frac{\tilde{l}_{i,\frac{N}{2}-\frac{1}{2}}}{\tilde{\Lambda}} ; N \right).
\end{aligned}
\end{equation}
We assume no slippage of CLs along the filaments so that the stress-free contour lengths of every FS in the undeformed and the deformed configurations are equal to $\tilde{\Lambda}$. Following Sections \ref{sec:SubsecIncreasingNInDiscreteModel} and \ref{sec:SubsecPreStressAsFunctionOfN}, we can write the axial forces as $\tilde{f}(r;N) = N_{c} \tilde{\mathcal{F}}(r)/N$ and express the elastic energy stored in a single FS at the undeformed end-to-end distance $\tilde{R}$ as
\begin{equation}\label{eqn:ExprEnergyFSRef}
\tilde{e} (\xi; N) = \int\limits_{\tilde{R}_{sf}}^{\tilde{R}} \frac{N_{c}}{N} \tilde{\mathcal{F}} \left(\frac{\tilde{s}}{\tilde{\Lambda}} \right) d \tilde{s},
\end{equation}
where $\xi$ is defined in \eqref{eqn:DefOfXi} and we have
\begin{equation}\label{eqn:EnergyIsIntOfForceTensile}
\tilde{e} \left( \frac{\tilde{r}}{\tilde{\Lambda}}; N \right) = \int\limits_{\tilde{R}_{sf}}^{\tilde{r}} \frac{N_{c}}{N} \tilde{\mathcal{F}} \left(  \frac{\tilde{s}}{\tilde{\Lambda}} \right) d \tilde{s} = \tilde{e}(\xi,N) + \int\limits_{\tilde{R}}^{\tilde{r}} \frac{N_{c}}{N} \tilde{\mathcal{F}} \left(  \frac{\tilde{s}}{\tilde{\Lambda}} \right) d \tilde{s}.
\end{equation}
To derive the strain energy density in Section \ref{sec:UpscalingAndContinuum}, it is also useful to introduce the elastically stored energy at an arbitrary CL $(i,j)$ as
\begin{equation}\label{eqn:EnergyAtCLij}
\tilde{e}_{i,j}(N) = \frac{1}{2} \left( \tilde{e} \left( \frac{\tilde{l}_{i-\frac{1}{2},j}}{\tilde{\Lambda}} ,N \right) + \tilde{e} \left( \frac{\tilde{l}_{i+\frac{1}{2},j}}{\tilde{\Lambda}},N \right) + \tilde{e} \left( \frac{\tilde{l}_{i,j-\frac{1}{2}}}{\tilde{\Lambda}} ,N \right) + \tilde{e} \left( \frac{\tilde{l}_{i,j+\frac{1}{2}}}{\tilde{\Lambda}} ,N \right) \right),
\end{equation}
where the factor $1/2$ accounts for the fact that the tensile energy stored in any FS corresponds to two CLs rather than just one. We further define $\tilde{D}_{sf} = N \tilde{R}_{sf}$. For any $r = \tilde{r}/\tilde{\Lambda}$ we can then using integration by substitution ($\tilde{s}/\tilde{\Lambda} = N \tilde{s}/\tilde{L} = t$) write
\begin{equation}\label{eqn:MacroscaleEnergyE}
\tilde{e}(r,N) = \frac{N_{c}}{N} \int\limits_{\tilde{R}_{sf}}^{\tilde{\Lambda}r} \tilde{\mathcal{F}}\left( \frac{\tilde{s}}{ \tilde{\Lambda}} \right) d \tilde{s} = \displaystyle \frac{\tilde{L} N_{c}}{N^2} \int\limits_{\tilde{D}_{sf}/\tilde{L}}^{r} \tilde{\mathcal{F}}(t) dt =  \left( \frac{N_{c}}{N} \right)^2 \tilde{\mathcal{E}}(r),
\end{equation}
which defines the energy $\tilde{\mathcal{E}}(r)$ stored in a FS in a situation representative of cytoskeleton ($N=N_{c}$). Defining $\xi_{sf} = \tilde{D}_{sf}/\tilde{L}$, this can be split into the energy due to pre-stress $\tilde{\mathcal{E}}_{P}$ and that supplied with the deformation $\tilde{\mathcal{E}}_{D}$ as
\begin{equation}\label{eqn:SplittingEnergyPrestressDef}
\tilde{\mathcal{E}}(r) = \frac{\tilde{D}}{\xi N_{c}} \int\limits_{\xi_{sf}}^{\xi} \tilde{\mathcal{F}}(t)dt + \frac{\tilde{D}}{\xi N_{c}} \int\limits_{\xi}^{r} \tilde{\mathcal{F}}(t)dt = \tilde{\mathcal{E}}_{P} + \tilde{\mathcal{E}}_{D}(r).
\end{equation}

\subsection{Decreasing the mesh spacing}
\label{sec:SubsecIncreasingNInDiscreteModel}
\subsubsection{Scaling geometric parameters with $N$}

In Section \ref{sec:Upscaling} we upscale the discrete force balance into a continuum problem as $N \to \infty$ ($\tilde{\Lambda}/\tilde{L} \to 0$). For the continuum limit to be a good approximation of the discrete model at the baseline setup representative of the cytoskeleton (i.e. using $N=N_{c}=100$; for default values of all parameters see Section \ref{sec:SuppDefaultDimensionalParametersActinVimentin}), we need to ensure that all model parameters are appropriately scaled as $N \to \infty$. As $N$ increases we have $\tilde{R} = N_{c}/N \times \tilde{R}_{c}$ and $\tilde{\Lambda} = N_{c}/N \times \tilde{\Lambda}_{c}$ where $\tilde{R}_{c}$ and $\tilde{\Lambda}_{c}$ are representative of the cytoskeleton. Note that the total length of the network increases without bounds as $N \to \infty$. Assuming constant density for the material of the filament, the total mass is a constant multiple of the total volume $2 (N-1) \tilde{L} \pi \tilde{b}^2$ and to keep this $O(1)$ as $N \to \infty$, we assume that $\tilde{b} = \tilde{b}_{c}\sqrt{N_{c}/N}$ where $\tilde{b}_{c}$ is a representative radius of the filament.

\subsubsection{Scaling forces (including those due to pre-stress) with $N$}

\label{sec:SubsecPreStressAsFunctionOfN}

Next we need to ensure that we get $O(1)$ tensile pre-stress in the $N \to \infty$ limit. Substituting the above scalings into \eqref{eqn:Blundell} and switching to $\varepsilon_{c}=1/N_{c}$, we get in the undeformed configuration
\begin{equation}\label{eqn:BlundellReferenceWithScalings}
\frac{\tilde{D}}{\tilde{L}} = \left(1 + \frac{N \varepsilon_{c} \tilde{f}_p}{\pi \tilde{Y} \tilde{b}_{c}^2} \right) \left( 1 - \sqrt{\frac{\tilde{k}_B \tilde{T}}{ \displaystyle \pi \tilde{\Lambda}_p \left( \tilde{f}_p + \frac{\pi^2 \tilde{k}_B \tilde{T} \tilde{\Lambda}_p N^2}{\tilde{L}^2} \right)}} \right).   
\end{equation}
Using a dimensionless force due to pre-stress $f_p = \tilde{f}_p/(\pi \tilde{Y} \tilde{b}_{c}^2)$ (as per Section \ref{sec:DiscrToCont}), equation \eqref{eqn:BlundellReferenceWithScalings} becomes
\begin{equation}\label{eqn:TerentjevReferenceDimlessVaryN}
\frac{\tilde{D}}{\tilde{L}} = \left( 1 + \varepsilon_{c} N f_p \right) \left( 1 - \sqrt{\frac{\tilde{k}_B \tilde{T}}{ \displaystyle \pi \tilde{\Lambda}_p \left( \pi \tilde{Y} \tilde{b}_{c}^2 f_p + \frac{\pi^2 \tilde{k}_B \tilde{T} \tilde{\Lambda}_p N^2}{\tilde{L}^2} \right) }} \right).
\end{equation}
Deriving an explicit relationship for $f_p(N)$ would be cumbersome and the situation is further complicated by the fact that $\tilde{L}$ depends on $f_p$. However, in order to keep the right-hand side of \eqref{eqn:TerentjevReferenceDimlessVaryN} $O(1)$ in $N \to \infty$ limit, $f_p$ must scale as $1/N$ for large $N$. We thus get $f_p(N) = \mathcal{F}_p/(\varepsilon_{c} N)$ with $\mathcal{F}_p = O(1)$, which ensures that the total elastic energy stored in the pre-stressed domain stays $O(1)$ as $N \to \infty$ and we arrive at a finite (and non-zero) pre-stress in the continuum limit. 

\begin{figure}[ht]
    \centering
    \subfloat[$N=10$]{\begin{overpic}[width=0.49\textwidth,tics=10]{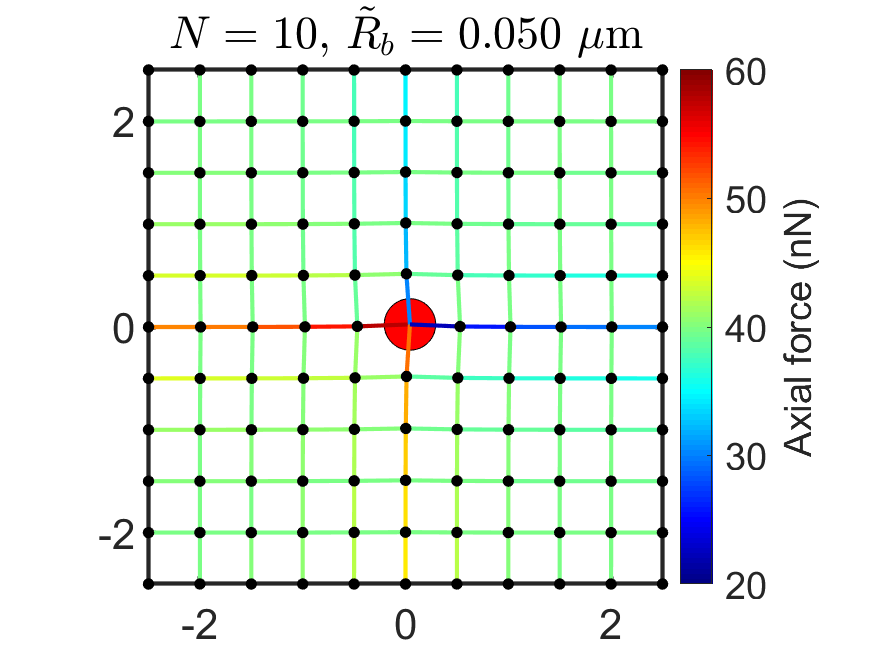}
    \end{overpic}\label{fig:PaperProfileN10}}
    \subfloat[$N=20$]{\begin{overpic}[width=0.49\textwidth,tics=10]{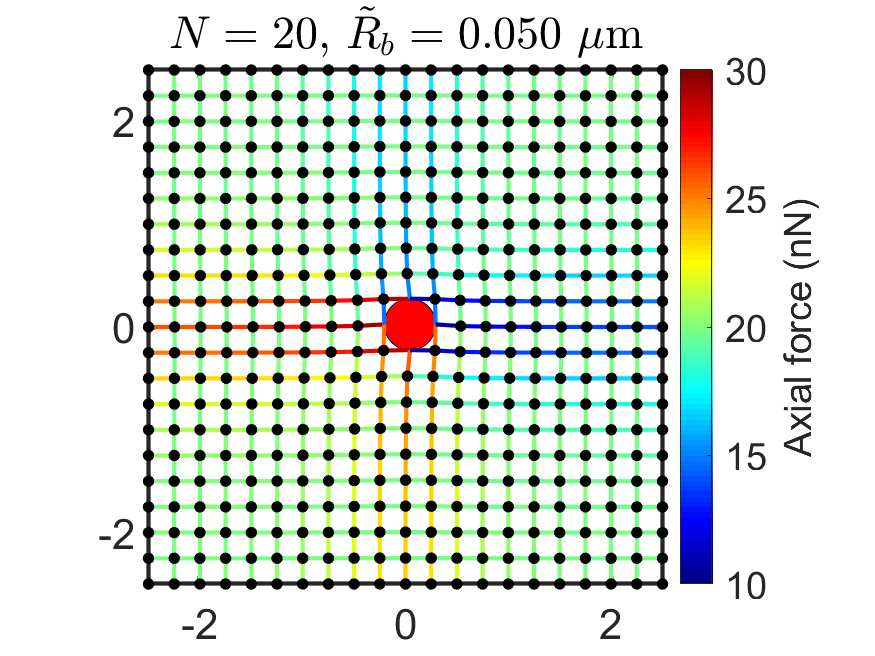}
    \end{overpic}\label{fig:PaperProfileN20}}
    \caption{The solution profiles for (a) $N=10$ and (b) $N=20$ using default model parameters for vimentin. Note that due to the smallness of $N$ we show profiles in full domains (not only in the vicinity of the bead as we did in the main body) and that the indicated lengths are in microns.}
    \label{fig:PaperProfileVaryN}
\end{figure}

Similarly, we must have $f(r;N) = \mathcal{F}(r)/(\varepsilon_{c} N)$. To demonstrate the central idea behind this scaling, resulting force distributions for $N=10$ and $20$ (and otherwise default parameters for vimentin, as described in Section \ref{sec:SuppDefaultDimensionalParametersActinVimentin}) are shown in Figure \ref{fig:PaperProfileVaryN}. Notice how the colorbar ranges vary with increasing $N$, which reflects the force scaling. In other words, for the discrete simulations to converge onto an $O(1)$ force response in the continuum ($N \to \infty$) limit, forces must scale as $1/N$ and these forces are thus in physiologically realistic range only for $N=N_{c}$.

\subsection{Implicit and explicit microscale constitutive laws for vimentin and actin}
\label{sec:SubSecExplicitPreStretchPreStress}

\subsubsection{Explicit microscale constitutive laws for fixed $N$}

\begin{figure}[ht]
    \centering
    \subfloat[Actin]{\begin{overpic}[width=0.49\textwidth,tics=10]{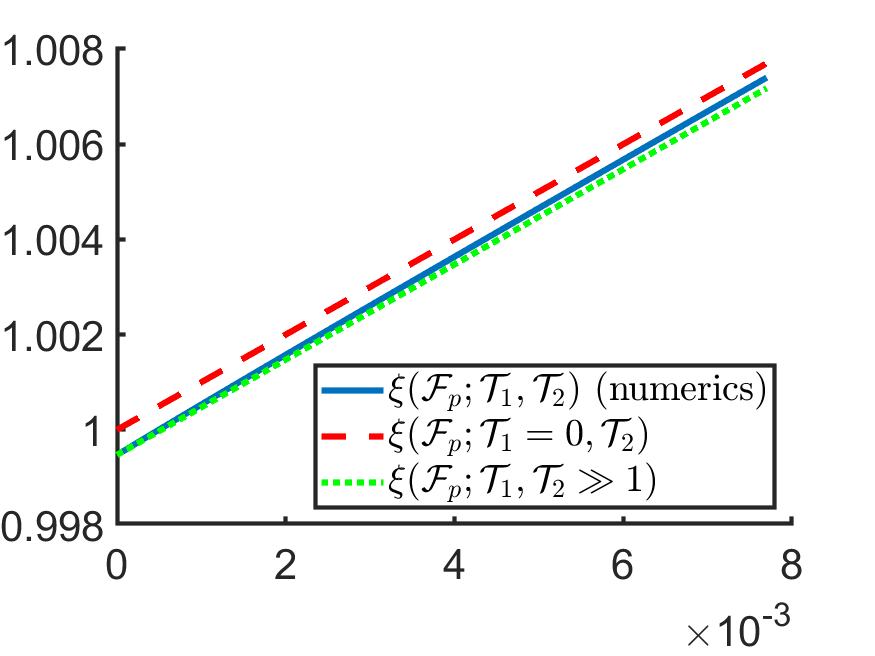}
    \Large{
    \put(77,8){$\mathcal{F}_p$}
    \put(13,75){$\xi$}
    }
    \end{overpic}\label{fig:ActinPrestressPrestretchFixedN}}
    \subfloat[Vimentin]{\begin{overpic}[width=0.49\textwidth,tics=10]{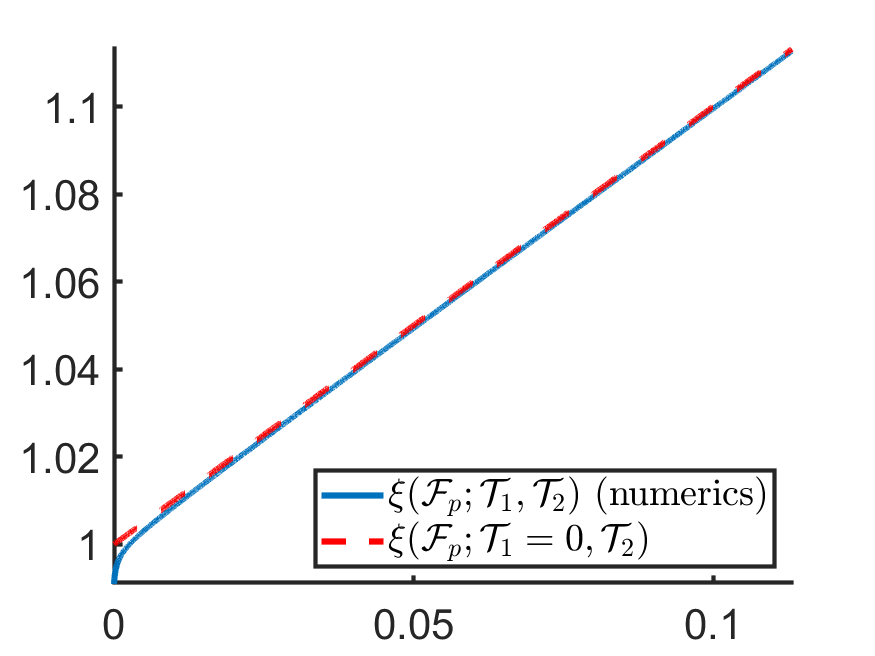}
    \Large{
    \put(65,2){$\mathcal{F}_p$}
    \put(13,75){$\xi$}
    }
    \end{overpic}\label{fig:VimentinPrestressPrestretchFixedN}}   
    \caption{Numerically computed $\xi(\mathcal{F}_p; \mathcal{T}_1,\mathcal{T}_2)$ (solid blue) as compared with its $\mathcal{T}_1=0$ approximation (dashed red) for (a) actin and (b) vimentin. The green dotted line in Figure (a) presents the approximation \eqref{eqn:ActinLargeCalT2ApproximationPrestressPrestretch} valid for actin.}
    \label{fig:PreStressPreStretchFixedN}
\end{figure}

\begin{figure}[ht]
    \centering
    \subfloat[Actin]{\begin{overpic}[width=0.49\textwidth,tics=10]{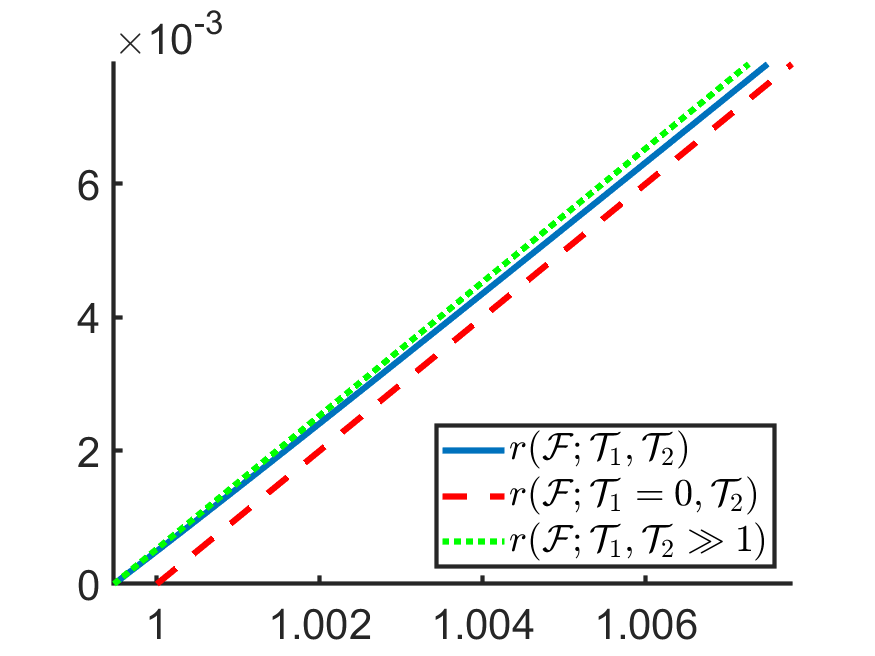}
    \Large{
    \put(77,8){$r$}
    \put(13,75){$\mathcal{F}$}
    }
    \end{overpic}\label{fig:ActinForceDistanceFixedNAndPrestress}}
    \subfloat[Vimentin]{\begin{overpic}[width=0.49\textwidth,tics=10]{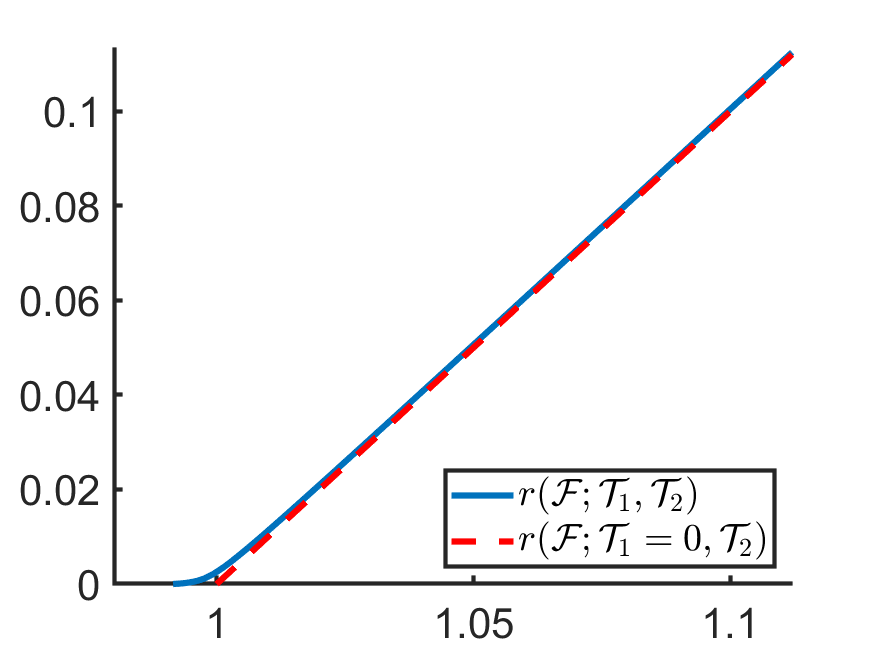}
    \Large{
    \put(65,2){$r$}
    \put(13,75){$\mathcal{F}$}
    }
    \end{overpic}\label{fig:VimentinForceDistanceFixedNAndPrestress}}   
    \caption{$r(\mathcal{F}; \mathcal{T}_1,\mathcal{T}_2)$ from \eqref{eqn:dimlessBlundellBestReference} using numerically calculated $\xi$ corresponding to default pre-stress (solid blue) as compared with its $\mathcal{T}_1=0$ approximation (dashed red) for (a) actin and (b) vimentin. The green dotted line in Figure (a) presents the approximation \eqref{eqn:ActinLargeCalT2ApproximationForceDistance} valid for actin.}
    \label{fig:ForceDistanceFixedNAndPrestress}
\end{figure}

We observe that $\mathcal{T}_1$ is very small for considered filaments - $\mathcal{T}_1 \approx 1.0 \times 10^{-9}$ for actin and $\mathcal{T}_1 \approx 1.9 \times 10^{-8}$ for vimentin. Substituting $\mathcal{T}_1 \ll 1$ into \eqref{eqn:dimlessBlundellBestReference} we get \eqref{eqn:TwoTermApproximationPreStretchPrestress} which provides an explicit relationship between the force due to pre-stress and pre-stretch. While this formula provides a good approximation to the implicit formula \eqref{eqn:dimlessBlundellBestReference} for vimentin (provided the pre-stress is not too small), a non-negligible gap exists in the approximation for actin - see Figure \ref{fig:PreStressPreStretchFixedN}. Similarly, when we plot the microscale constitutive law \eqref{eqn:dimlessBlundellBest} (using $\xi$ corresponding to the default pre-stress as found numerically from \eqref{eqn:dimlessBlundellBestReference}) and compare it with the $\mathcal{T}_1 = 0$ approximation \eqref{eqn:LinearSpringsDimlessSmallTau1NoCompression}, we again observe good agreement for vimentin but a clear gap for actin (see Figure \ref{fig:ForceDistanceFixedNAndPrestress}).
The gaps for actin can be explained by a combination of considered force range (the maximum considered value of $\mathcal{F}$ is small for actin compared to vimentin, due to the small tensile strength of the former) and the size of $\mathcal{T}_2$ ($\approx 170$ for actin, which means that the spacing between neighbouring CLs is much shorter than the persistence length of actin). Having noticed the largeness of $\mathcal{T}_2$ we (assume $\mathcal{T}_1 = O(1)$ and) substitute the ansatz
\begin{equation}\label{eqn:LargeTau2PreStretchPreStressAnsatz}
\xi(\mathcal{F}_p;\mathcal{T}_1, \mathcal{T}_2, \varepsilon_{c}, N) = \xi_{0}(\mathcal{F}_p; \mathcal{T}_1, \varepsilon_{c}, N) + \frac{1}{\mathcal{T}_2} \xi_1(\mathcal{F}_p;\mathcal{T}_1, \varepsilon_{c}, N) + O(\mathcal{T}_2^{-2})
\end{equation}
into \eqref{eqn:dimlessBlundellBestReference} and get
$$ \xi_{0} + \frac{1}{\mathcal{T}_2} \xi_1 + O(\mathcal{T}_2^{-2}) = \left(1+\mathcal{F}_p \right) \left(1 - \sqrt{\frac{\mathcal{T}_1}{ \mathcal{F}_p/(\varepsilon_{c} N) + 4 \pi^3 \left(\varepsilon_{c} N \mathcal{T}_2 \right)^2 \mathcal{T}_1 \left( \xi_{0}^2+ O(\sqrt{\mathcal{T}_2^{-1}}) \right)}} \right) .$$
At the leading order in $\mathcal{T}_2$ ($O(1)$) we again get
$$ \xi_{0} = 1 + \mathcal{F}_p$$
and at $O(\mathcal{T}_2^{-1})$ we conclude
$$ \xi_1 = - \frac{1+\mathcal{F}_p}{2 \pi^{3/2} \varepsilon_{c} N \xi_0^2} = - \frac{1}{2 \pi^{3/2} \varepsilon_{c} N \left( 1 + \mathcal{F}_p \right)}.$$
Retaining the first two terms, the approximation reads
\begin{equation}\label{eqn:ActinLargeCalT2ApproximationPrestressPrestretch}
\xi = 1+ \mathcal{F}_p - \frac{1}{\mathcal{T}_2} \frac{1}{2 \left( 1+ \mathcal{F}_p \right) \pi^{3/2} \varepsilon_{c} N}.    
\end{equation}
For actin, this provides a good approximation (without a significant gap) to \eqref{eqn:dimlessBlundellBestReference}, as shown in Figure \ref{fig:ActinPrestressPrestretchFixedN}. Similarly, we can expand \eqref{eqn:dimlessBlundellBest} for $\mathcal{T}_2 \gg 1$ and retaining $O(\mathcal{T}_2^{-1})$ terms we get
\begin{equation}\label{eqn:ActinLargeCalT2ApproximationForceDistance}
r = \left( 1+ \mathcal{F} \right) \left( 1- \frac{1}{2 \pi^{3/2} \varepsilon_{c} N \xi \mathcal{T}_2} \right).
\end{equation}
Using the approximation for $\xi$ from \eqref{eqn:ActinLargeCalT2ApproximationPrestressPrestretch} in \eqref{eqn:ActinLargeCalT2ApproximationForceDistance}, we again recover an excellent approximation without a gap, see Figure \ref{fig:ActinForceDistanceFixedNAndPrestress}. 

\paragraph{Redimensionalized microscale constitutive laws}

For the sake of completeness, we also state the approximate microscale constitutive laws in their dimensional forms (dimensionalizing both the force and the end-to-end distance). For vimentin, dimensionless microscale constitutive law \eqref{eqn:LinearSpringsDimlessSmallTau1NoCompression} can be redimensionalized using \eqref{eqn:TwoTermApproximationPreStretchPrestress} to give an expression in terms of the dimensional parameters of the approximate model (and $N$) which reads
\begin{equation}\label{eqn:RedimedMicroscaleConstitutiveLawVimentin}
\tilde{f} = \text{max} \left\{0, \frac{\pi \tilde{Y} \tilde{b}_{c}^2}{\tilde{R}_{c}} \left[ \left( 1+ \frac{\tilde{\mathcal{F}}_p}{\pi \tilde{Y} \tilde{b}_{c}^2}\right) \tilde{r} - \frac{\tilde{D}}{N}  \right]\right\}.
\end{equation}
From \eqref{eqn:ActinLargeCalT2ApproximationPrestressPrestretch}, we can deduce for actin
\begin{equation}\label{eqn:TwoTermApproximationPreStretchPrestressDimensional}
\tilde{\Lambda} = \frac{\tilde{R}}{\displaystyle 1+ \mathcal{F}_p - \frac{1}{\mathcal{T}_2} \frac{1}{2 \left( 1+ \mathcal{F}_p \right) \pi^{3/2} \varepsilon_{c} N}}
\end{equation}
and eventually conclude from \eqref{eqn:ActinLargeCalT2ApproximationForceDistance} the redimensionalized microscale constitutive law in the form
\begin{equation}\label{eqn:RedimedMicroscaleConstitutiveLawActin}
\tilde{f} =  \text{max} \left\{0, \frac{\pi \tilde{Y} \tilde{b}_{c}^2}{\tilde{R}_{c}} \left[ \frac{\xi}{\displaystyle 1 - \frac{\tilde{D}}{\pi^{3/2} \xi N \tilde{\Lambda}_p}} \tilde{r} - \frac{\tilde{D}}{N} \right] \right\},
\end{equation}
where $\xi$ is given in \eqref{eqn:ActinLargeCalT2ApproximationPrestressPrestretch}.

\subsubsection{Scaling of $\xi$ with $N$ for fixed $\mathcal{F}_p$}

\begin{figure}[ht]
    \centering
    \subfloat[Actin]{\begin{overpic}[width=0.49\textwidth,tics=10]{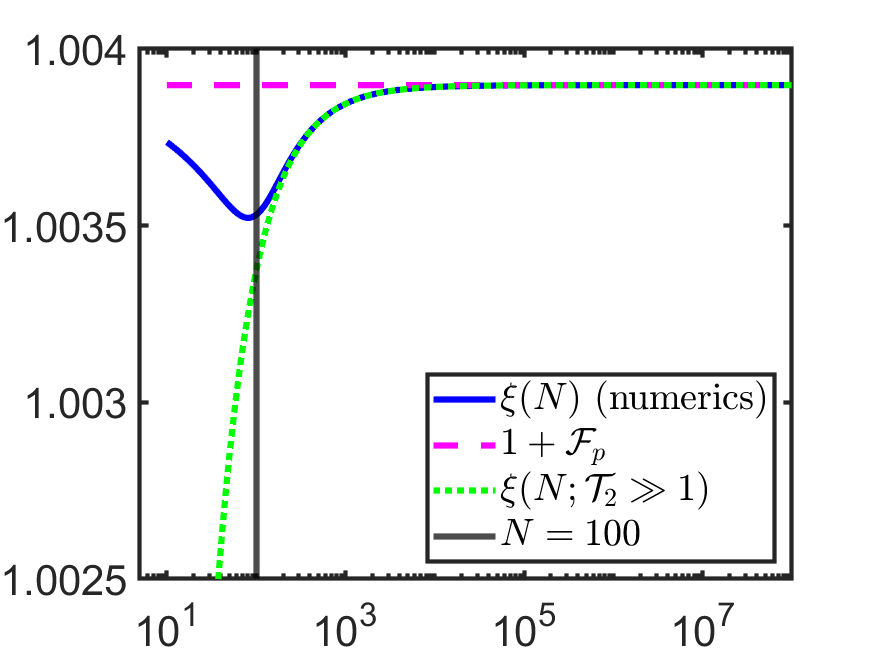}
    \Large{
    \put(68,2){$N$}
    \put(15,73){$\xi$}
    }
    \end{overpic}\label{fig:ActinXiVsNDefaultPrestress}}
    \subfloat[Vimentin]{\begin{overpic}[width=0.49\textwidth,tics=10]{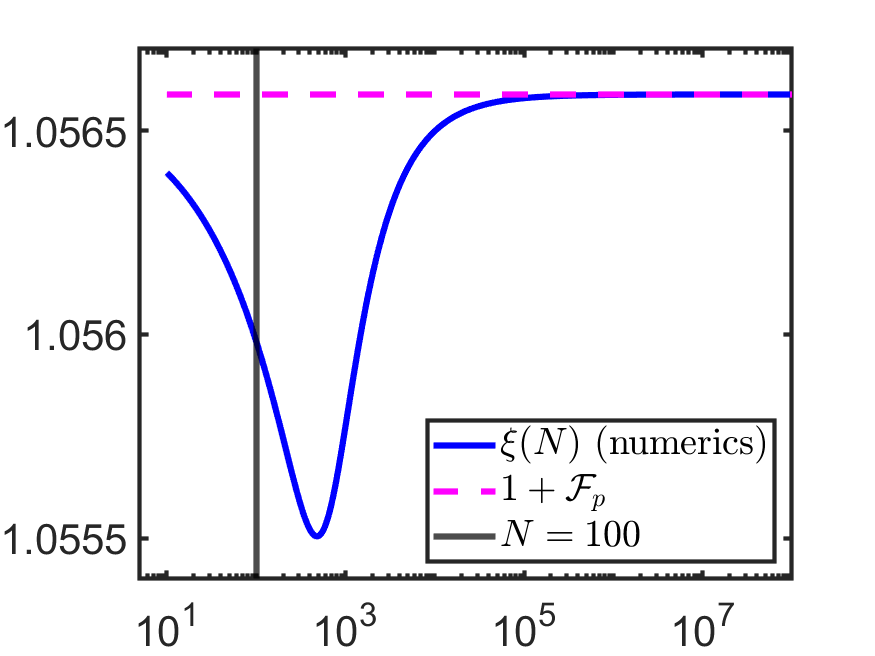}
    \Large{
    \put(68,2){$N$}
    \put(15,73){$\xi$}
    }
    \end{overpic}\label{fig:VimentinXiVsNDefaultPrestress}}   
    \caption{As $N \to \infty$, for default model parameters (including $\mathcal{F}_p$) the numerically computed $\xi$ (solid blue) converges to $1+\mathcal{F}_p$ (dashed magenta) for both actin (a) and vimentin (b). Even though the convergence is not yet attained at $N=1/\varepsilon_{c}=100$, the dependence on $N$ is weak. The green dotted curve in panel (a) corresponds to the approximation \eqref{eqn:ActinLargeCalT2ApproximationPrestressPrestretch}.}
    \label{fig:XiVsNDefaultPrestress}
\end{figure}

We observed in Section \ref{sec:SubsecPreStressAsFunctionOfN} that the force due to pre-stress must scale as $f_p(N) = \mathcal{F}_p/(\varepsilon_{c}N),$ with $\mathcal{F}_p = O(1)$. In order to study how $\xi$ scales with $N$ and determine whether this scaling is consistent for both the implicit model and the explicit approximations, we fix all parameters at their default value for both actin and vimentin (including $\mathcal{F}_p$; see Section \ref{sec:SuppDefaultDimensionalParametersActinVimentin}) and find the root $\xi$ of \eqref{eqn:dimlessBlundellBestReference} as function of $N$ numerically. The log-plots in Figure \ref{fig:XiVsNDefaultPrestress} show that while $\xi$ has not yet converged to its $N \to \infty$ limit $1+\mathcal{F}_p$ for $N=1/\varepsilon_{c}=100$ (used as default throughout this work and indicated by vertical black lines in the figure), the dependence is very weak for both actin and vimentin. Moreover, \eqref{eqn:ActinLargeCalT2ApproximationPrestressPrestretch} provides a good approximation for $\xi(N)$ near $N=1/\varepsilon_{c}=100$ for actin (see dotted green curve in Figure \ref{fig:ActinXiVsNDefaultPrestress}).

\subsection{Discrete model simulations for actin}
\label{sec:SubsecBrittlenessActin}
We simulated the discrete model for actin using the microscale constitutive law \eqref{eqn:RedimedMicroscaleConstitutiveLawActin} with $\xi$ given by \eqref{eqn:ActinLargeCalT2ApproximationPrestressPrestretch}. As in the main body, we plot the steady-state solutions in the vicinity of the bead.
\begin{figure}[ht]
    \centering
    \begin{overpic}[width=0.32\textwidth,tics=10]{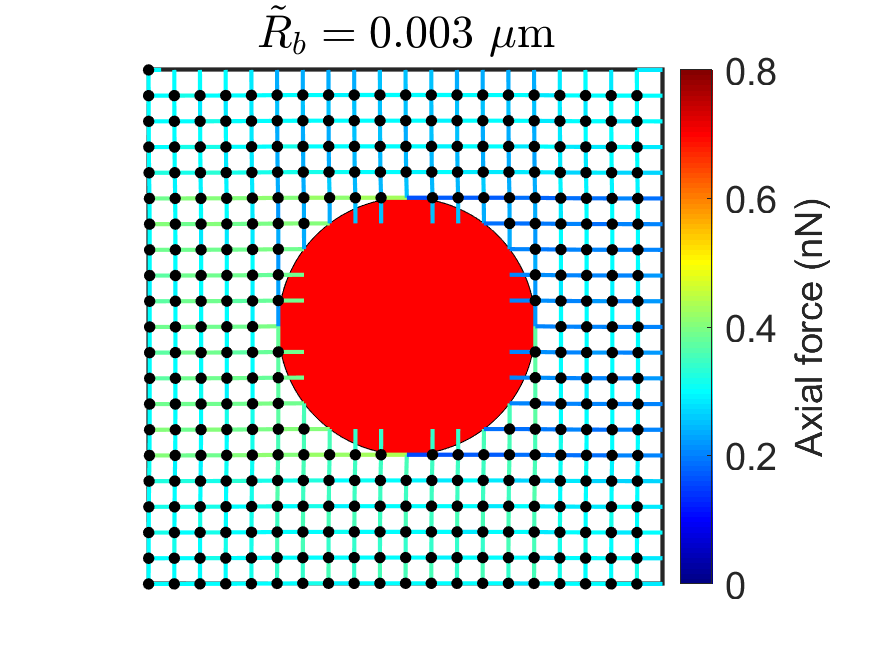}\end{overpic}\label{fig:ActinDisplacemenPoint003}
    \begin{overpic}[width=0.32\textwidth,tics=10]{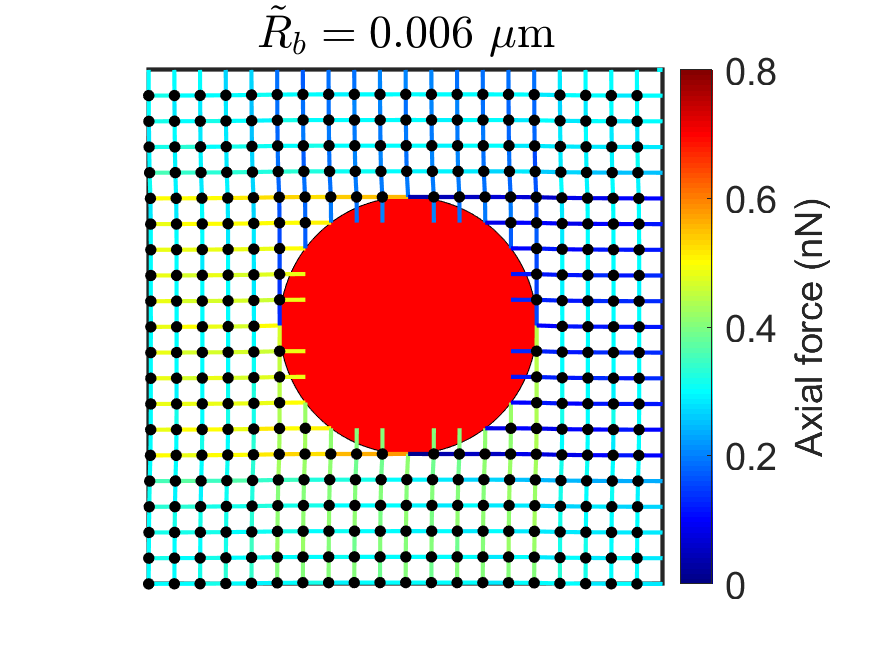}
\end{overpic}\label{fig:ActinDisplacementPoint006}
    \begin{overpic}[width=0.32\textwidth,tics=10]{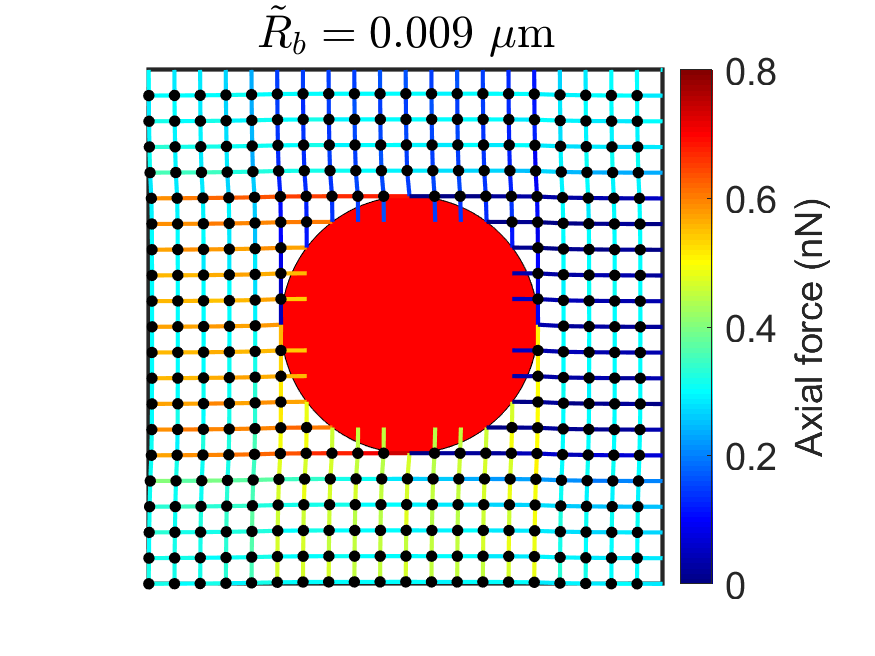}
\end{overpic}\label{fig:ActinDisplacementPoint009}
    \caption{Steady-state solutions using default actin parameters with $N=1/\varepsilon_{c} = 100$ for increasing $\tilde{R}_b$ (zoomed-in onto the bead).}
    \label{fig:ActinBrittle}
\end{figure}
For bead displacements $\tilde{R}_b$ as small as (roughly) one tenth of the mesh size ($\tilde{R} = 0.05 \mu$m), actin FSs start experiencing tensile forces beyond the upper limit of their tensile strength, $600$ pN (see Figure \ref{fig:ActinBrittle}). In particular, FSs in the wake of the bead motion whose undeformed orientation has a significant component in its direction would typically be broken for very small bead displacement, by being stretched beyond the tensile strength. Moreover, best-studied actin CLs like filamin or alpha-actinin typically unbind at even lower rupture forces of $40-80$ pN \cite{Ferrer2008} and such unbinding and rebinding is thought to play a crucial role in viscoelastic response of cytoskeletal networks. At present, our modelling framework is stationary and does not account for dynamic CLs. We thus acknowledge that the model as it stands is not yet suitable for realistic description of crosslinked actin networks and postpone such considerations for future work. 

\newgeometry{left=0.7cm,right=2.0cm,top=2.0cm,bottom=2.0cm}

\section{Upscaling and continuum model}
\label{sec:UpscalingAndContinuum}

\subsection{Details of discrete-to-continuum upscaling}
\label{sec:appDiscreteToCont}

\subsubsection{Upscaling the force balance}

Denoting partial derivatives with subscripts, we can express the relevant finite differences using Taylor expansions as
\begin{equation}
\begin{aligned}\label{eqn:TaylorExpandDiscreteFirst}
& x_{i+1,j} - x_{i,j} = x(X_{i+1},Y_{j}) - x(X_i,Y_j) = (X_{i+1}-X_i) x_X(X_i,Y_j) + \\ & \frac{(X_{i+1}-X_i)^2}{2} x_{XX}(X_i,Y_j) + O((X_{i+1}-X_i)^3) = \varepsilon x_X(X_i,Y_j) + \frac{\varepsilon^2}{2} x_{XX}(X_i,Y_j) + O(\varepsilon^3).
\end{aligned}
\end{equation}
For other differences involving the deformed coordinate $x$ occuring in our equations, we get
\begin{equation}\label{eqn:TaylorExpandDiscreteRest}
\begin{aligned}
& x_{i-1,j} - x_{i,j} = - \varepsilon x_X(X_i,Y_j) + \frac{\varepsilon^2}{2} x_{XX}(X_i,Y_j)+ O(\varepsilon^3) \\
& x_{i,j+1} - x_{i,j} = \varepsilon x_Y(X_i,Y_j) + \frac{\varepsilon^2}{2} x_{YY}(X_i,Y_j)+ O(\varepsilon^3) \\ 
& x_{i,j-1} - x_{i,j} = - \varepsilon x_Y(X_i,Y_j) + \frac{\varepsilon^2}{2} x_{YY}(X_i,Y_j)+ O(\varepsilon^3),
\end{aligned}
\end{equation}
and analogous equations hold for $y$. For convenience, we omit the point at which the derivatives are evaluated from now on. In what follows we assume that all relevant partial derivatives of $x$ and $y$ are $O(1)$. Assuming $\varepsilon \ll 1$, we apply \eqref{eqn:TaylorExpandDiscreteFirst}-\eqref{eqn:TaylorExpandDiscreteRest} to the dimensionless force balance \eqref{eqn:DimlessForceBalance} and bringing back the $1/(\varepsilon_c N) = \varepsilon/\varepsilon_c$ factor, we get
\begin{equation}\label{eqn:dimensionlessGeneralFullUpscale}
\begin{aligned}
& \bigg\{ \mathcal{F} \left( \xi \sqrt{\left(-x_X + \frac{\ve}{2} x_{XX} + O(\ve^2) \right)^2 + \left(-y_X + \frac{\ve}{2} y_{XX} + O(\ve^2)\right)^2} \right) \frac{\left( -x_X + \frac{\ve}{2} x_{XX} + O(\ve^2), -y_X + \frac{\ve}{2} y_{XX} + O(\ve^2) \right)}{\sqrt{\left(-x_X + \frac{\ve}{2} x_{XX} + O(\ve^2) \right)^2 + \left(-y_X + \frac{\ve}{2} y_{XX} + O(\ve^2)\right)^2}} + \\
& \mathcal{F} \left( \xi \sqrt{\left(x_X + \frac{\ve}{2} x_{XX} + O(\ve^2) \right)^2 + \left(y_X + \frac{\ve}{2} y_{XX} + O(\ve^2)\right)^2} \right) \frac{ \left( x_X + \frac{\ve}{2} x_{XX} + O(\ve^2), y_X + \frac{\ve}{2} y_{XX} + O(\ve^2) \right)}{\sqrt{\left(x_X + \frac{\ve}{2} x_{XX} + O(\ve^2) \right)^2 + \left(y_X + \frac{\ve}{2} y_{XX} + O(\ve^2)\right)^2}} + \\
& \mathcal{F} \left( \xi \sqrt{\left(-x_ Y+ \frac{\ve}{2} x_{YY} + O(\ve^2) \right)^2 + \left(-y_Y + \frac{\ve}{2} y_{YY} + O(\ve^2)\right)^2} \right) \frac{ \left( -x_Y + \frac{\ve}{2} x_{YY} + O(\ve^2), -y_Y + \frac{\ve}{2} y_{YY} + O(\ve^2) \right)}{\sqrt{\left(-x_Y + \frac{\ve}{2} x_{YY} + O(\ve^2) \right)^2 + \left(-y_Y + \frac{\ve}{2} y_{YY} + O(\ve^2)\right)^2}} + \\
& \mathcal{F} \left( \xi \sqrt{\left(x_Y + \frac{\ve}{2} x_{YY} + O(\ve^2) \right)^2 + \left(y_Y + \frac{\ve}{2} y_{YY} + O(\ve^2)\right)^2} \right) \frac{ \left( x_Y + \frac{\ve}{2} x_{YY} + O(\ve^2), y_Y + \frac{\ve}{2} y_{YY} + O(\ve^2) \right)}{\sqrt{\left(x_Y + \frac{\ve}{2} x_{YY} + O(\ve^2) \right)^2 + \left(y_Y + \frac{\ve}{2} y_{YY} + O(\ve^2)\right)^2}} \bigg\} \frac{\ve}{\varepsilon_{c}} = \boldsymbol{0}.
\end{aligned}
\end{equation}
Assuming $\varepsilon \ll 1$, we Taylor expand the denominators according to
\begin{equation}\label{eqn:TaylorExpandOneOverSqrt}
\frac{1}{\sqrt{A+B \varepsilon + C \ve^2 + O(\varepsilon^3)}} = \frac{1}{\sqrt{A}} - \frac{B \varepsilon}{2 A^{3/2}} + \ve^2 \frac{3 B^2 - 4 A C}{8 A^{5/2}} + O(\varepsilon^3)
\end{equation}
and those in the discrete bending terms (with $\ve^2$ cancelling out as a common term in both the numerator and denominator) according to
\begin{equation}\label{eqn:TaylorExpandOneOver}
\frac{1}{A+B \varepsilon + O(\varepsilon^2)} = \frac{1}{A} - \frac{B \varepsilon}{A^2} + O(\varepsilon^2).
\end{equation}
The nonlinear force terms are expanded as
\begin{equation}\label{eqn:expandNonlinearForce}
\mathcal{F}(\psi(\ve)) = \mathcal{F}(\psi(0)) + \ve \mathcal{F}'(\psi(0)) \psi'(0) + O(\ve^2)
\end{equation}
which holds for sufficiently smooth functions $\mathcal{F}$ and $\psi$. We denote
$$\lambda^X(X,Y) = \xi \sqrt{x_X^2 + y_X^2} \qquad \lambda^Y(X,Y) = \xi \sqrt{x_Y^2 + y_Y^2}$$
using which the $X-$component of the force balance can be simplified to
\begin{equation}
\begin{aligned}
& \bigg\{ \left( -x_X + \frac{\ve}{2} x_{XX} + O(\ve^2) \right) \left( \frac{1}{\sqrt{x_X^2 + y_X^2}} + \frac{\ve (x_X x_{XX}+y_X y_{XX})}{2 (x_X^2 + y_X^2)^{3/2}} + O(\ve^2) \right) \left( \mathcal{F} \left( \lambda^X \right) - \ve \xi \mathcal{F}' \left(\lambda^X \right) \frac{x_X x_{XX} + y_X y_{XX}}{2 \sqrt{x_X^2 + y_X^2}} + O(\ve^2) \right) + \\
& \left(x_X + \frac{\ve}{2} x_{XX} + O(\ve^2) \right) \left( \frac{1}{\sqrt{x_X^2 + y_X^2}} - \frac{\ve (x_X x_{XX}+y_X y_{XX})}{2 (x_X^2 + y_X^2)^{3/2}} + O(\ve^2) \right) \left( \mathcal{F} \left( \lambda^X \right) + \ve \xi \mathcal{F}' \left(\lambda^X \right) \frac{x_X x_{XX} + y_X y_{XX}}{2 \sqrt{x_X^2 + y_X^2}} + O(\ve^2) \right) + \\
& \left( -x_Y + \frac{\ve}{2} x_{YY} + O(\ve^2) \right) \left( \frac{1}{\sqrt{x_Y^2 + y_Y^2}} + \frac{\ve (x_Y x_{YY}+y_Y y_{YY})}{2 (x_Y^2 + y_Y^2)^{3/2}} + O(\ve^2) \right) \left( \mathcal{F} \left( \lambda^Y \right) - \ve \xi \mathcal{F}' \left(\lambda^Y \right) \frac{x_Y x_{YY} + y_Y y_{YY}}{2 \sqrt{x_Y^2 + y_Y^2}} + O(\ve^2) \right) + \\
& \left( x_Y + \frac{\ve}{2} x_{YY} + O(\ve^2) \right) \left( \frac{1}{\sqrt{x_Y^2 + y_Y^2}} - \frac{\ve (x_Y x_{YY}+y_Y y_{YY})}{2 (x_Y^2 + y_Y^2)^{3/2}} + O(\ve^2) \right) \left( \mathcal{F} \left( \lambda^Y \right) + \ve \xi \mathcal{F}' \left(\lambda^Y \right) \frac{x_Y x_{YY} + y_Y y_{YY}}{2 \sqrt{x_Y^2 + y_Y^2}} + O(\ve^2) \right) \bigg\} \frac{\ve}{\varepsilon_{c}} = 0
\end{aligned}
\end{equation}
and the $Y-$component to
\begin{equation}
\begin{aligned}
& \bigg\{ \left( -y_X + \frac{\ve}{2} y_{XX} + O(\ve^2) \right) \left( \frac{1}{\sqrt{x_X^2 + y_X^2}} + \frac{\ve (x_X x_{XX}+y_X y_{XX})}{2 (x_X^2 + y_X^2)^{3/2}} + O(\ve^2) \right) \left( \mathcal{F} \left( \lambda^X \right) - \ve \xi \mathcal{F}' \left(\lambda^X \right) \frac{x_X x_{XX} + y_X y_{XX}}{2 \sqrt{x_X^2 + y_X^2}} + O(\ve^2) \right) + \\
& \left(y_X + \frac{\ve}{2} y_{XX} + O(\ve^2) \right) \left( \frac{1}{\sqrt{x_X^2 + y_X^2}} - \frac{\ve (x_X x_{XX}+y_X y_{XX})}{2 (x_X^2 + y_X^2)^{3/2}} + O(\ve^2) \right) \left( \mathcal{F} \left( \lambda^X \right) + \ve \xi \mathcal{F}' \left(\lambda^X \right) \frac{x_X x_{XX} + y_X y_{XX}}{2 \sqrt{x_X^2 + y_X^2}} + O(\ve^2) \right) + \\
& \left( -y_Y + \frac{\ve}{2} y_{YY} + O(\ve^2) \right) \left( \frac{1}{\sqrt{x_Y^2 + y_Y^2}} + \frac{\ve (x_Y x_{YY}+y_Y y_{YY})}{2 (x_Y^2 + y_Y^2)^{3/2}} + O(\ve^2) \right) \left( \mathcal{F} \left( \lambda^Y \right) - \ve \xi \mathcal{F}' \left(\lambda^Y \right) \frac{x_Y x_{YY} + y_Y y_{YY}}{2 \sqrt{x_Y^2 + y_Y^2}} + O(\ve^2) \right) + \\
& \left( y_Y + \frac{\ve}{2} y_{YY} + O(\ve^2) \right) \left( \frac{1}{\sqrt{x_Y^2 + y_Y^2}} - \frac{\ve (x_Y x_{YY}+y_Y y_{YY})}{2 (x_Y^2 + y_Y^2)^{3/2}} + O(\ve^2) \right) \left( \mathcal{F} \left( \lambda^Y \right) + \ve \xi \mathcal{F}' \left(\lambda^Y \right) \frac{x_Y x_{YY} + y_Y y_{YY}}{2 \sqrt{x_Y^2 + y_Y^2}} + O(\ve^2) \right)  \bigg\} \frac{\ve}{\varepsilon_{c}} = 0.
\end{aligned}
\end{equation}
At $O(1)$ and $O(\ve)$, the balance is automatically satisfied. Returning to vector form, at $O(\ve^2)$ we get 
$$ \boldsymbol{0} = \frac{1}{\varepsilon_{c}} \left\{ \mathcal{F}(\lambda^X) \left( \frac{(x_X,y_X)}{\sqrt{x_X^2 + y_X^2}} \right)_X + \xi \frac{x_X x_{XX} + y_X y_{XX}}{x_X^2 + y_X^2} \mathcal{F}'(\lambda^X)(x_X,y_X) + \mathcal{F}(\lambda^Y) \left( \frac{(x_Y,y_Y)}{\sqrt{x_Y^2 + y_Y^2}} \right)_Y + \right.  $$
$$ \left.  \xi \frac{x_Y x_{YY} + y_Y y_{YY}}{x_Y^2 + y_Y^2} \mathcal{F}'(\lambda^Y) (x_Y, y_Y)  \right\}.$$
Recalling the definitions of $\lambda^X$ and $\lambda^Y$ and multiplying by $\varepsilon_c$, we conclude the macroscale force balance \eqref{eqn:EffectiveGeneralNonlinearFilamentNoBending}.

\subsubsection{Deriving strain energy density}

We further deduce the strain energy function for the derived continuum problem. As before, we nondimensionalize the lengths with respect to $\tilde{D}$ and the forces with $\pi \tilde{Y} \tilde{b}_{c}^2$ (so that the energies are nondimensionalized with $\pi \tilde{D} \tilde{Y} \tilde{b}_{c}^2$), Taylor expand the dimensional energy stored at CL $(i,j)$ \eqref{eqn:EnergyAtCLij} and upon further simplifications get
$$ e_{i,j}(\ve) = \frac{\ve^2}{\varepsilon_{c}^2} \left( \mathcal{E} \left( \xi \sqrt{x_X^2 + y_X^2} \right) + \mathcal{E} \left( \xi \sqrt{x_Y^2 + y_Y^2} \right) + O(\ve) \right), $$
where $\mathcal{E}$ is the dimensionless counterpart of the dimensional $\tilde{\mathcal{E}}$ defined in \eqref{eqn:MacroscaleEnergyE}. To arrive at macroscale strain energy density $W$, we divide by an area corresponding to one CL in the undeformed configuration $\ve^2$ and sending $\ve \to 0$ ($N \to \infty$) obtain
\begin{equation}\label{eqn:StrainEnergy1}
W =   \frac{1}{\varepsilon_{c}^2} \left( \mathcal{E} \left( \xi \sqrt{x_X^2 + y_X^2} \right) + \mathcal{E} \left( \xi \sqrt{x_Y^2 + y_Y^2} \right) \right).
\end{equation}

\subsection{Deducing the continuum problem under the linearized constitutive law}
\label{sec:SuppDeduceContinuumLinearSprings}

Under the linearized microscale constitutive law for vimentin \eqref{eqn:LinearSpringsDimlessSmallTau1NoCompression},
the continuum problem \eqref{eqn:EffectiveGeneralNonlinearFilamentNoBending} takes (upon dividing by $\xi$) the form
\begin{equation}\label{eqn:EffectiveContinuumDivergence}
\left( (x_X,y_X) \max \left\{0, 1 - \frac{1}{\xi \sqrt{x_X^2+y_X^2}} \right\} \right)_X + \left( (x_Y,y_Y) \max \left\{0, 1 - \frac{1}{\xi \sqrt{y_Y^2+x_Y^2}} \right\} \right)_Y = \boldsymbol{0}.
\end{equation}
We further state the dimensional stress tensor \eqref{eqn:StressTensor1} under \eqref{eqn:LinearSpringsDimlessSmallTau1NoCompression} to be
\begin{gather}\label{eqn:StressTensor1Linear}
\tilde{\boldsymbol{S}} = \frac{\pi \tilde{Y} \tilde{b}_{c}^2}{\tilde{\Lambda}_{c}}
\begin{pmatrix}
  \displaystyle \tilde{x}_{\tilde{X}} \max \left\{0, 1 - \frac{1}{\xi \sqrt{\tilde{x}_{\tilde{X}}^2+\tilde{y}_{\tilde{X}}^2}} \right\}  &
  \displaystyle  \tilde{y}_{\tilde{X}} \max \left\{0, 1 - \frac{1}{\xi \sqrt{\tilde{x}_{\tilde{X}}^2+\tilde{y}_{\tilde{X}}^2}} \right\}  \\
  \displaystyle \tilde{x}_{\tilde{Y}} \max \left\{0, 1 - \frac{1}{\xi \sqrt{\tilde{x}_{\tilde{Y}}^2 + \tilde{y}_{\tilde{Y}}^2}} \right\}  &
  \displaystyle \tilde{y}_{\tilde{Y}} \max \left\{0, 1 - \frac{1}{\xi \sqrt{\tilde{x}_{\tilde{Y}}^2 + \tilde{y}_{\tilde{Y}}^2}} \right\}
\end{pmatrix},
\end{gather}
and the dimensional counterpart of \eqref{eqn:EffectiveContinuumDivergence} then reads
\begin{equation}\label{eqn:EffectiveContinuumDivergenceDimensional}
\begin{aligned}
& \frac{\pi \tilde{Y} \tilde{b}_{c}^2}{\tilde{\Lambda}_{c}} \left\{ \left( (\tilde{x}_{\tilde{X}},\tilde{y}_{\tilde{X}}) \max \left\{0, 1 - \frac{1}{\xi \sqrt{\tilde{x}_{\tilde{X}}^2+\tilde{y}_{\tilde{X}}^2}} \right\} \right)_{\tilde{X}} + \left( (\tilde{x}_{\tilde{Y}},\tilde{y}_{\tilde{Y}}) \max \left\{0, 1 - \frac{1}{\xi \sqrt{\tilde{y}_{\tilde{Y}}^2+\tilde{x}_{\tilde{Y}}^2}} \right\} \right)_{\tilde{Y}} \right\} = \boldsymbol{0},
\end{aligned}
\end{equation}
which is used in continuum simulations. Using the displacement field
$$ \tilde{u}(\tilde{X},\tilde{Y}) = \tilde{x}(\tilde{X},\tilde{Y}) - \tilde{X} \qquad  \tilde{v}(\tilde{X},\tilde{Y}) = \tilde{y}(\tilde{X},\tilde{Y}) - \tilde{Y},$$
this can be rewritten as
\begin{equation}\label{eqn:EffectiveContinuumDivergenceDimensional_Displacements}
\begin{aligned}
& \frac{\pi \tilde{Y} \tilde{b}_{c}^2}{\tilde{\Lambda}_{c}} \left\{ \left( (1 + \tilde{u}_{\tilde{X}},\tilde{v}_{\tilde{X}}) \max \left\{0, 1 - \frac{1}{\xi \sqrt{\left( 1 + \tilde{u}_{\tilde{X}} \right)^2+\tilde{v}_{\tilde{X}}^2}} \right\} \right)_{\tilde{X}} + \left( (\tilde{u}_{\tilde{Y}}, 1+ \tilde{v}_{\tilde{Y}})  \max \left\{0, 1 - \frac{1}{\xi \sqrt{\tilde{u}_{\tilde{Y}}^2 + \left( 1+ \tilde{v}_{\tilde{Y}} \right)^2}} \right\} \right)_{\tilde{Y}} \right\} = \boldsymbol{0}.
\end{aligned}
\end{equation}

\subsection{Strain energy density and nominal stress tensor}
\label{sec:StrainEnergyFiberReinforced}

\subsubsection{Relation to nonlinear elasticity models for fiber-reinforced materials} 

\begin{figure}[ht]
\centering
\begin{overpic}[width=1.0\textwidth,tics=10]{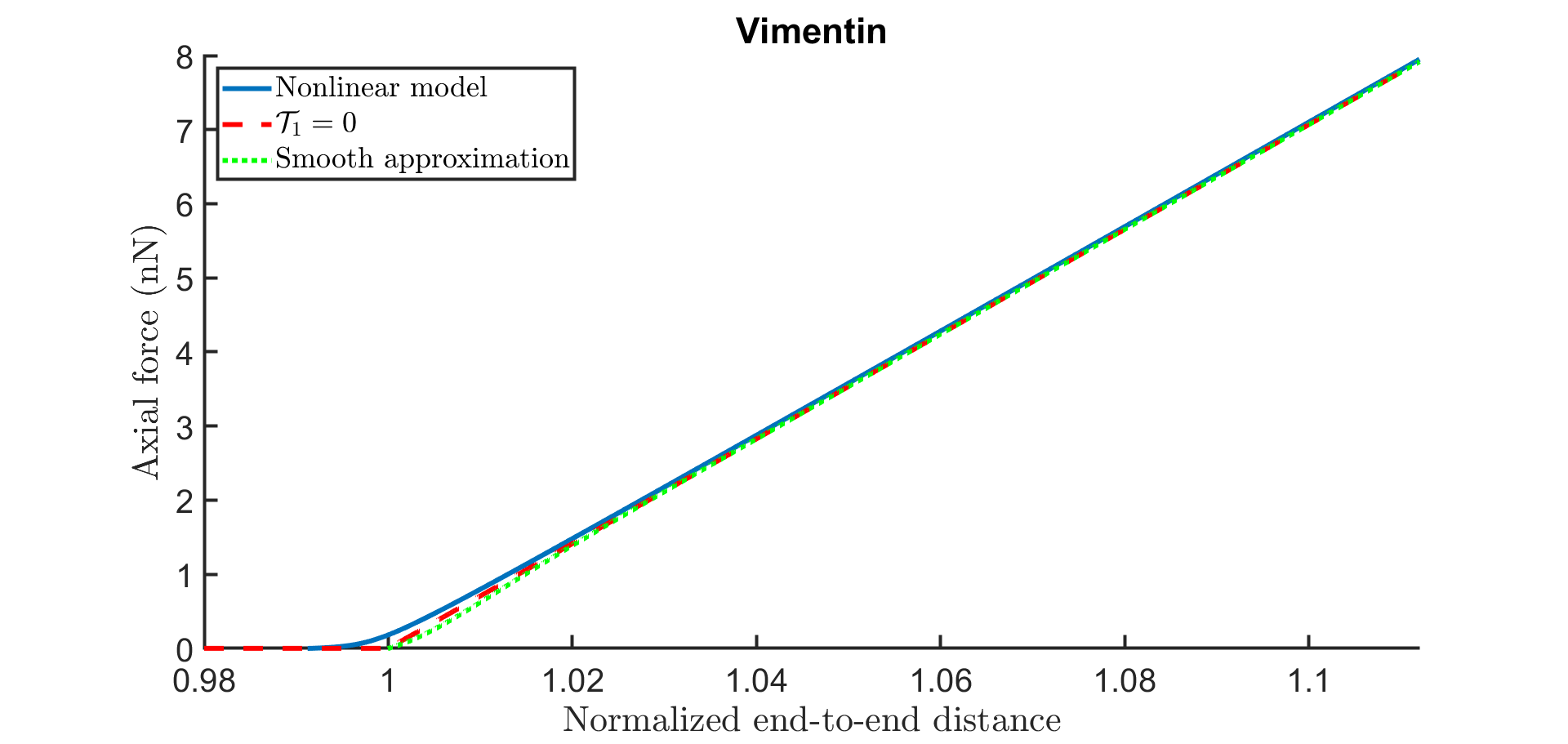}
\end{overpic}
\caption{The smooth approximation \eqref{eqn:SmoothApproximation} with $\kappa=200$ yields negligible changes to the microscale constitutive law for vimentin.}
\label{fig:SmoothingWorks}
\end{figure}

Redimensionalizing \eqref{eqn:StrainEnergy1} (in order to facilitate comparison with the standard results on fiber-reinforced materials), the strain energy density can be rewritten in terms of the deformation gradient tensor $\boldsymbol{F}$ with components $F_{kl} = \partial \tilde{x}_i / \partial \tilde{X}_{j}$ or in terms of the right Cauchy-Green deformation tensor $\boldsymbol{C} = \boldsymbol{F}^{T} \boldsymbol{F}$ as
\begin{equation}\label{eqn:StrainEnergyAsFunctionOfDefGradGeneral}
\tilde{W} = \frac{ \displaystyle \tilde{\mathcal{E}}  \left( \xi \sqrt{F_{11}^2 + F_{21}^2} \right) + \tilde{\mathcal{E}}  \left( \xi \sqrt{F_{12}^2 + F_{22}^2} \right)}{\tilde{R}_{c}^2} = \frac{ \tilde{\mathcal{E}} \left( \xi \sqrt{C_{11}} \right) + \tilde{\mathcal{E}} \left( \xi \sqrt{C_{22}} \right) }{\tilde{R}_{c}^2}.
\end{equation}
Introducing $\boldsymbol{M} = (1,0)$ and $\boldsymbol{M'} = (0,1)$ as the two directions of filaments in the undeformed configuration and employing the theory of fiber-reinforced materials, two invariants corresponding to these directions take forms $I_4 = \boldsymbol{M} \cdot (\boldsymbol{C} \boldsymbol{M}) = C_{11}$ and $I_6 = \boldsymbol{M'} \cdot (\boldsymbol{C} \boldsymbol{M'}) = C_{22}$. We can therefore express $\tilde{W}$ also in terms of the invariants of $\boldsymbol{C}$ as \eqref{eqn:StrainEnergyAsFunctionOfCInvariants}, thus establishing connection to the rich literature on constitutive modelling of fiber-reinforced materials. Using \eqref{eqn:SplittingEnergyPrestressDef}, we can express this strain energy density as
\begin{equation}\label{eqn:StrainEnergyAsFunctionOfCInvariants_Forces}
\tilde{W}(\boldsymbol{C})=  \frac{\displaystyle 2 \int\limits_{\xi_{sf}}^{\xi} \tilde{\mathcal{F}}(t) dt + \int\limits_{\xi}^{\xi \sqrt{I_4(\boldsymbol{C})}} \tilde{\mathcal{F}}(t) dt + \int\limits_{\xi}^{\xi \sqrt{I_6(\boldsymbol{C})}} \tilde{\mathcal{F}}(t) dt}{\xi \tilde{R}_{c}},
\end{equation}
where $\xi_{sf}$ denotes the normalized stress-free end-to-end distance so that the first integral represents the strain energy stored in the undeformed domain due to pre-stress (noting that in the undeformed configuration one has $\boldsymbol{F} = \boldsymbol{I}$ and $I_4(\boldsymbol{C}) = I_6(\boldsymbol{C}) = 1$) and the last two integrals represent the elastically-stored energy supplied with the deformation. Finally, assuming the approximation \eqref{eqn:LinearSpringsDimlessSmallTau1NoCompression}, the strain energy density \eqref{eqn:StrainEnergyAsFunctionOfCInvariants} can be written as
\begin{equation}\label{eqn:StrainEnergyLinearSprings}
\tilde{W}(\boldsymbol{C}) = \frac{\pi \tilde{Y} \tilde{b}_{c}^2 \xi}{2 \tilde{R}_{c}} \left[ \max \left(0, \sqrt{I_4 (\boldsymbol{C})} - \frac{1}{\xi} \right)^2 + \max \left(0, \sqrt{I_6 (\boldsymbol{C})} - \frac{1}{\xi} \right)^2\right],
\end{equation}
where $\xi$ is approximated using \eqref{eqn:TwoTermApproximationPreStretchPrestress}. The problem \eqref{eqn:EffectiveContinuumDivergenceDimensional_Displacements} can thus be re-formulated in the framework of nonlinear elasticity as minimization of the strain energy \eqref{eqn:StrainEnergyLinearSprings}. We further propose a smooth approximation to the microscale constitutive law and using
\begin{equation}\label{eqn:SmoothApproximation}
\max{\left(0,x \right)} \approx \frac{x}{1 + e^{-\kappa x}},
\end{equation}
which provides a good approximation to the maximum function for large $\kappa$, we get a smooth approximation to the strain energy
\begin{equation}\label{eqn:StrainEnergyLinearSpringsSmooth}
\tilde{W}(\boldsymbol{C}) = \frac{\pi \tilde{Y} \tilde{b}_{c}^2 \xi}{2 \tilde{R}_{c}} \left[ \left( \frac{\displaystyle \sqrt{I_4 (\boldsymbol{C})} - 1/\xi}{ 1+e^{ -\kappa \left(\sqrt{I_4 (\boldsymbol{C})} - 1/\xi \right)} } \right)^2 + \left( \frac{\displaystyle \sqrt{I_6 (\boldsymbol{C})} - 1/\xi}{ 1+e^{ -\kappa \left(\sqrt{I_6 (\boldsymbol{C})} - 1/\xi \right)} } \right)^2 \right],
\end{equation}
the minimization of which is implemented in our FEniCS code. Figure \ref{fig:SmoothingWorks} documents that the approximation \eqref{eqn:SmoothApproximation} with $\kappa =200$ leads to only negligible changes in the microscale constitutive law for vimentin.

\subsubsection{Stress tensor}

Next, we deduce the components of the nominal stress tensor $\tilde{\boldsymbol{S}}$ from $\tilde{S}_{kl} = \partial \tilde{W}/\partial F_{ji}$ getting \eqref{eqn:StressTensor1}. Storm et al. \cite{Storm2005Supp} applied the Doi-Edwards construction \cite{Doi1988} to a crosslinked network with an arbitrary distribution $\tilde{\Psi}$ of end-to-end separation vectors $\tilde{\boldsymbol{r}}$ and arrived at an averaged Cauchy stress tensor $\tilde{\boldsymbol{\sigma}}$ of the form
$$ \tilde{\sigma}_{kl}^T = \frac{\tilde{\varrho}}{\text{det}(\boldsymbol{F})} \left\langle \tilde{f}(\abs{\boldsymbol{F} \tilde{\boldsymbol{r}}}) \frac{F_{il} \tilde{r}_l F_{jk} \tilde{r}_k}{\abs{\boldsymbol{F} \tilde{\boldsymbol{r}}}} \right\rangle_{\tilde{\Psi}(\tilde{\boldsymbol{r}})}$$
where $\tilde{\varrho}$ denotes the number of FSs per unit volume (otherwise their notation coincides with ours). Note that the microscale force was given as function of the dimensional end-to-end distance $\tilde{r}$, as opposed to the dimensionless distance $r$. Letting $\tilde{\delta}(\tilde{\boldsymbol{r}})$ denote the Dirac delta function centered at $\tilde{\boldsymbol{r}}= \boldsymbol{0}$ and applying this formula to our two-dimensional case with 
$$\tilde{\Psi}(\tilde{\boldsymbol{r}}) =  \frac{1}{4} \left\{ \tilde{\delta}(\tilde{\boldsymbol{r}} - (\tilde{R},0)) + \tilde{\delta}(\tilde{\boldsymbol{r}} - (-\tilde{R},0)) + \tilde{\delta}(\tilde{\boldsymbol{r}} - (0,\tilde{R})) + \tilde{\delta}(\tilde{\boldsymbol{r}} - (0,-\tilde{R})) \right\} $$
reflecting the undeformed orientations of filaments in our geometry and with $\tilde{\varrho} = 2 N^2/\tilde{D}^2$, we arrive at a result identical to the one obtained when the connection $\tilde{\boldsymbol{\sigma}}^T = \text{det}(\boldsymbol{F})^{-1} \tilde{\boldsymbol{S}}^T \boldsymbol{F}^T$ from nonlinear elasticity \cite{Ogden1997Supp} is applied to \eqref{eqn:StressTensor1}, which further certifies the correctness of our results. 

\restoregeometry

\section{Details of the small-deformations and small-bead analysis}
\label{sec:SmallDefAndSmallBeadDetails}
\subsection{Deriving small-deformations limit}
\label{sec:appDeriveSmallDeform}
Substituting \eqref{eqn:SmallDeform} into \eqref{eqn:EffectiveGeneralNonlinearFilamentNoBending} we get
\begin{adjustwidth}{-10pt}{1pt}
\begin{equation}\label{eqn:EffectiveGeneralNonlinearFilamentNoBendingXYExplicit2}
\begin{aligned}
& \boldsymbol{0} = \left( \mathcal{F} \left( \xi \sqrt{(1 + R_{b} \hat{x}_X)^2 +(R_{b} \hat{y}_X)^2 + O(R_{b}^2)} \right)  \frac{(1+R_{b} \hat{x}_X + O(R_{b}^2), R_{b} \hat{y}_X + O(R_{b}^2))}{\sqrt{(1+R_{b} \hat{x}_X)^2 + (R_{b} \hat{y}_X)^2 +O(R_{b}^2)}} \right)_X +  \\
& \left( \mathcal{F} \left( \xi \sqrt{(R_{b} \hat{x}_Y)^2 +(1+ R_{b} \hat{y}_Y)^2 + O(R_{b}^2)} \right)  \frac{(R_{b} \hat{x}_Y + O(R_{b}^2), 1 + R_{b} \hat{y}_Y + O(R_{b}^2))}{\sqrt{(R_{b} \hat{x}_Y)^2 + (1+ R_{b} \hat{y}_Y)^2 + O(R_{b}^2)}} \right)_Y .
\end{aligned}
\end{equation}
\end{adjustwidth}
Taylor expanding $\mathcal{F}$ as well as the denominators for $R_{b} \ll 1$, we get 
\begin{equation}\label{eqn:EffectiveGeneralNonlinearFilamentNoBendingXYExplicit3}
\begin{aligned}
& \boldsymbol{0} = \left(\left( \mathcal{F}(\xi) + R_{b} \xi \mathcal{F}'(\xi) \hat{x}_X + O(R_{b}^2)  \right)  (1+R_{b} \hat{x}_X + O(R_{b}^2), R_{b} \hat{y}_X + O(R_{b}^2)) \left(1 - R_{b} \hat{x}_X + O(R_{b}^2) \right) \right)_X +  \\
& \left( \left( \mathcal{F}(\xi) + R_{b} \xi \mathcal{F}'(\xi) \hat{y}_Y + O(R_{b}^2) \right) (R_{b} \hat{x}_Y + O(R_{b}^2), 1 + R_{b} \hat{y}_Y + O(R_{b}^2)) \left(1 - R_{b} \hat{y}_Y + O(R_{b}^2) \right) \right)_Y.
\end{aligned}
\end{equation}
The balance at $O(1)$ is trivially satisfied. At $O(R_{b})$, we get \eqref{eqn:EffectiveGeneralNonlinearFilamentNoBendingXODelta}-\eqref{eqn:EffectiveGeneralNonlinearFilamentNoBendingYODelta} in the main text.

\subsection{Details of small-bead asymptotics}
\label{sec:App_SmallBeadAsymptotics}
\subsubsection{$\hat{x}$ problem}

We split the analysis into the inner (boundary) region characterized by
$$ \frac{X}{a} = \bar{X} = O(1) \quad \frac{Y}{a} = \bar{Y} = O(1)$$
where we use the ansatz for inner solution
$$\hat{x}^I(\bar{X},\bar{Y},a) = \hat{x}_0^I(\bar{X},\bar{Y}) + \frac{1}{\ln{(1/a)}} \hat{x}_1^I(\bar{X},\bar{Y}) + \frac{1}{\ln^2{(1/a)}} \hat{x}_2^I(\bar{X},\bar{Y}) + O \left( \frac{1}{\ln^3{(1/a)}}\right)$$
satisfying the boundary condition at $\bar{X}^2 + \bar{Y}^2 = 1$, and the outer region $X = O(1) = Y$ with the outer solution
$$\hat{x}^O(X,Y,a) = \hat{x}_0^O(X,Y) + \frac{1}{\ln{(1/a)}} \hat{x}_1^O(X,Y) + \frac{1}{\ln^2{(1/a)}} \hat{x}_2^O(\bar{X},\bar{Y}) + O \left( \frac{1}{\ln^3{(1/a)}}\right)$$
satisfying the Dirichlet condition at the outer boundary $\hat{x}^O = 0$. The rationale behind the logarithmic terms in the expansions will become apparent in the course of the analysis. In the inner region, we transform the $\bar{Y}$ coordinate according to $\bar{Y} = \sqrt{\omega} \bar{Z}$ so that we get Laplace's equation for the leading-order inner solution
$$ \hat{x}^I_{0,\bar{X} \bar{X}} + \hat{x}^I_{0,\bar{Z} \bar{Z}} = 0 $$
on a domain with an ellipse cut out of it as depicted in Figure \ref{fig:ellipticalGeo}. We introduce elliptical coordinates
\begin{equation}\label{eqn:ElliptCoordinates}
\bar{X} = c \sinh(\mu) \sin(\nu) \qquad \bar{Z} = c \cosh(\mu) \cos(\nu)
\end{equation}
where $c = \sqrt{(1-\omega)/\omega}$ is the (linear) eccentricity of the inner ellipse, $(\mu, \nu) \in (\mu_1, \infty) \times [0, 2 \pi]$ so that $\mu= \mu_1 = \cosh^{-1}{\left((1-\omega)^{-1/2}\right)}$ represents the elliptical (inner) boundary. Note that even if our outer domain boundary is a circle in $(X,Y)$ coordinates (a square domain being even less amenable to analysis), in $(\bar{X},\bar{Z})$ coordinates it transforms into an ellipse with the same eccentricity as the inner (bead) elliptic boundary and therefore it cannot be simply characterized by $\mu = \mu_2$ for some $\mu_2 > \mu_1$ (because the eccentricity of ellipses given by $\mu = $ constant in elliptical coordinates strictly decreases with $\mu$). This causes the full problem (with $a>0$) to be analytically intractable and forced us to only study the $a \ll 1$ limit.
Writing $\Phi_0^I(\mu, \nu) = \hat{x}_0^I(\bar{X},\bar{Z})$ we then have $\Phi_0^I(\mu_1,\nu) = \cos{(\varphi_*)}$ for all $\nu$ and
$$ \frac{1}{c^2 (\cosh^2(\mu) + \sin^2(\nu))} \left( \Phi_{0, \mu \mu}^I + \Phi_{0, \nu \nu}^I \right) = 0  $$
and thus
$$\Phi_{0, \mu \mu}^I + \Phi_{0, \nu \nu}^I  = 0.$$
We assume $\Phi_0^I$ to be $2 \pi-$periodic in $\nu$. As nothing drives the variation in $\nu$ in the inner layer, we search for a solution in the form $\Phi_0^I(\mu,\nu) = \Phi_0^I(\mu)$ solving $\Phi_{0, \mu \mu}^I = 0$. The solution reads $\Phi_0^I = A_0 \mu + B_0$. Following the same line of reasoning, we get that the higher-order terms $\Phi_i^I(\mu,\nu) = \hat{x}_i^I(\bar{X},\bar{Z})$ are of the same form $\Phi_i^I = A_i \mu + B_i$. We transform back to Cartesian coordinates \cite{Sun2017} so that
\begin{equation}\label{eqn:ElliptCoordMu}
\mu = \frac{1}{2} \ln{\left( 1- 2q (\bar{X}, \bar{Z}) + 2 \sqrt{q^2(\bar{X}, \bar{Z}) - q(\bar{X}, \bar{Z})} \right)}
\end{equation}
with
$$q(\bar{X}, \bar{Z}) = \frac{- (\bar{X}^2 + \bar{Z}^2 - c^2) - \sqrt{(\bar{X}^2 + \bar{Z}^2 - c^2)^2 + 4 c^2 \bar{X}^2}}{2 c^2}.$$
In $(\bar{X},\bar{Y})$ variables and expressed using $\omega$, $q$ reads \eqref{eqn:qAsFtionOfXY}. Next we wish to write $\hat{x}_0^I$ in the outer coordinates $(X,Z)$ for the purposes of matching with the outer layer. To do this, we first rewrite $q$ in these variables and expand in $a$ as
$$q = \frac{- (X/a)^2 - (Z/a)^2 + c^2 - \sqrt{\left((X/a)^2 + (Z/a)^2 - c^2 \right)^2 + 4 c^2 (X/a)^2}}{2 c^2} = $$
$$ \frac{- (X/a)^2 - (Z/a)^2 + c^2 - 1/a^2 \times \sqrt{(X^2+Z^2)^2 + a^2 c^2 (2X^2 - 2 Z^2) + a^4 c^4} }{2 c^2} = \frac{1}{a^2} \left( - \frac{X^2 + Z^2}{c^2} \right) + O(1)$$
and then substitute it into \eqref{eqn:ElliptCoordMu} to get
$$ \mu = \frac{1}{2} \ln{\left( 1 + \frac{2(X^2 + Z^2)}{a^2 c^2} + O(1) + 2 \sqrt{\left( \frac{X^2 + Z^2}{a^2 c^2} \right)^2 + O\left(\frac{1}{a^2}\right) + \frac{X^2 + Z^2}{a^2 c^2} + O(1) } \right)} =$$
$$ \frac{1}{2} \ln{\left( \frac{4(X^2 + Z^2)}{a^2 c^2} +O(1) \right)} = \ln{\left( \frac{1}{a} \right)} + \ln{\left( \frac{2}{c} \sqrt{X^2 + Z^2} \right)} + O(a^2). $$
Thus we see that $A_0$ must equal $0$ because otherwise the matching would require a contribution of order $\ln{(1/a)} \gg 1$ to exist in the outer solution. The inner boundary condition at the leading order then enforces $B_0 = \cos(\varphi_*)$. Our inner approximation thus so far reads
\begin{equation}\label{eqn:InnerApproximationLeadingOrderDone}
\hat{x}^I = \cos(\varphi_*) + \frac{1}{\ln{(1/a)}} \left( A_1 \mu + B_1 \right) + \frac{1}{\ln^2{(1/a)}} \left( A_2 \mu + B_2 \right)   + O\left(\frac{1}{\ln^3{(1/a)}}\right)
\end{equation}
and writing this in outer variables we get
\begin{equation}\label{eqn:smalla_innerInOuterVar}
\cos(\varphi_*) + A_1 + O\left(\frac{1}{\ln{(1/a)}}\right).
\end{equation}
The leading order outer solution satisfies Laplace's equation (in $(X,Z)$ variables) and Dirichlet boundary condition $\hat{x}_0^O = 0$ at the outer boundary. Irrespective of whether we assume this outer boundary to be a circle or a square in the original - i.e. $(X,Y)$ - variables, the only admissible constant solution to this problem is $\hat{x}_0^O \equiv 0$. Comparing this with \eqref{eqn:smalla_innerInOuterVar} we conclude that the matching requires $A_1 = - \cos(\varphi_*)$. Finally, to satisfy the inner boundary condition at $O\left(\frac{1}{\ln{(1/a)}}\right)$, we must have $B_1 = - A_1 \mu_1 = \cos{(\varphi_*)} \cosh^{-1}\left( (1-\omega)^{-1/2} \right)$. Substituting $A_1$ and $B_1$ back into \eqref{eqn:InnerApproximationLeadingOrderDone} we get
\begin{equation}\label{eqn:InnerApproximationLeadingOrderDone2}
\hat{x}^I = \cos(\varphi_*) + \frac{\cos(\varphi_*) \left(  - \mu + \cosh^{-1}\left( (1-\omega)^{-1/2} \right) \right)}{\ln{(1/a)}} + \frac{1}{\ln^2{(1/a)}} \left( A_2 \mu + B_2 \right)   + O\left(\frac{1}{\ln^3{(1/a)}}\right).
\end{equation}
Writing this in outer variables we get
\begin{equation}\label{eqn:smalla_innerInOuterVarOneUp}
0 + \frac{1}{\ln{(1/a)}} \left( \cos(\varphi_*) \left(- \ln{(2/c)} - \ln{\left(\sqrt{X^2+Z^2}\right)} + \cosh^{-1}\left( (1-\omega)^{-1/2} \right)  \right) + A_2 \right).
\end{equation}
Note that to match the $\ln{\left( \sqrt{X^2 + Z^2} \right)}$ behaviour, the first-order correction in the outer solution must satisfy
$$ \hat{x}_{1 \, XX}^O + \hat{x}_{1 \, ZZ}^O = -2 \pi \cos{(\varphi_*)} \delta_{(0,0)}, $$
where $\delta_{(0,0)}$ denotes the Dirac delta function (centered at the origin), and must vanish at the outer boundary. Matching further requires $A_2 = \cos{(\varphi_*)} \left( \ln{(2/c)} - \cosh^{-1}\left( (1-\omega)^{-1/2} \right) \right) $.
The inner boundary condition at this order implies $B_2 = - \cosh^{-1}\left( (1-\omega)^{-1/2} \right) \cos{(\varphi_*)} \left( \ln{(2/c)} - \cosh^{-1}\left( (1-\omega)^{-1/2} \right) \right) $. By induction, we can deduce the form of general $A_i$ and $B_i$, getting $A_0 = 0$, $B_0 = \cos{(\varphi_*)}$ and for $i \geq 1$
\begin{equation}\label{eqn:GeneralAiBi}
A_i = - \cos{(\varphi_*)} \left( \cosh^{-1}{\left( (1-\omega)^{-1/2} \right)} - \ln{(2/c)} \right)^{i-1} \qquad B_i = -\cosh^{-1}{\left( (1-\omega)^{-1/2} \right)} A_i.
\end{equation}
The inner expansion thus reads
$$ \hat{x}^I = \cos{(\varphi_*)} + \sum\limits_{i=1}^{\infty} \frac{A_i \mu + B_i}{\ln^{i}{(1/a)}} + O(a) $$
which using \eqref{eqn:GeneralAiBi} and the formula for the sum of an infinite geometric series gives
$$ \hat{x}^I = \cos{(\varphi_*)} \left(1 + \frac{\cosh^{-1}{\left( (1-\omega)^{-1/2} \right)} - \mu}{\ln{(1/a)}} \sum\limits_{i=1}^{\infty} \left( \frac{\cosh^{-1}{\left( (1-\omega)^{-1/2} \right)} - \ln{(2/c)}}{\ln{(1/a)}} \right)^{i-1} \right) + O(a) =$$
$$ = \cos{(\varphi_*)} \left( 1 + \frac{\cosh^{-1}{\left( (1-\omega)^{-1/2} \right)} - \mu}{\ln{(1/a)} + \ln{(2/c)} - \cosh^{-1}{\left( (1-\omega)^{-1/2} \right)}} \right) + O(a). $$
Substituting $\mu$ from \eqref{eqn:ElliptCoordMu} and expressing $c$ in terms of $\omega$, we arrive at \eqref{eqn:Smalla_InnerApproximationx}.

\subsubsection{$\hat{y}$ problem}

The equation for $\hat{y}$ becomes Laplace's equation after transforming the $\bar{X}$ coordinate according to $\bar{X} = \sqrt{\omega} \bar{W}$ (keeping $\bar{Y}$) and we have the boundary condition $\hat{y} = \sin{(\varphi_*)}$ at the inner ellipse. We then must transform to elliptical coordinates (ellipses now being oriented along the $\bar{W}$ axis rather than $\bar{Z}$ axis) as
\begin{equation}\label{eqn:ElliptCoordinates_yProblem}
\bar{Y} = c \sinh(\mu) \sin(\nu) \qquad \bar{W} = c \cosh(\mu) \cos(\nu)
\end{equation}
where again $c = \sqrt{(1-\omega)/\omega}$ is the (linear) eccentricity of the inner ellipse and $\mu_1 = \cosh^{-1}{\left((1-\omega)^{-1/2}\right)}$ denotes the (inner) elliptical boundary. The solutions to the resulting inner problems again read $C_i \mu + D_i$ and we again have 
\begin{equation}\label{eqn:ElliptCoordMuForYProblem}
\mu = \frac{1}{2} \ln{\left( 1- 2q_2 (\bar{W}, \bar{Y}) + 2 \sqrt{q_2^2(\bar{W}, \bar{Y}) - q_2(\bar{W}, \bar{Y})} \right)}
\end{equation}
where we now have
$$q_2(\bar{W}, \bar{Y}) = \frac{- (\bar{W}^2 + \bar{Y}^2 - c^2) - \sqrt{(\bar{W}^2 + \bar{Y}^2 - c^2)^2 + 4 c^2 \bar{Y}^2}}{2 c^2}.$$
In $(\bar{X},\bar{Y})$ variables and expressed using $\omega$, $q_2$ reads \eqref{eqn:q2AsFtionOfXY}. As before, we write the inner solutions in terms of the outer variable and conclude that in order to match we must have $C_0 = 0$ and inner boundary condition at the leading order gives $D_0 = \sin{(\varphi_*)}$. At the higher orders, we (analogous to the $\hat{x}$ case) conclude
\begin{equation}\label{eqn:GeneralCiDi}
C_i = - \sin{(\varphi_*)} \left( \cosh^{-1}{\left( (1-\omega)^{-1/2} \right)} - \ln{(2/c)} \right)^{i-1} \qquad D_i = -\cosh^{-1}{\left( (1-\omega)^{-1/2} \right)} C_i
\end{equation}
which leads to the inner expansion \eqref{eqn:Smalla_InnerApproximationy} for $\hat{y}$.

\subsection{Leading-order approximations to the strain fields for small beads}

\label{sec:SuppLeadingOrderStrainFields}

We get
\begin{equation}\label{eqn:hatxInner}
\begin{aligned}
\hat{x}_{X/Y}^I = &- \frac{\cos{(\varphi_*)}}{2 \ln{(1/a)} + \ln{(4 \omega/(1-\omega))} - 2 \cosh^{-1}{\left( (1- \omega)^{-1/2} \right)}} \frac{1}{1 - 2 q + 2 \sqrt{q^2-q}} \left( -2 + \frac{2q-1}{\sqrt{q^2 - q}} \right) \frac{1}{a} \frac{\partial q}{\partial \bar{X}/\bar{Y}} +O(1) = \\
& \frac{1}{a} \frac{\cos{(\varphi_*)}}{2 \ln{(1/a)} + \ln{(4 \omega/(1-\omega))} - 2 \cosh^{-1}{\left( (1- \omega)^{-1/2} \right)}} \frac{1}{\sqrt{q^2 - q}} \frac{\partial q}{\partial \bar{X}/\bar{Y}} +O(1)
\end{aligned}
\end{equation}
\begin{equation}\label{eqn:hatyInner}
\begin{aligned}
& \hat{y}_{X/Y}^I = \frac{1}{a} \frac{\sin{(\varphi_*)}}{2 \ln{(1/a)} + \ln{(4 \omega/(1-\omega))} - 2 \cosh^{-1}{\left( (1- \omega)^{-1/2} \right)}} \frac{1}{\sqrt{q_2^2 - q_2}} \frac{\partial q_2}{\partial \bar{X}/\bar{Y}} +O(1)
\end{aligned}
\end{equation}
where $q(\bar{X},\bar{Y})$ and $q_2(\bar{X},\bar{Y})$ are given by \eqref{eqn:qAsFtionOfXY} and \eqref{eqn:q2AsFtionOfXY}. We have
\begin{equation}\label{eqn:dq12dXY_Aux}
\begin{aligned}
& \frac{\partial q}{\partial \bar{X}} = - \frac{\omega \bar{X}}{1-\omega} \left( 1 + \frac{\omega \bar{X}^2 + \bar{Y}^2 + (1- \omega)}{\sqrt{(\omega \bar{X}^2 + \bar{Y}^2 - (1-\omega))^2 + 4 (1-\omega) \omega \bar{X}^2}} \right), \\
& \frac{\partial q}{\partial \bar{Y}} = - \frac{\bar{Y}}{1-\omega} \left( 1 + \frac{\omega \bar{X}^2 + \bar{Y}^2 - (1- \omega)}{\sqrt{(\omega \bar{X}^2 + \bar{Y}^2 - (1-\omega))^2 + 4 (1-\omega) \omega \bar{X}^2}} \right), \\
& \frac{\partial q_2}{\partial \bar{X}} = - \frac{\bar{X}}{1-\omega} \left( 1 + \frac{\bar{X}^2 + \omega \bar{Y}^2 - (1- \omega)}{\sqrt{( \bar{X}^2 + \omega \bar{Y}^2 - (1-\omega))^2 + 4 (1-\omega) \omega \bar{Y}^2}} \right),\\
& \frac{\partial q_2}{\partial \bar{Y}} = - \frac{\omega \bar{Y}}{1-\omega} \left( 1 + \frac{\bar{X}^2 + \omega \bar{Y}^2 + (1- \omega)}{\sqrt{( \bar{X}^2 + \omega \bar{Y}^2 - (1-\omega))^2 + 4 (1-\omega) \omega \bar{Y}^2}} \right).
\end{aligned}
\end{equation}

\subsection{Calculating net force exerted on a small bead}
\label{sec:SuppEvaluateNetForce}
Parameterizing the circle as $(\bar{X},\bar{Y}) = (\cos{(\varphi)}, \sin{(\varphi)})$, we get the unit normal vector $\boldsymbol{N} = (\cos{(\varphi)}, \sin{(\varphi)})$ (pointing into the material) in the undeformed configuration and we calculate the total force exerted on the bead by the material as a line integral over the circle of dimensional radius $a \tilde{D}$ and thus
\begin{gather} \nonumber
\tilde{\boldsymbol{F}}_b = \frac{ \pi \tilde{Y} \tilde{b}_{c}^2}{\tilde{R}_{c}} \int\limits_{0}^{2 \pi}
\begin{pmatrix}
  \displaystyle \mathcal{F}(\xi) + R_{b} \xi \mathcal{F}'(\xi)  \hat{x}_X^I + O(R_{b}^2) &
  \displaystyle R_{b} \mathcal{F}(\xi) \hat{x}_Y^I + O(R_{b}^2)  \\
  \displaystyle R_{b} \mathcal{F}(\xi) \hat{y}_X^I  + O(R_{b}^2)  &
  \displaystyle \mathcal{F}(\xi) + R_{b} \xi \mathcal{F}'(\xi)  \hat{y}_Y^I  + O(R_{b}^2)
\end{pmatrix} 
\begin{pmatrix}
  \displaystyle \cos{(\varphi)}  \\
  \displaystyle \sin{(\varphi)}
\end{pmatrix} 
 a \tilde{D} d \varphi = \\
 \frac{\pi \tilde{Y} \tilde{b}_{c}^2 a  R_{b} \xi \mathcal{F}'(\xi)}{\varepsilon_{c}}\int\limits_{0}^{2 \pi}
\begin{pmatrix}
  \displaystyle  \hat{x}_X^I &
  \displaystyle  \omega \hat{x}_Y^I  \\
  \displaystyle  \omega \hat{y}_X^I  &
  \displaystyle  \hat{y}_Y^I
\end{pmatrix} 
\begin{pmatrix}
  \displaystyle \cos{(\varphi)}  \\
  \displaystyle \sin{(\varphi)}
\end{pmatrix} 
 d \varphi  + O(R_{b}^2).
\end{gather}
Notice that the leading-order contributions (i.e. $O(1)$ in $R_{b}$) cancel out. We calculate the leading-order approximation for the $X-$ and $Y-$components of the net force using \eqref{eqn:hatxInner}-\eqref{eqn:hatyInner} as
\begin{equation}\label{eqn:ForcesAnalyticalTwoComponents}
\begin{aligned}
\tilde{F}_b^X =& \frac{\cos(\varphi_*) \pi \tilde{Y} \tilde{b}_{c}^2 R_{b} \xi \mathcal{F}'(\xi)}{\varepsilon_{c} \left(2 \ln{(1/a)} + \ln{(4 \omega/(1-\omega))} - 2 \cosh^{-1}{\left( (1- \omega)^{-1/2} \right)}\right)} \int\limits_0^{2 \pi}  \frac{1}{\sqrt{q^2 - q}} \left( \frac{\partial q}{\partial \bar{X}} \cos{(\varphi)} + \omega \frac{\partial q}{\partial \bar{Y}} \sin{(\varphi)} \right) d \varphi + O(a) \\
\tilde{F}_b^Y =& \frac{\sin(\varphi_*) \pi \tilde{Y} \tilde{b}_{c}^2 R_{b} \xi \mathcal{F}'(\xi)}{\varepsilon_{c} \left(2 \ln{(1/a)} + \ln{(4 \omega/(1-\omega))} - 2 \cosh^{-1}{\left( (1- \omega)^{-1/2} \right)}\right)} \int\limits_0^{2 \pi}  \frac{1}{\sqrt{q_2^2 - q_2}} \left( \omega  \frac{\partial q_2}{\partial \bar{X}} \cos{(\varphi)} + \frac{\partial q_2}{\partial \bar{Y}} \sin{(\varphi)} \right) d \varphi +O(a)
\end{aligned}
\end{equation}

\subsubsection{Evaluating the integrands at the bead}

Evaluating \eqref{eqn:qAsFtionOfXY} at the bead $(\bar{X},\bar{Y}) = (\cos{(\varphi)}, \sin{(\varphi)})$ and using $\sin^2{(\varphi)} = 1 - \cos^2{(\varphi)}$, we get
\begin{equation}\label{eqn:qAtBeadAux}
\begin{aligned}
q(\varphi) = & \frac{- \omega \cos^2{(\varphi)} - \sin^2{(\varphi)} + (1-\omega) - \sqrt{(\omega \cos^2{(\varphi)} +\sin^2{(\varphi)} - (1-\omega))^2 + 4(1-\omega)\omega \cos^2{(\varphi)}} }{2(1-\omega)} \\
& = \frac{ (1-\omega) \cos^2{(\varphi)} - \omega - \sqrt{(\omega-(1-\omega)\cos^2{(\varphi)})^2 + 4(1-\omega)\omega \cos^2{(\varphi)}} }{2(1-\omega)} = - \frac{\omega}{1-\omega}
\end{aligned}
\end{equation}
and conclude
$$ \frac{1}{\sqrt{q^2 -q}} = \frac{1-\omega}{\sqrt{\omega}}.$$
Analogously, it is easy to show that $q_2 = q$ at the bead and thus
$$ \frac{1}{\sqrt{q_2^2 -q_2}} = \frac{1-\omega}{\sqrt{\omega}}$$
Similarly, using \eqref{eqn:dq12dXY_Aux}, the same trigonometric identity and our knowledge on what the square root term in \eqref{eqn:qAtBeadAux} simplifies into, we get at the bead
$$ \frac{\partial q}{\partial \bar{X}} \cos{(\varphi)} + \omega \frac{\partial q}{\partial \bar{Y}} \sin{(\varphi)} =$$
$$ - \frac{\omega}{1-\omega} \left\{ \cos^2{(\varphi)} \left( 1 + \frac{\omega \cos^2{(\varphi)} + \sin^2{(\varphi)} + (1-\omega)}{\sqrt{(\omega \cos^2{(\varphi)} + \sin^2{(\varphi)} - (1-\omega))^2 + 4 \omega (1- \omega) \cos^2{(\varphi)} }} \right) +  \right. $$
$$ \left. \sin^2{(\varphi)} \left( 1 + \frac{\omega \cos^2{(\varphi)} + \sin^2{(\varphi)} - (1-\omega)}{\sqrt{(\omega \cos^2{(\varphi)} + \sin^2{(\varphi)} - (1-\omega))^2 + 4 \omega (1- \omega) \cos^2{(\varphi)} }} \right) \right\} = $$
$$- \frac{\omega}{1-\omega} \left\{ 1+  \frac{\cos^2{(\varphi)} (2-\omega + (\omega-1) \cos^2{(\varphi)}) + (1-\cos^2{(\varphi)})(\omega + (\omega-1) \cos^2{(\varphi)})}{\omega + (1-\omega)\cos^2{(\varphi)}}  \right\} = - \frac{2 \omega}{1-\omega}$$
and following the same steps also
$$ \omega \frac{\partial q_2}{\partial \bar{X}} \cos{(\varphi)} + \frac{\partial q_2}{\partial \bar{Y}} \sin{(\varphi)} = - \frac{2 \omega}{1-\omega}.$$
Substituting back into \eqref{eqn:ForcesAnalyticalTwoComponents} we get 
\begin{equation}\label{eqn:SmallBeadSmallDefAsymptoticsForceVectorSupp}
\tilde{\boldsymbol{F}}_b = - (\cos{(\varphi_*)}, \sin{(\varphi_*)}) \tilde{F}_b^0 + O(a),
\end{equation}
where
\begin{equation}\label{eqn:SmallBeadSmallDefAsymptotics_ForceMagnitude}
\tilde{F}_b^0 = \frac{ 2 \pi  R_{b}/\varepsilon_{c} \sqrt{\xi \mathcal{F}(\xi) \mathcal{F}'(\xi)}}{\ln{(1/a)} + \ln{(2 \sqrt{\omega/(1-\omega)})} - \cosh^{-1}((1-\omega)^{-\frac{1}{2}})}  \pi \tilde{Y} \tilde{b}_{c}^2.
\end{equation}
Using the constitutive law \eqref{eqn:LinearSpringsDimlessSmallTau1NoCompression} and simplifying, the leading-order dimensionless net force can be written as
\begin{equation}\label{eqn:magnitudeOfNetForceLinearSprings2}
\begin{aligned}
F_b^0 & = \frac{ 2 \pi  R_{b}/\varepsilon_{c} \sqrt{\mathcal{F}_p \left( 1 + \mathcal{F}_p \right)}}{\ln{(2 \sqrt{\mathcal{F}_p}/a)} - \cosh^{-1}(\sqrt{1+ \mathcal{F}_p})} = \frac{ 2 \pi  R_{b}/\varepsilon_{c} \sqrt{\mathcal{F}_p \left( 1 + \mathcal{F}_p \right)}}{\ln{(2 \sqrt{\mathcal{F}_p}/a)} - \ln{(\sqrt{1+\mathcal{F}_p}+\sqrt{\mathcal{F}_p})}} \\
& \frac{ 2 \pi  R_{b}/\varepsilon_{c} \sqrt{\mathcal{F}_p \left( 1 + \mathcal{F}_p \right)}}{\displaystyle \ln{\left(\frac{2 \sqrt{\mathcal{F}_p}}{a \left( \sqrt{1+\mathcal{F}_p}+\sqrt{\mathcal{F}_p}\right)}\right)}} = \frac{ 2 \pi  R_{b}/\varepsilon_{c} \sqrt{\mathcal{F}_p \left( 1 + \mathcal{F}_p \right)}}{\displaystyle \ln{\left(\frac{2 \left( \sqrt{\mathcal{F}_p(1+\mathcal{F}_p)}-\mathcal{F}_p\right) }{a}\right)}},
\end{aligned}
\end{equation}
which gives \eqref{eqn:SmallBeadSmallDefAsymptoticsForceVector}-\eqref{eqn:magnitudeOfNetForceLinearSprings}.

\bibliography{ReferencesSupplementary} 
\bibliographystyle{plain} 

\makeatletter\@input{xx.tex}\makeatother